\documentclass[useAMS,usenatbib]{mn2e} 

\usepackage{natbib}

\usepackage{latexsym,graphicx}
\usepackage{color}
\usepackage{fixltx2e}
\usepackage{verbatim}
\usepackage{float}
\usepackage{amsmath,amssymb}
\usepackage{times}
\usepackage{soul}

\usepackage{setspace}

\newcommand\hii{H\,{\sc ii} \,}

\def\apgt{\ {\raise-.5ex\hbox{$\buildrel>\over\sim$}}\ }
\def\aplt{\ {\raise-.5ex\hbox{$\buildrel<\over\sim$}}\ }

\newcommand{\degree}{\ensuremath{^\circ}}

\title[On the observability of bow shocks of Galactic runaway OB stars]{On the observability of bow shocks of Galactic runaway OB stars}

\author[D. M.-A.~Meyer et al.]
       {D. M.-A.~Meyer,$^{1}$\thanks{E-mail: dominique.meyer@uni-tuebingen.de} A.-J.~van~Marle,$^{2,3}$ R.~Kuiper$^{1}$ and W.~Kley$^{1}$  
       \\
       $^{1}$Institut f\" ur Astronomie und Astrophysik, Universit\" at T\" ubingen,  Auf der Morgenstelle 10, 72076 T\" ubingen, Germany \\
       $^{2}$KU Leuven, Centre for mathematical Plasma Astrophysics, Celestijnenlaan 200B, B-3001 Leuven, Belgium \\       
       $^{3}$Laboratoire AstroParticule et Cosmologie - Universit\' e Paris 7 Diderot - 10 rue Alice Domon et L\' eonie Duquet, Paris, 75013, France \\              
       }

\begin{document}

\date{Received January 18  2015; accepted Month day, 2015}

\maketitle

\label{firstpage}

\begin{abstract}
Massive stars \textcolor{black}{that have been} ejected from their parent cluster and supersonically sailing 
away through the interstellar medium (ISM) are classified as 
exiled. They generate circumstellar bow shock nebulae that can be observed. 
We present two-dimensional, axisymmetric hydrodynamical simulations of 
a representative sample of stellar wind bow shocks from Galactic OB stars 
in an ambient medium of densities ranging from $n_{\rm ISM }=0.01$ up to 
$10.0\, \rm cm^{-3}$. 
Independently of their location in the Galaxy, we confirm that the infrared is the 
most appropriated waveband to search for bow shocks from massive stars. 
Their spectral energy distribution is the convenient tool to analyze them 
since their emission does not depend on the temporary effects which could 
affect unstable, thin-shelled bow shocks.  
Our numerical models of Galactic bow shocks generated by high-mass ($\approx 40\, \rm M_{\odot}$) 
runaway stars yield H$\alpha$ fluxes which could be observed by facilities such as the 
{\it SuperCOSMOS H-Alpha Survey}. The brightest bow shock nebulae are produced in the 
denser regions of the ISM. 
We predict that bow shocks {\it in the field} observed at H$\alpha$ by means of 
Rayleigh-sensitive facilities are formed around stars of initial mass larger than 
about $20\, \rm M_{\odot}$. 
\textcolor{black}{
Our models of bow shocks from OB stars have the emission maximum in the wavelength range 
$3\, \leq\, \lambda\, \leq\, 50\, \mu \rm m$ which can be up to several orders of magnitude 
brighter than the runaway stars themselves, particularly for stars of initial mass larger 
than $20\, \rm M_{\odot}$.
}
%
%
\end{abstract}



\section{Introduction}
\label{sect:introduction}

The estimate of massive star feedback is a crucial question in the 
understanding of the Galaxy's 
functioning~\citep{langer_araa_50_2012}. Throughout their 
short lives, they release strong winds~\citep{holzer_araa_8_1970} and ionising 
radiation~\citep{diazmiller_apj_501_1998} which modify their ambient medium. This 
results in diaphanous \hii regions~\citep{dyson_ass_35_1975}, parsec-scale bubbles 
of stellar wind~\citep{weaver_apj_218_1977}, 
inflated~\citep{petrovic_aa_450_2006} or 
shed~\citep{woosley_rvmp_74_2002,garciasegura_1996_aa_305} stellar envelopes 
that impact their close surroundings and which can alter the propagation of 
their subsequent supernova shock 
wave~\citep{vanveelen_phd,meyer_mnras_450_2015}. Understanding the formation 
processes of these circumstellar structures allows us to constrain the impact of 
massive stars, e.g. on the energetics or the chemical evolution of the 
interstellar medium (ISM). Moreover, it links studies devoted to the dynamical 
evolution of supernova remnants expanding into the \textcolor{black}{perturbed}  
ISM~\citep{rozyczka_mnras_261_1993} with works focusing on the physics of the 
star forming ISM~\citep{peters_apj_711_2010}. 

\textcolor{black}{
While bow-shock-like structures can develop around any astrophysical object that moves 
supersonically with respect to its ambient medium~\citep[see e.g.][]{2016arXiv160107799T}, 
it particularly affects the surroundings of bright stars running through the ISM~\citep{blaauw_bain_15_1961}. 
}
These arc-like structures of 
swept-up stellar wind material and ISM gas \textcolor{black}{are the result of} the distortion of their 
stellar wind bubble by the bulk motion of their central 
star~\citep{weaver_apj_218_1977}. 
Their size and their morphology are governed by 
their stellar wind mass loss, \textcolor{black}{the wind velocity,} the bulk motion of the runaway star and 
their local ambient medium properties~\citep{comeron_aa_338_1998}. These 
distorted wind bubbles have been first noticed in optical [O{\sc iii}] $\lambda 
\, 5007$ spectral emission line around the Earth's closest runaway star,   
the OB star $\zeta$ Ophiuchi~\citep{gull_apj_230_1979}. Other noticeable  
fast-moving massive stars producing a stellar wind bow shock are, e.g.  the blue 
supergiant Vela-X1~\citep{kaper_apj_475_1997}, the red supergiant 
Betelgeuse~\citep{noriegacrespo_aj_114_1997} and the very massive star 
BD+43$\degree$365 running away from Cygnus OB2~\citep{comeron_aa_467_2007}.

\begin{table*}
	\centering
	\caption{
	\textcolor{black}{
	 Mass $M_{\star}$ (in $\rm M_{\odot}$), luminosity $L_{\star}$ 
	 (in $L_{\odot}$), mass loss $\dot{M}$ (in $\rm M_{\odot}\, \rm yr^{-1}$) and wind 
	 velocity $v_{\rm w}$ (in $\rm km\, \rm s^{-1}$) of our runaway stars at the beginning 
	 of the simulations, at a time $t_{\rm start}$ ($\rm Myr$) after the zero-age main-sequence time. 
	 $T_{\rm eff}$ (in $\rm K$) is the effective temperature of the star at 
	 $t_{\rm start}$. The number of ionizing photons released per unit time $S_{\star}$ 
	 (in $\rm{photon}\, \rm{s}^{-1}$) is taken from~\citet{diazmiller_apj_501_1998}.  
	 Finally, $t_{\rm MS}$ is the main-sequence timescale of the star (in $\rm{Myr}$).
	 }   
	 }
	\begin{tabular}{cccccccc}
	\hline
	\hline
	$M_{\star}\, (\rm M_{\odot})$  &  $t_{\mathrm{ start}}\, (\rm Myr)$
				   &   $\textcolor{black}{\log(L_{\star}/\rm L_{\odot})}$                              
			           &   $\log(\dot{M}/\rm M_{\odot}\, \rm yr^{-1})$
			           &   $v_{\rm w}\, (\mathrm{km}\, \mathrm{s}^{-1})$			           
			           &   $T_{\rm eff}\, (\mathrm{K})$
				   &   $S_{\star} (\mathrm{photon}\, \mathrm{s}^{-1})$
				   &   $t_{\rm MS} (\mathrm{Myr})$
			\\ \hline   
	$10$ &  $5.0$  &  $3.80$ 		    &  $-9.52$   & $1082$ & $25200$    & $10^{45}$    & $22.5$  \\        
	$20$ &  $3.0$  &  $4.74$                    &  $-7.38$   & $1167$ & $33900$    & $10^{48}$    & $\,\,\,8.0$  \\        
  	$40$ &  $0.0$  &  $5.34$  	            &  $-6.29$   & $1451$ & $42500$    & $10^{49}$    & $\,\,\,4.0$  \\ 
	\hline 
	\end{tabular}
\label{tab:lum_stars}
\end{table*}

Analysis of data from the {\it Infrared Astronomical Satellite} 
facility~\citep[{\it IRAS},][]{neugebauer_278_apj_1984} later extended to 
measures taken with the {\it Wide-Field Infrared Satellite Explorer}~\citep[{\it 
WISE},][]{wright_aj_140_2010} led to the compilation of bow shock 
records, see e.g.~\citep{buren_apj_329_1988}. Soon arose the speculation that those isolated 
nebulae can serve \textcolor{black}{as a tool independent on assumptions regarding 
the internal} physics of these stars, to constrain the still highly 
debated mass loss of massive stars~\citep{gull_apj_230_1979} and/or their 
ambient medium density~\citep{huthoff_aa_383_2002}. This also raised 
questions related to the ejection mechanisms of OB stars from young 
stellar clusters~\citep{hoogerwerf_aa_365_2001}. 
More recently, multi-wavelengths data led to the publication of the E-BOSS 
catalog of stellar wind bow shocks~\citep{peri_aa_538_2012,2015arXiv150404264P}.

Early simulations discussed the general morphology of the 
bow shocks around OB stars~\citep[][, and references therein]{brighenti_mnras_277_1995}, their 
(in)stability~\citep{blondin_na_57_1998} and the general uncompatibility of the 
shape of stellar wind bow shocks with analytical approximations such as the one 
of~\citet{wilkin_459_apj_1996}, see in~\citet{comeron_aa_338_1998}. However, 
observing massive star bow shocks remains difficult and they are mostly 
serendipitously noticed in infrared observations of the neighbourhood of stellar 
clusters~\citep{gvaramadze_aa_490_2008}. Moreover, their optical emission may be 
screened by the \hii region which surrounds the driving star and this may affect 
their H$\alpha$ observations~\citep{brown_aa_439_2005}. We are particularly 
interested in the prediction of the easiest bow shocks to 
observe, their optical emission properties and their location in the Galaxy. 

In the present study, we extend our numerical investigation of the 
circumstellar medium of runaway massive stars~\citep[][hereafter Paper~I]{meyer}. 
\textcolor{black}{Note that our approach is primarily focussed on exploring the 
parameter space, rather than a series of simulations tailored to a specific  
bow shock nebula. Our parameter study} explores 
the effects of the ambient medium density on the emission properties of the bow-like nebulae 
around the most common runaway stars, in the spirit of works on bow shocks generated by low-mass 
stars~\citep[][and references therein]{villaver_apj_748_2012}. \textcolor{black}{The underlying 
assumptions are the same as in our previous purely hydrodynamical models, i.e. we neglect the 
magnetisation of the ISM, ignore any intrinsic inhomogeneity in the ISM density 
field and consider that both the wind and the ISM gas are a perfect gas. 
%
%
}

Our paper is organised as follows. In Section~\ref{sect:method} we present the 
numerical methods and the microphysics that is included in our models. The 
resulting numerical simulations are presented and discussed in 
Section~\ref{sect:results}. We then analyze and discuss the emission properties 
of our bow shock models in Section~\ref{sect:emission}. Finally, we formulate 
our conclusions in Section~\ref{section:cc}.


\section{Method}
\label{sect:method}


\subsection{Governing equations}
\label{subsect:goveq}

\textcolor{black}{
The bow shock problem is described in our work by the Euler equations of 
hydrodynamics. It is a set of equations \textcolor{black}{for mass conservation}, 
\begin{equation}
	   \frac{\partial \rho}{\partial t}  + 
	   \bmath{\nabla}  \cdot (\rho\bmath{v}) =   0,
\label{eq:euler1}
\end{equation}
\textcolor{black}{conservation of linear momentum},
\begin{equation}
	   \frac{\partial \rho \bmath{v} }{\partial t}  + 
           \bmath{\nabla} \cdot ( \bmath{v} \otimes \rho \bmath{v}) 	      + 
           \bmath{\nabla}p 			      =   \bmath{0},
\label{eq:euler2}
\end{equation}
and \textcolor{black}{conservation of energy},
\begin{equation}
	  \frac{\partial E }{\partial t}   + 
	  \bmath{\nabla} \cdot(E\bmath{v})   +
	  \bmath{\nabla} \cdot (p \bmath{v})   =	   
	  \itl{\Phi}(T,\rho) +
	  \bmath{\nabla} \cdot \bmath{{F}_{\rm c}},
\label{eq:euler3}
\end{equation}
where,
\begin{equation}
	E = \frac{p}{(\gamma - 1)} + \frac{\rho v^{2}}{2},
\label{eq:energy}
\end{equation}
and where $\rho$ is the mass density of the gas, $p$ its pressure, 
$\bmath{v}$ the vector velocity. The temperature of the gas is given by,
\begin{equation}
	T =  \mu \frac{ m_{\mathrm{H}} }{ k_{\rm{B}} } \frac{p}{\rho},
\label{eq:temperature}
\end{equation}
where $k_{\rm B}$ is the Boltzmann constant and $\mu$ is the mean molecular 
weight, such that $\rho=\mu n m_{\rm H}$ with $n$ the total number density of the 
fluid and $m_{\rm H}$ the mass of \textcolor{black}{a hydrogen} atom. The adiabatic index of 
the gas is $\gamma=5/3$.  
The Eq.~(\ref{eq:euler3}) includes (i) the effects of both the cooling and 
the heating of the gas by optically-thin radiative processes and 
(ii) the transport of heat by electronic thermal conduction (see 
Section~\ref{subsect:phys}). 
}

\subsection{Hydrodynamical simulations}
\label{subsect:hydrosim}

We run two-dimensional, axisymmetric, hydrodynamical numerical simulations using the {\sc 
pluto} code~\citep{mignone_apj_170_2007, migmone_apjs_198_2012} in axisymmetric, 
cylindrical coordinates on a uniform grid $[z_{\rm min},z_{\rm 
max}]\times[O,R_{\rm max}]$ of spatial resolution $\Delta=2.25\times 10^{-4}\, 
\rm{pc}\, \rm{cell}^{-1}$ minimum. The stellar wind is injected into the 
computational domain filling a circle of radius 20 cells centered onto the origin 
$O$~\citep[see e.g.,][and references 
therein]{comeron_aa_338_1998,meyer_mnras_2013}. The interaction with the ISM is 
calculated in the reference frame of the moving 
star~\citep{vanmarle_aa_469_2007,vanmarle_apj_734_2011,vanmarle_aa_561_2014}. 
Inflowing ISM gas mimicing the stellar motion is set at the $z=z_{\rm max}$ 
boundary whereas semi-permeable boundary conditions are set at $z=z_{\rm min}$ and at 
$R=R_{\rm max}$. Wind material is distinguished from the ISM using a passive 
tracer $Q$ that is advected with the gas and initially set to $Q=1$ in the 
stellar wind and to $Q=0$ in the ISM. 
The ISM composition is assumed to be solar~\citep{asplund_araa_47_2009}.

\textcolor{black}{
The stellar models are calculated using the stellar evolution code 
of~\citet{heger_apj_626_2005,yoon_443_aa_2005,petrovic_aa_435_2005} also 
described in~\citet{brott_aa_530_2011a}. It includes the mass loss recipe 
of~\citet{kudritzki_aa_219_1989} for the main sequence-phase and 
of~\citet{dejager_aas_72_1988} for the red supergiant phase. 
%
The stellar properties are displayed in fig.~3 of Paper~I while we summarize the wind 
properties at the beginning of our simulations in our Table~\ref{tab:lum_stars}. 
%
}

\begin{table}
	\centering
	\caption{The hydrodynamical models.   
	 Parameters $M_{\star}$ (in $\rm M_{\odot}$), $v_{\star}$ (in $\mathrm{km}\, \mathrm{s}^{-1}$) 
	 and $n_{\rm ISM}$ (in $\mathrm{cm}^{-3}$) are the initial mass of the considered moving star, 
	 its space velocity and its local ISM density, respectively.   
	 }
	\begin{tabular}{lccccr}
	\hline
	\hline
	${\rm {Model}}$ &   $M_{\star}\, (\rm M_{\odot})$                              
 			&   $v_{\star}\, (\mathrm{km}\, \mathrm{s}^{-1})$
			&   $n_{\rm ISM}\, (\mathrm{cm}^{-3})$   
			\\ \hline   
	MS1020n0.01   &  $10$  &  $20$  &  $0.01$     \\        
	MS1040n0.01   &  $10$  &  $40$  &  $0.01$     \\        
	MS1070n0.01   &  $10$  &  $70$  &  $0.01$     \\ 
	MS2040n0.01   &  $20$  &  $40$  &  $0.01$     \\        
	MS2070n0.01   &  $20$  &  $70$  &  $0.01$     \\ 
	\hline
	MS1020n0.1    &  $10$  &  $20$  &  $0.10$     \\             
	MS1040n0.1    &  $10$  &  $40$  &  $0.10$     \\        
	MS1070n0.1    &  $10$  &  $70$  &  $0.10$     \\ 
	MS2020n0.1    &  $20$  &  $20$  &  $0.10$     \\             
	MS2040n0.1    &  $20$  &  $40$  &  $0.10$     \\        
	MS2070n0.1    &  $20$  &  $70$  &  $0.10$     \\  
	MS4070n0.1    &  $40$  &  $70$  &  $0.10$     \\	
	\hline
	MS1020n10     &  $10$  &  $20$  &  $10.0$     \\             
	MS1040n10     &  $10$  &  $40$  &  $10.0$     \\        
	MS1070n10     &  $10$  &  $70$  &  $10.0$     \\ 
	MS2020n10     &  $20$  &  $20$  &  $10.0$     \\             
	MS2040n10     &  $20$  &  $40$  &  $10.0$     \\        
	MS2070n10     &  $20$  &  $70$  &  $10.0$     \\  
	MS4020n10     &  $40$  &  $20$  &  $10.0$    \\         
	MS4040n10     &  $40$  &  $40$  &  $10.0$     \\     
	MS4070n10     &  $40$  &  $70$  &  $10.0$     \\
	\hline 
	\end{tabular}
\label{tab:models}
\end{table}

\subsection{Microphysics}
\label{subsect:phys}

In order to proceed on our previous bow shock studies~\citep[Paper~I,][]{meyer_mnras_450_2015},
we include the same microphysics in our simulations of the circumstellar medium of
runaway, massive stars, i.e. we take into account losses and gain of internal energy
by optically-thin cooling and heating together with electronic thermal
conduction. Optically-thin radiative processes are included into the
model using the cooling and heating laws established for a fully ionized medium 
in Paper~I. \textcolor{black}{They are the right-hand term $\itl{\Phi}(T,\rho)$ of 
Eq.~(\ref{eq:euler3}) which is given by,
\begin{equation}  
	 \itl \Phi(T,\rho)  =  n_{\mathrm{H}}\itl{\Gamma}(T)   
		   		 -  n^{2}_{\mathrm{H}}\itl{\Lambda}(T),
\label{eq:dissipation}
\end{equation}
where $\itl{\Gamma}(T)$ and $\itl{\Lambda}(T)$ are the heating and cooling laws, respectively, 
and $n_{\mathrm{H}}$ is the hydrogen number density.} 
It mainly consist of cooling contributions from hydrogen and
helium for temperatures $T<10^{6}\, \rm K$ whereas it is principally due to metals 
for temperatures $T \ge 10^{6}\, \rm K$~\citep{wiersma_mnras_393_2009}.  
A term representing the cooling from collisionally excited forbidden 
lines~\citep{henney_mnras_398_2009} incorporates the effects of, among other, the
[O{\sc iii}] $\lambda \, 5007$ line emission. The heating contribution includes the
reionisation of recombining hydrogen atoms by the starlight~\citep{osterbrock_1989,hummer_mnras_268_1994}. 
All our models include electronic thermal conduction~\citep{cowie_apj_211_1977}. 
\textcolor{black}{The divergence term in Eq.~(\ref{eq:euler3}) represents the flux of heat, 
\begin{equation}
      \bmath{{F}_{\rm c}} = \kappa  \bmath{ \nabla } T,
\label{eq:tc}
\end{equation}
where $\kappa$ is the heat conduction coefficient~\citep{spitzer_1962}. More details 
about thermal conduction is also given in Paper~I.}

\begin{figure}
	\centering
	\begin{minipage}[b]{0.41\textwidth}
		\includegraphics[width=1.0\textwidth]{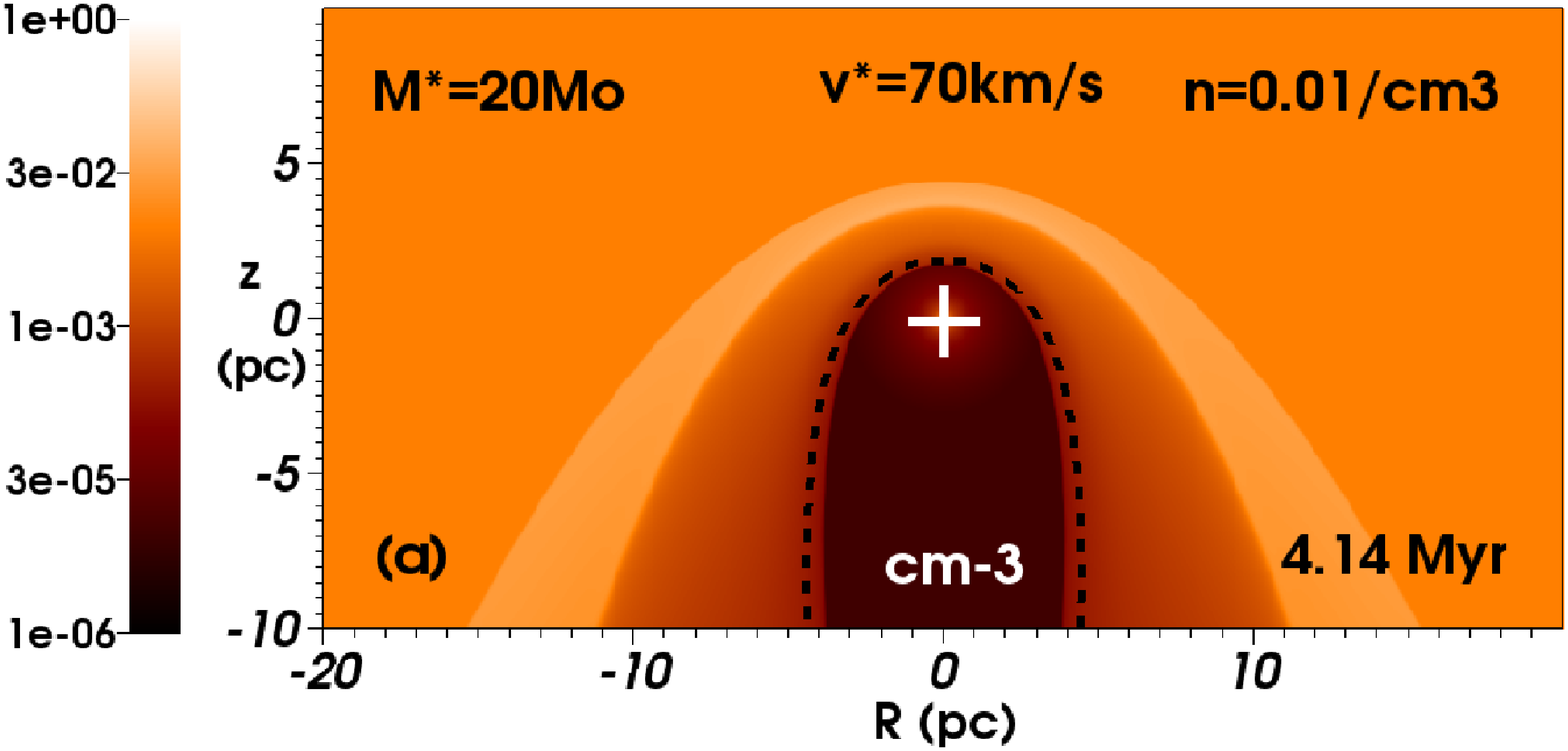}
	\end{minipage} \\
	\begin{minipage}[b]{0.41\textwidth}
		\includegraphics[width=1.0\textwidth]{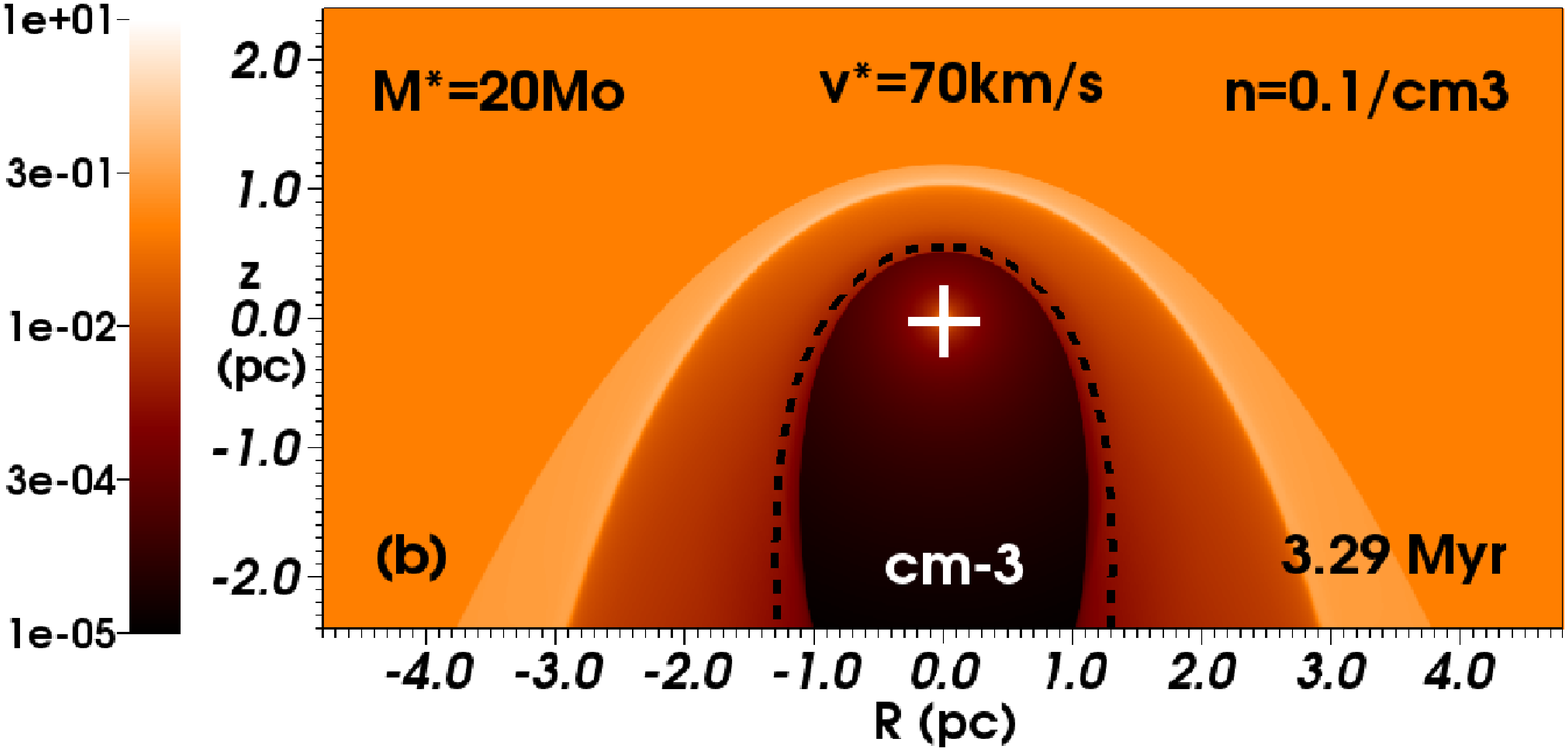}
	\end{minipage} \\
	\begin{minipage}[b]{0.41\textwidth}
		\includegraphics[width=1.0\textwidth]{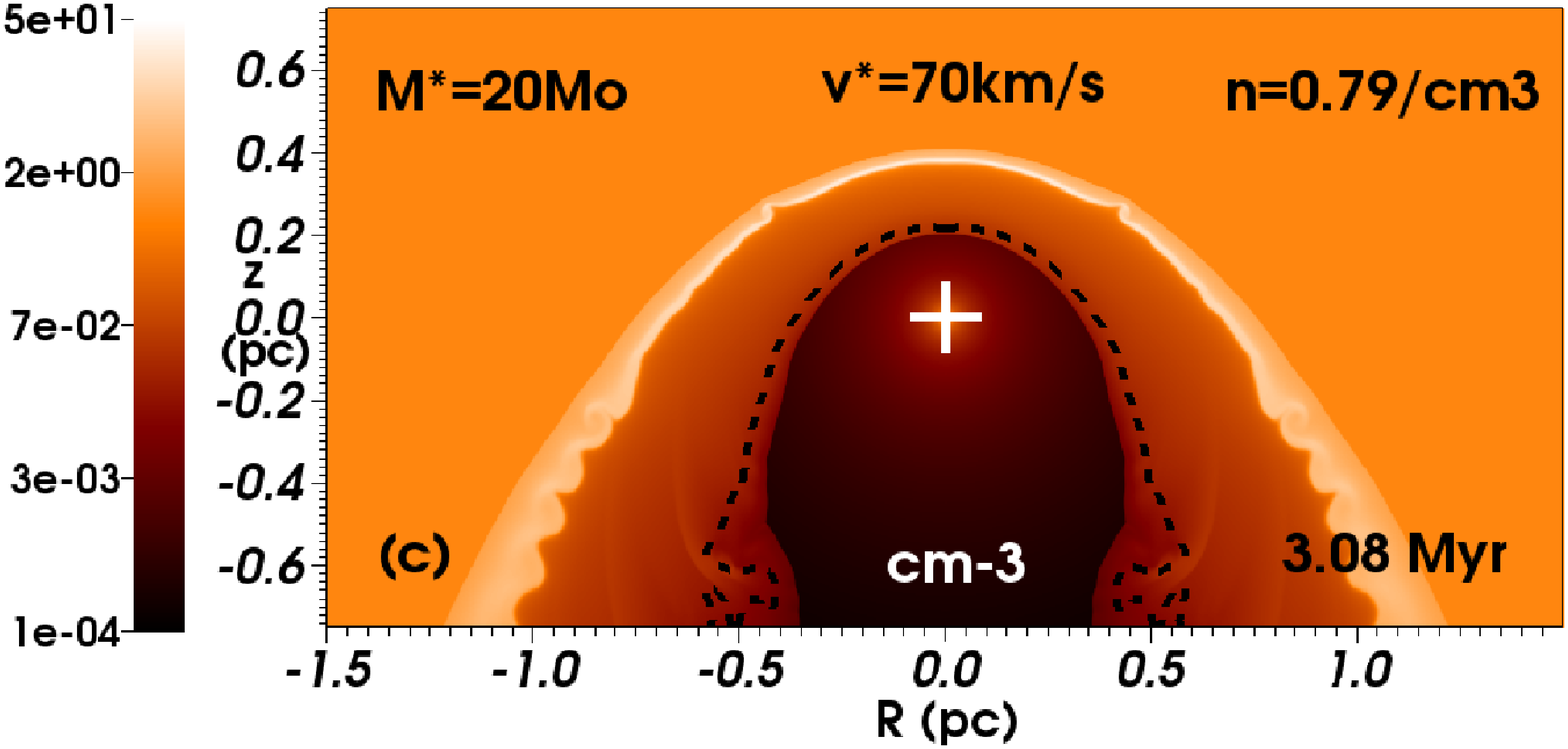}
	\end{minipage} \\
	\begin{minipage}[b]{0.41\textwidth}
		\includegraphics[width=1.0\textwidth]{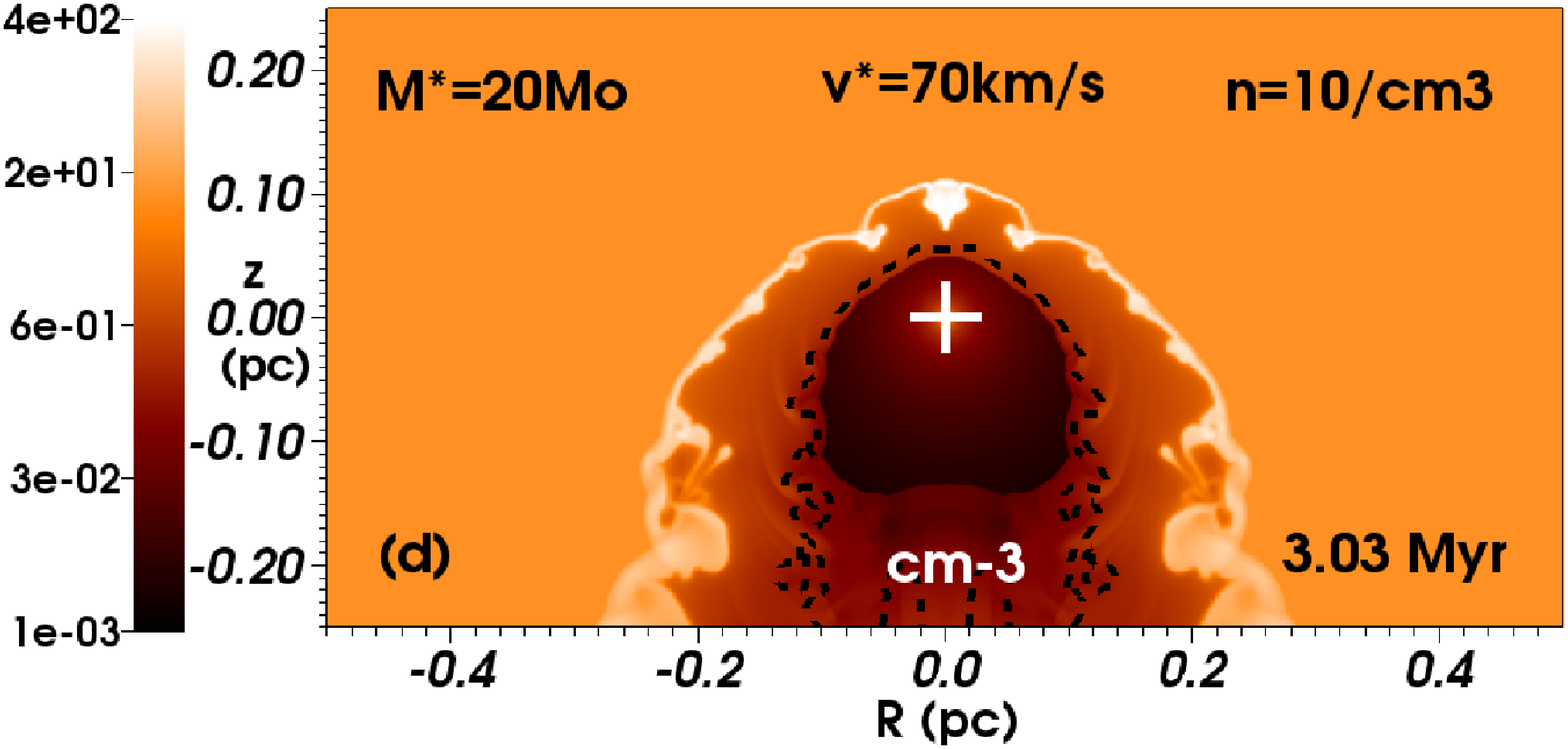}
	\end{minipage}	
	\caption{ 
	         Stellar wind bow shocks from the main sequence phase of
the $20\,  \rm M_{\odot}$ \textcolor{black}{ZAMS} star moving with velocity $70\, \mathrm{km}\,
\mathrm{s}^{-1}$ as a function of the ISM density, with $n_{\rm ISM}=0.01$ (a), 
$0.1$ (b), $0.79$ (c) and $10.0\, \mathrm{cm}^{-3}$ (d). 
The gas number density (in $\rm cm^{-3}$) is shown in the logarithmic 
scale. The dashed black contour traces the boundary between
wind and ISM material. The cross indicates the position 
of the runaway star. The $R$-axis represents the radial
direction and the $z$-axis the direction of stellar motion (in $\mathrm{pc}$). 
Only part of the computational domain is shown. 
		 }
	\label{fig:grid_density}  
\end{figure}

\subsection{Parameter range}
\label{subsect:para}

This work consists of a parameter study extending our previous 
investigation of stellar wind bow shock (Paper~I) to regions of the Galaxy where the 
ISM has either lower or higher densities. 
\textcolor{black}{
We perform re-runs of the models in Paper~I that correspond to the main-sequence phase 
of our stars, but assume a different ISM background density and only consider the cases 
where the main-sequence lifetime of the stars is larger than the four crossing times of 
the gas $|z_{\rm max}-z_{\rm min}|/v_{\star}$ through the whole computational domain 
which are necessary to model steady state bow shocks. 
}
The boundary conditions are unchanged, i.e. we consider runaway 
stars of $10$, $20$ and $40\, \rm M_{\odot}$ \textcolor{black}{zero-age main-sequence 
(ZAMS)} star moving with velocity $v_{\star}=20$, 
$40$ and $70\, \rm km\, \rm s^{-1}$, respectively. Differences come from the 
chosen ISM number density that ranges from $n_{\rm ISM}=0.01$ to $10.0\, \rm 
cm^{-3}$ whereas our preceeding work exclusively focused on bow 
shocks models with $n_{\rm ISM}=0.79\, \rm cm^{-3}$.
\textcolor{black}{
The simulation labels are summarised in Table~\ref{tab:models}. The analysis of 
our simulations include the main-sequence models with $n_{\rm ISM}=0.79\, \rm cm^{-3}$ 
of Paper~I and we refer to them using their original labels. 
}


\section{Bow shocks morphology}
\label{sect:results}


\subsection{Bow shocks structure}
\label{subsect:structure}

In Fig.~\ref{fig:grid_density} we show the density fields in our hydrodynamical 
simulations of our $20\, \rm M_{\odot}$  \textcolor{black}{ZAMS} star moving with velocity $v_{\star}=70\, \rm 
km\, \rm s^{-1}$ in a medium of number density $n_{\rm ISM}=0.01$ (panel a, model 
MS2070n0.01), $0.1$ (panel b, model MS2070n0.1), $0.79$ (panel c, model 
MS2070) and $10.0\, \rm cm^{-3}$ (panel d, model MS2070n10), respectively. 
\textcolor{black}{
The figures correspond to times about $4.14$, $3.29$, $3.08$ and $3.03\, \rm 
Myr$ after the beginning of the main-sequence phase. 
Our bow shocks have the typical structure of a circumstellar nebulae generated 
by a fast-moving OB star undergoing the conjugated effects of both cooling and 
heating by optically-thin radiative processes and thermal conduction and the 
thickness of the shock layer depends on whether the shocks are adiabatic or 
radiative, which in their turn depends on the shock conditions, 
see~\citet{comeron_aa_338_1998}, Paper~I and the references therein. 
All our bow shock simulations have such a structure. 
}

\subsection{Bow shocks size}
\label{subsect:scaling}

The bow shocks have a stand-off distance $R(0)$, i.e. the distance separating 
them from the star along the direction of motion predicted by~\citet{wilkin_459_apj_1996}. 
It decreases as a function of (i) $v_{\star}$, 
(ii) $\dot{M}$ (c.f. Paper~I) and (iii)  $n_{\rm ISM}$ 
since $R(0) \propto n_{\rm ISM}^{-1/2}$. A dense ambient medium produces a large 
ISM ram pressure $n_{\rm ISM}v_{\star}^{2}$ which results in a compression of the whole bow 
shock and consequently in a reduction of $R(0)$. As an example, our simulations 
involving a $20\, \rm M_{\odot}$  \textcolor{black}{ZAMS} star with $v_{\star}=70\, 
\mathrm{km}\, \mathrm{s}^{-1}$ has $R(0)\approx 3.80$, $1.14$, $0.38$ and 
$0.07\rm pc$ when the driving star moves in $n_{\rm ISM}=0.01$, $0.1$, $0.79$ and 
$10\, \rm cm^{-3}$, respectively (Fig.~\ref{fig:grid_density}a-d), which is reasonably 
in accordance with~\citet{wilkin_459_apj_1996}. All our measures of $R(0)$ are taken at the 
contact discontinuity, because it is appropriate measure to compare models with 
Wilkin's analytical solution~\citep{mohamed_aa_541_2012}. 

\begin{figure}
	\centering
	\begin{minipage}[b]{ 0.45\textwidth}
		\includegraphics[width=1.0\textwidth]{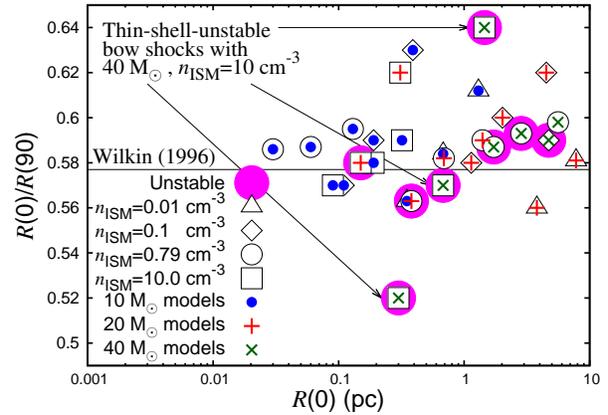}
	\end{minipage}	
	\caption{ 
	         Axis ratio $R(0)/R(90)$ of our bow shock models.   
	         The figure shows the ratio $R(0)/R(90)$ measured in the density field of our 
	         models measured at their contact discontinuity, as a function of their stand-off 
	         distance $R(0)$ (in $\rm pc$).  
		 Symbols distinguish models as a function of (i) the ISM ambient medium with 
		 $n_{\rm ISM}=0.01$ (triangles), $0.1$ (diamonds), $0.79$ (circles) and $10.0\, \rm cm^{-3}$ 
		 (squares) and (ii) of the initial mass of the star with $10\, \rm M_{\odot}$ (blue dots), 
		 $20\, \rm M_{\odot}$ (blue plus signs) and $40\, \rm M_{\odot}$ (dark green crosses), respectively.
		 The thin horizontal black line corresponds to the analytic solution 
		 $R(0)/R(90)= 1/\sqrt{3}\approx 0.58$ of~\citet{wilkin_459_apj_1996}. 
		 \textcolor{black}{The large purple dots highlight the unstable bow shocks. }
		 }	
	\label{fig:axis_ratio}  
\end{figure}

\textcolor{black}{
In Fig.~\ref{fig:axis_ratio} we plot the ratio $R(0)/R(90)$  as a function of 
the stand-off distance of the bow shocks, where $R(90)$ is the distance between the 
star and the bow shock measured along the direction perpendicular to the direction 
of motion. 
%
The internal structure of the bow shocks depends on the stellar wind and the bulk 
motion of the star (Paper~I) but are also sensible to the ISM density. 
Important variations in the stars' ambient local medium can produce 
large compression of the region of shocked ISM gas, which in its turn decreases up to 
form a thin layer of cold shocked ISM gas~\citep[see Fig.~\ref{fig:grid_density}b and fig.~1 
of][]{comeron_aa_338_1998}.
This phenomenon typically arises in simulations combining moving stars with 
strong mass loss such as our \textcolor{black}{$40\, \rm M_{\odot}$ ZAMS star}, together with 
velocity $v_{\star} \ge 40\, \rm km\, \rm s^{-1}$ and/or $n_{\rm ISM} \ge 1\, 
\rm cm^{-3}$ (see large magenta dots in Fig.~\ref{fig:axis_ratio}). Those thin-shells are more prone 
to develop non-linear instabilities~\citep{vishniac_apj_428_1994, garciasegura_1996_aa_305}.
%
}

\textcolor{black}{ 
Interestingly, analytic approximations of the overall shape of a bow shock often assume 
such an infinitely thin structure~\citep{wilkin_459_apj_1996} and predict 
that $R(0)/R(90)=1/\sqrt{3}\approx0.58$. Thin shells are very unstable and experience periodical large distortions 
which can make the shape of the bow shock inconsistent with Wilkin's prediction of 
$R(0)/R(90)$ (see black arrows in Fig.~\ref{fig:axis_ratio}). 
Most of the models are within $20$ per 
cent of Wilkin's solution (see horizontal black line at $R(0)/R(90) \approx 
1/\sqrt{3}$). Only a few models have $R(0)/R(90) \le 1/\sqrt{3}$ because their 
opening is smaller than predicted. However, some simulations with $v_{\star} = 
20\, \rm km\, \rm s^{-1}$ have large and spread bow shocks in which $R(0)/R(90) \geq 0.62$, 
see e.g. our simulation MS2020n0.1 with $R(90)\approx 4.51\, \rm pc$. 
}

\subsection{Non-linear instabilities and mixing of material}
\label{subsect:stability}

\begin{figure}
	\centering
	\begin{minipage}[b]{0.43\textwidth}
		\includegraphics[width=1.0\textwidth]{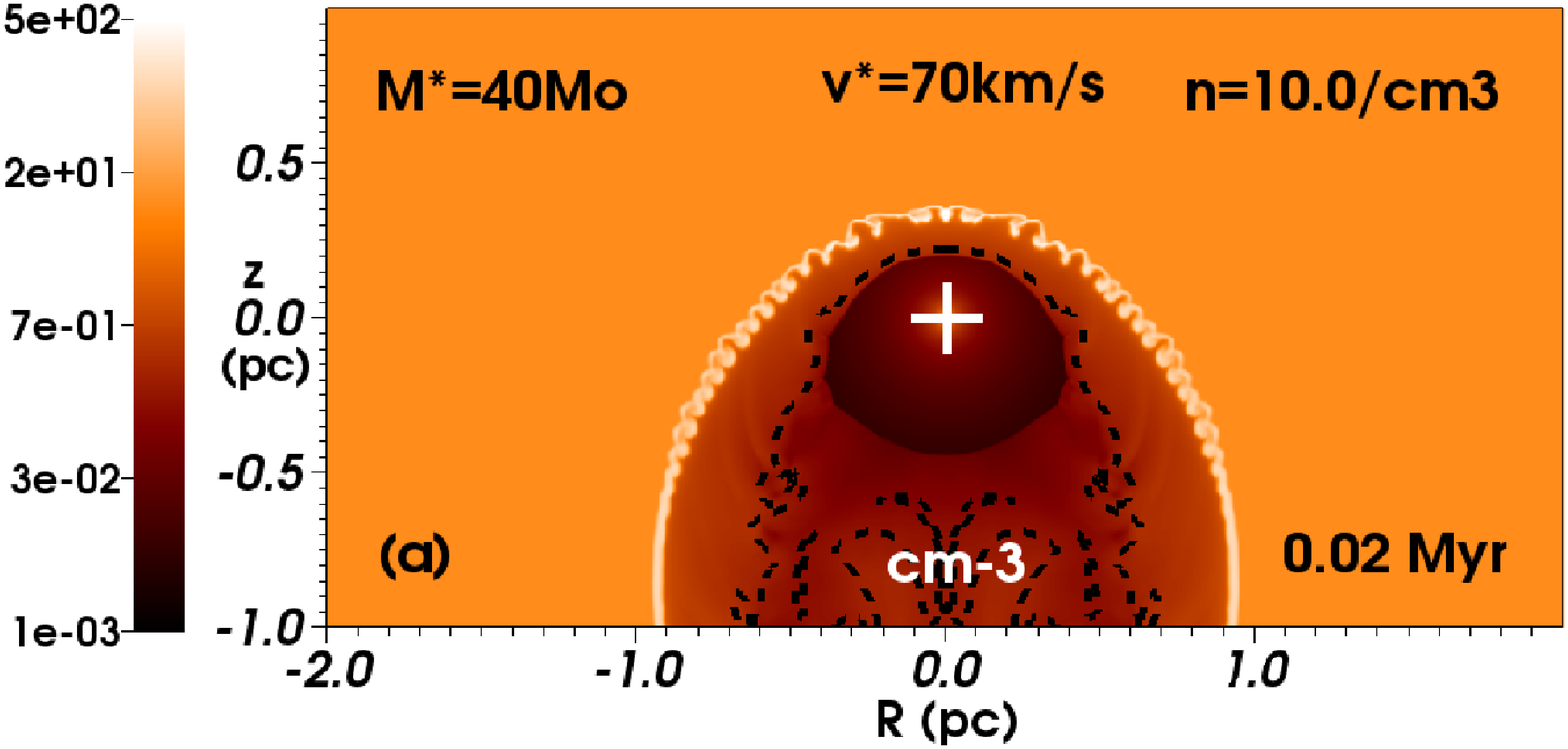}
	\end{minipage} 	\\
	\begin{minipage}[b]{0.43\textwidth}
		\includegraphics[width=1.0\textwidth]{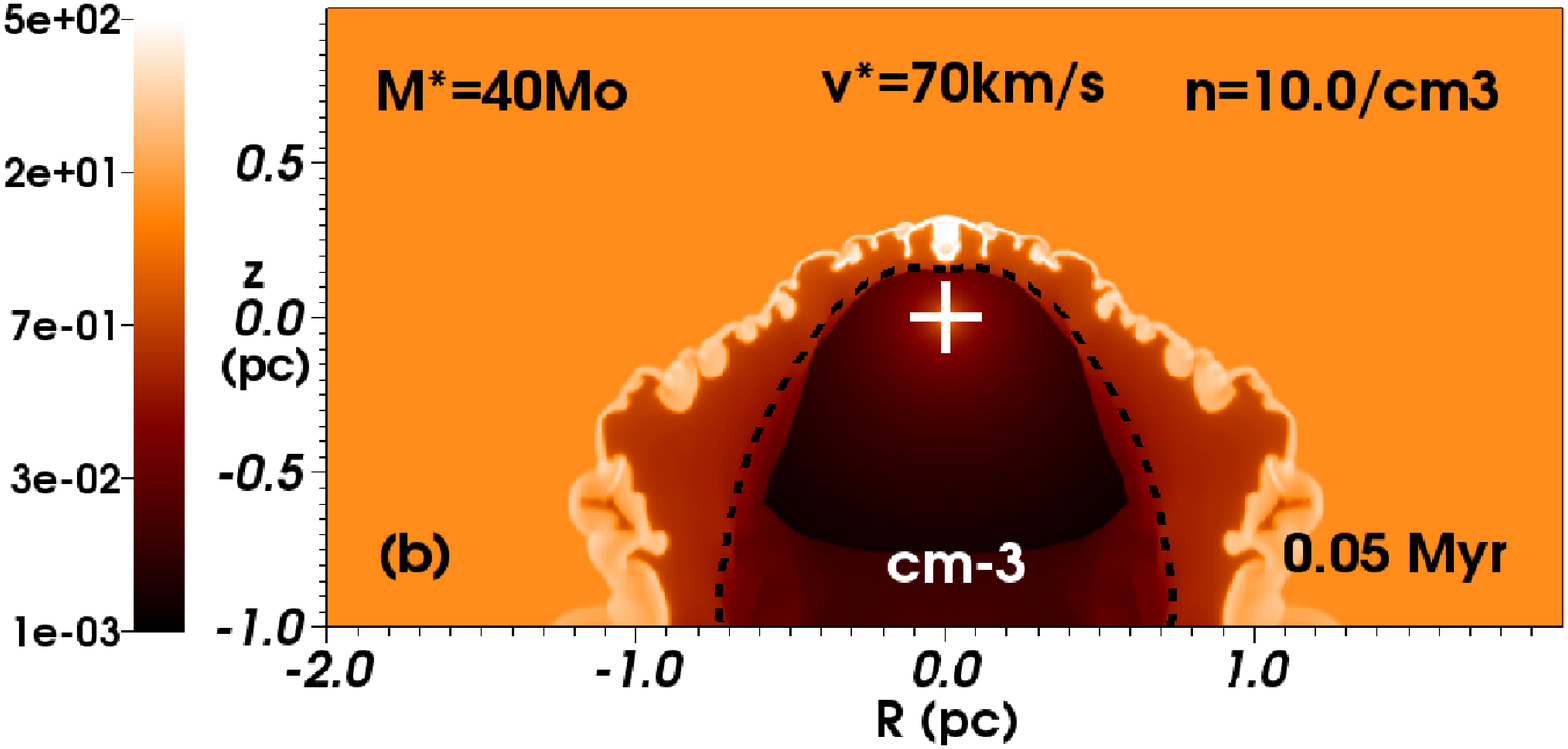}
	\end{minipage} 	\\	
	\begin{minipage}[b]{0.43\textwidth}
		\includegraphics[width=1.0\textwidth]{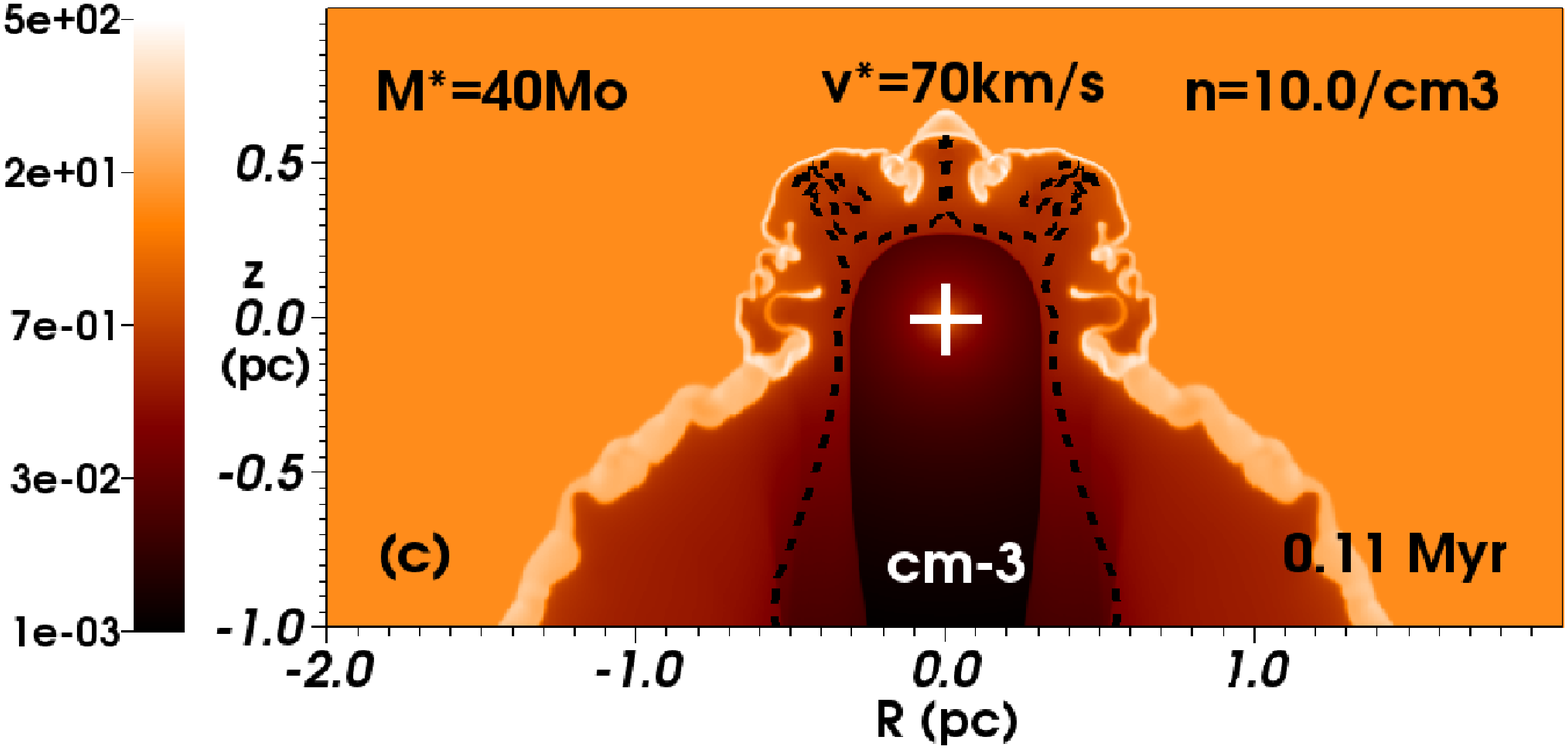}
	\end{minipage}\\	
	\begin{minipage}[b]{0.43\textwidth}
		\includegraphics[width=1.0\textwidth]{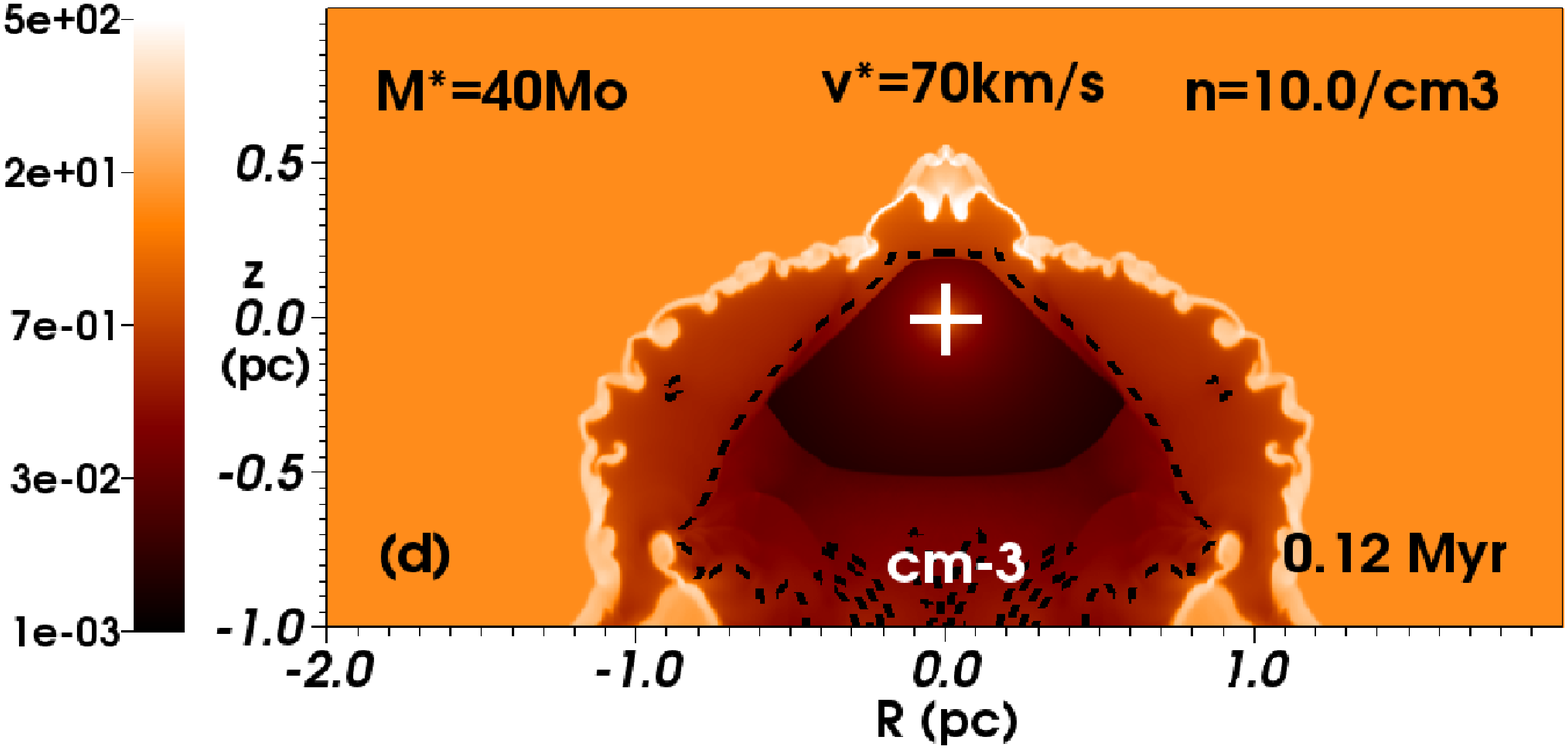}
	\end{minipage} 	
	\caption{ 
	Same as Fig.~\ref{fig:grid_density} for our $40\, \rm M_{\odot}$  \textcolor{black}{ZAMS} 
star moving through an ISM of density $n_{\rm ISM}=10.0\, \mathrm{cm}^{-3}$ with 
velocity $70\, \mathrm{km}\, \mathrm{s}^{-1}$ \textcolor{black}{(model MS4070n10)}. Figures are shown at times 
$0.02$ (a), $0.05$ (b), $0.11$ (c) and $0.12\, \rm Myr$ (d) after the 
beginning of the main sequence phase of the star, respectively. It illustrates 
the development of the non-linear thin-shell instability in the bow 
shock. 
	 }
	\label{fig:grid_velocity}  
\end{figure}

In Fig.~\ref{fig:grid_velocity} we show a time sequence evolution of the density 
field in hydrodynamical simulations of $40\, \rm M_{\odot}$ zero-age main-sequence  
star moving with velocity $v_{\star}=70\, \rm km\, \rm s^{-1}$ in a medium of 
number density $n=10.0\, \rm cm^{-3}$ (model MS4070n10). The figures are shown 
at times $0.02$ (a), $0.05$ (b), $0.11$ (c) and $0.12\, \rm Myr$ (d), respectively. 
After $0.02\, \rm Myr$ the whole shell is sparsed with small size clumps which 
are the seeds of non-linear instabilities (Fig.~\ref{fig:grid_velocity}b). The 
fast stellar motion ($v_{\star}=70\, \rm km\, \rm s^{-1}$) provokes a distortion 
of the bubble into an ovoid shape~\citep[see fig.~7 of][]{weaver_apj_218_1977} 
and the high ambient medium density ($n=10.0\, \rm cm^{-3}$) induces rapidly a 
thin shell after only about $0.01\, \rm Myr$. 

\begin{figure}
	\centering
	\begin{minipage}[b]{ 0.44\textwidth}
		\includegraphics[width=1.0\textwidth]{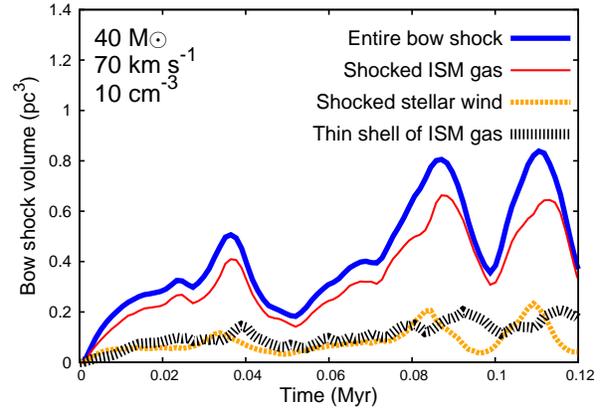}
	\end{minipage}	\\	
	\caption{ 
	         Bow shock volume ($z\ge0$) in our model \textcolor{black}{MS4070n10} (see Fig.~\ref{fig:grid_velocity}a-e). 
	         The figure shows the volume of perturbed material (in $\rm pc^{3}$) in the computational domain 
	         (thick solid blue line), together with the volume of shocked ISM gas (thin solid red line) 
	         and shocked stellar wind (thick dotted orange line), respectively, as 
	         function of time (in $\rm Myr$). 
		 The large dotted black line represents the volume of the thin shell of shocked ISM.   
		 }	
	\label{fig:volume}  
\end{figure}

The bow shock then experiences a series of cycles in which small scaled eddies 
grow in the shell (Fig.~\ref{fig:grid_velocity}b) and further distort its apex 
into wing-like structures (Fig.~\ref{fig:grid_velocity}c) which are pushed  
sidewards because of the transverse component of the stellar wind acceleration 
(Fig.~\ref{fig:grid_velocity}d). Our model MS4070n10 has both characteristics 
from the models E " High ambient density" and G " Instantaneous cooling" 
of~\citet{comeron_aa_338_1998}. 
Thin-shelled stellar wind bow shocks develop non-linear instabilities, in 
addition to the Kelvin-Helmholtz instabilities that typically affect interfaces 
between shearing flows of opposite directions, i.e. the outflowing stellar wind 
and the ISM gas penetrating the bow shocks~\citep{vishniac_apj_428_1994, 
garciasegura_1996_aa_305,vanmarle_aa_469_2007}. A detailed discussion of the development of 
such non-linearities affecting bow shocks generated by OB runaway stars is 
in~\citet{comeron_aa_338_1998}.

In Fig.~\ref{fig:volume} we plot the evolution of the volume of the bow shock in our 
model MS4070n10 (thick solid blue line), separating the volume of shocked ISM 
gas (thin dotted red line) from the volume of shocked stellar wind (thick dotted 
orange line) in the apex ($z\ge0$) of the bow shock. Such a discrimination of 
the volume of wind and ISM gas is possible because a passive scalar tracer is is 
numerically advected simultaneously with the flow. The figure further illustrates the 
preponderance of the volume of shocked ISM in the bow shock compared to the 
stellar wind material, regardless the growth of eddies. Interestingly, the 
volume of dense shocked ISM gas (large dotted black line) does not have large 
time variations (see Section~\ref{sect:emission}).


\section{Bow shock energetics and emission signatures}
\label{sect:emission}



\begin{figure}
          \centering 
	  \includegraphics[width=0.44\textwidth]{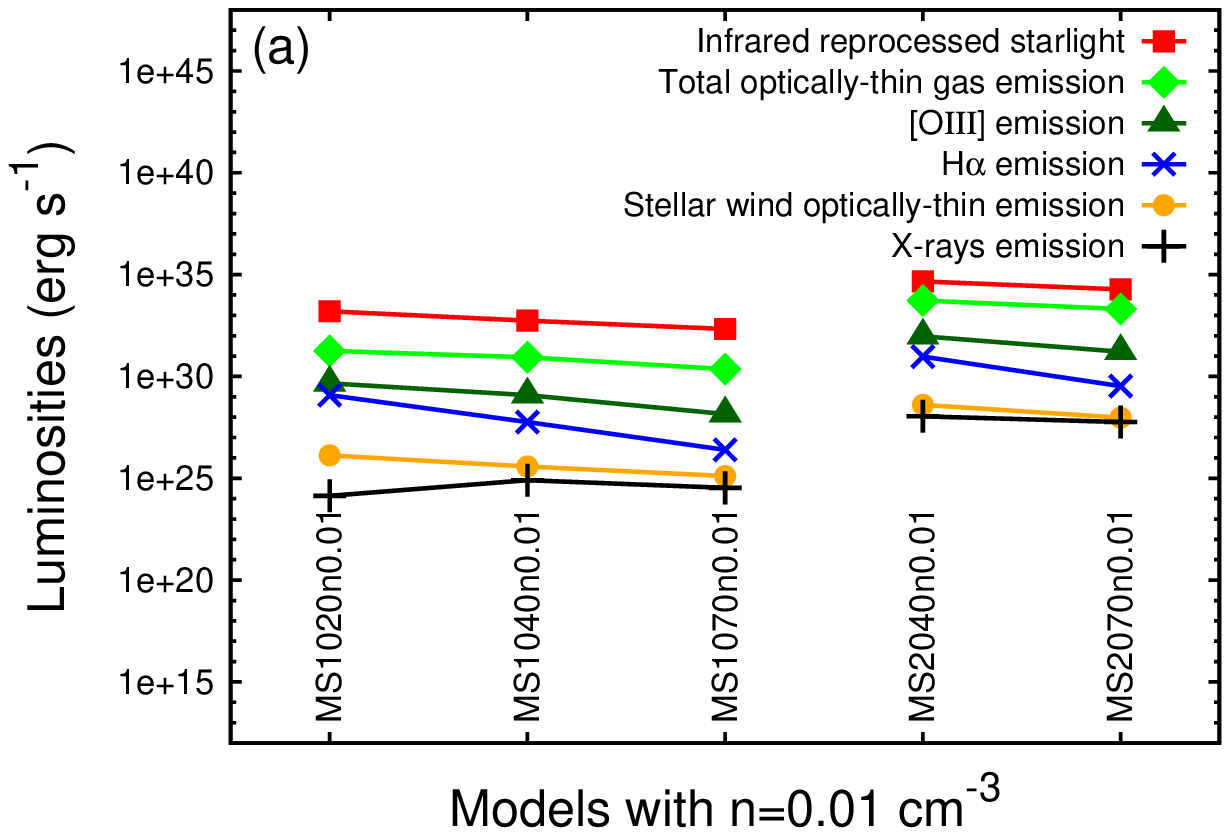}    	  
	  \includegraphics[width=0.44\textwidth]{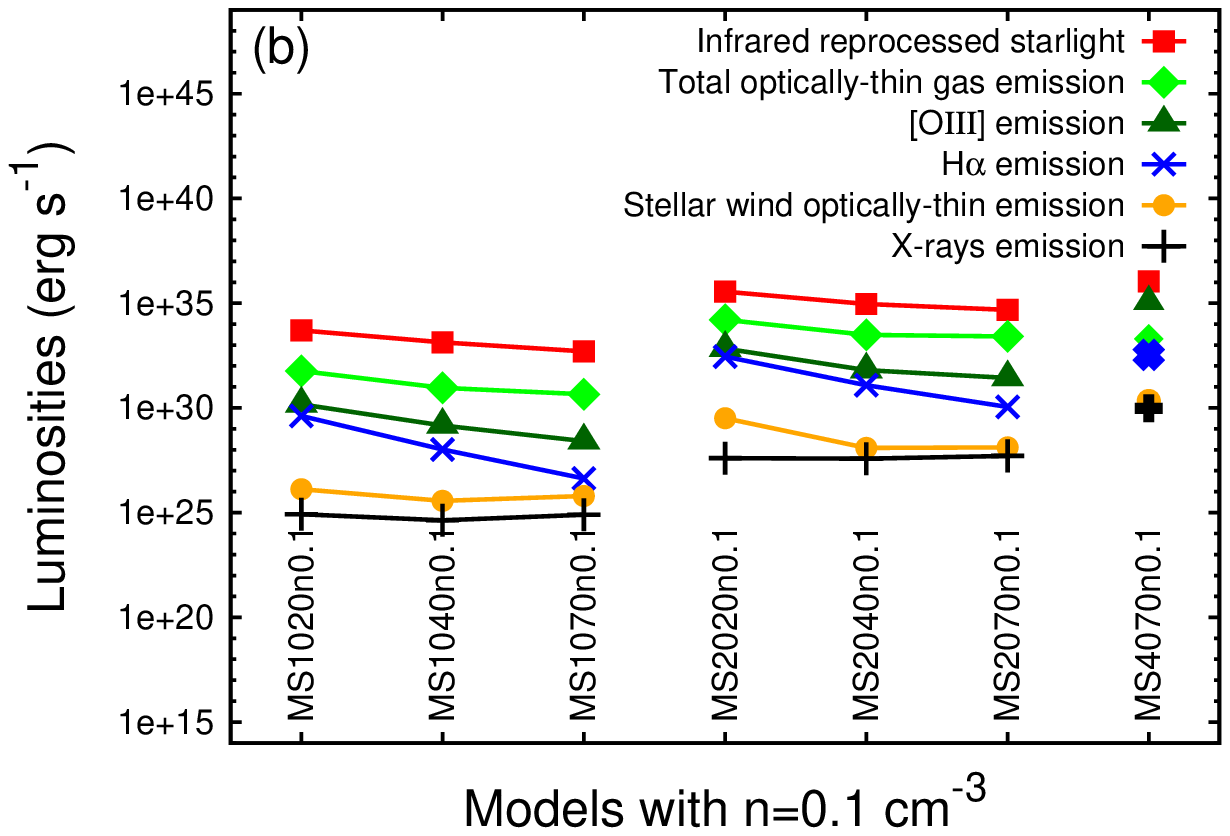}   	  
	  \includegraphics[width=0.44\textwidth]{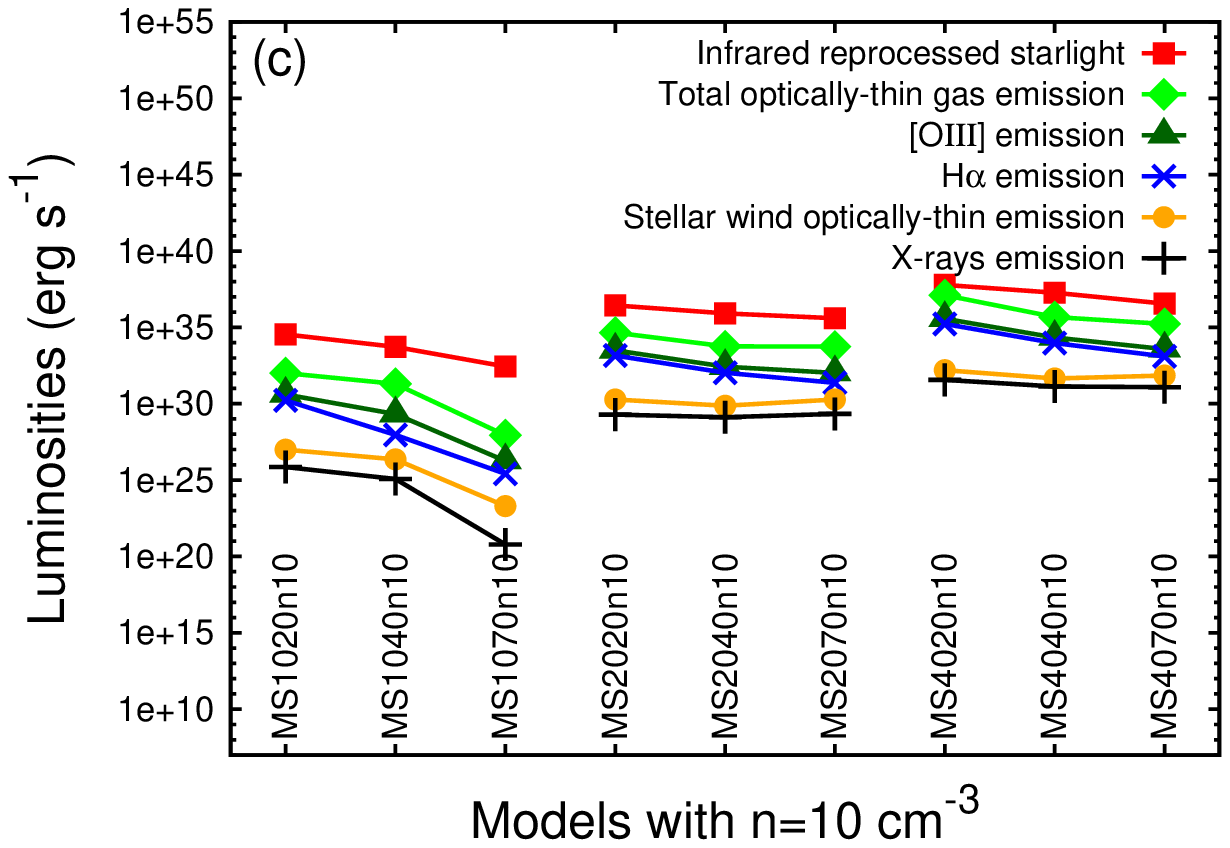}    	  
	  \caption{ 
	  Bow shocks luminosities. The 
	  panels correspond to models with an ISM density \textcolor{black}{$n_{\rm ISM}=0.01$ (a), 
	  $0.1$ (b) and $10.0\, \rm cm^{-3}$ (c).} 
The simulations labels are indicated under the corresponding values. 
	  }
          \label{fig:lum1}
\end{figure}

\subsection{Methods}
\label{subsect:methods}

In Fig.~\ref{fig:lum1} the total bow shock luminosity $L_{\rm total}$ (pale green diamonds) is 
calculated integrating the losses by optically-thin radiation in the $z \ge 0$ 
region of the computational domain~\citep[][Paper~I]{mohamed_aa_541_2012}. Shocked 
wind emission $L_{\rm wind}$ (orange dots) is discriminated from $L_{\rm total}$ 
with the help of the passive scalar $Q$ that is advected with the gas, 
\textcolor{black}{such that, 
\begin{equation}
	\itl{ L}_{\rm total} = 2\pi \iint_{\rm z \ge 0} \itl{\Lambda}(T) n_{\rm H}^{2} R~dR~dz,
        \label{eq:lum_ism}
\end{equation}
and, 
\begin{equation}
	\itl{ L}_{\rm wind} = 2\pi \iint_{\rm z \ge 0} \itl{\Lambda}(T) n_{\rm H}^{2} Q R~dR~dz,
        \label{eq:lum_wind}
\end{equation}
respectively. This allows us to isolate the stellar wind material in the bow shock.}
Additionaly, we compute $L_{\rm H\alpha}$ (blue crosses) 
and $L_{[\rm O{\sc III}]}$ (dark green triangles) which stand for the bow shock luminosities 
at H$\alpha$ and at [O{\sc iii}] $\lambda \, 5007$ spectral line emission using the prescriptions for the 
emission coefficients in~\citet{dopita_aa_29_1973} and~\citet{osterbrock_1989}, 
respectively. The overall X-ray luminosity $L_{\rm X}$ (black right crosses) is computed with emission 
coefficients generated with the {\sc xspec} program~\citep{arnaud_aspc_101_1996} 
with solar metalicity and chemical abundances from~\citet{asplund_araa_47_2009}. 
\textcolor{black}{
The total infrared emission $L_{\rm IR}$ (red squares) is estimated as a reemission of a 
fraction of the starlight bolometric flux on dust grains of gas-to-dust mass ratio $200$,  
which are assumed to be present in the bow shocks. We assume that all dust grains are spherical 
silicates particles of radius $a=5.0\, \rm nm$ only, which are mixed with the gas and continuously 
penetrate the bow shock as we assume that the star moves with a constant velocity. More details 
on the dust model and the infrared estimate of the bow shock emission is given in Appendix B 
of Paper~I.} 
%
%

\subsection{Results}
\label{subsect:results}

\subsubsection{Optical luminosities}
\label{subsect:luminosities}

\begin{figure}
	\centering
	\begin{minipage}[b]{ 0.44\textwidth}
		\includegraphics[width=1.0\textwidth]{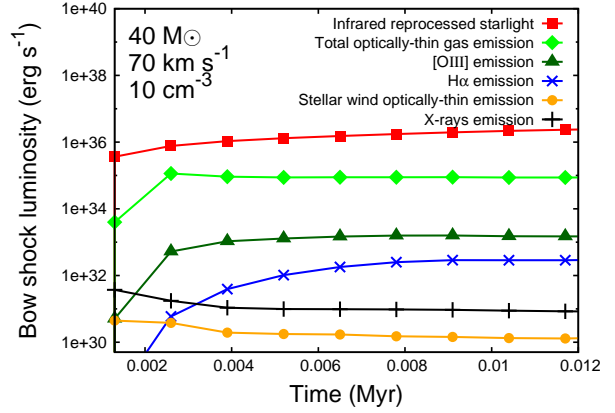}
	\end{minipage}	\\	
	\caption{ 
	         Luminosities of our bow shock simulation of a 
	         $40\, \rm M_{\odot}$  \textcolor{black}{ZAMS} star moving with velocity $v_{\star}=70\, \rm km\, \rm s^{-1}$ 
	         through a medium with $n_{\rm ISM}=10\, \rm cm^{-3}$ (see corresponding time-sequence 
	         evolution of its density field in Fig.~\ref{fig:grid_velocity}a-e). 
	         Plotted quantities and color-coding are similar to Fig.~\ref{fig:lum1} and are shown as 
	         function of time (in $\rm Myr$). 
		 }	
	\label{fig:lum3}  
\end{figure}

\begin{figure*}
	\centering
	\begin{minipage}[b]{ 0.47\textwidth}
		\includegraphics[width=1.0\textwidth]{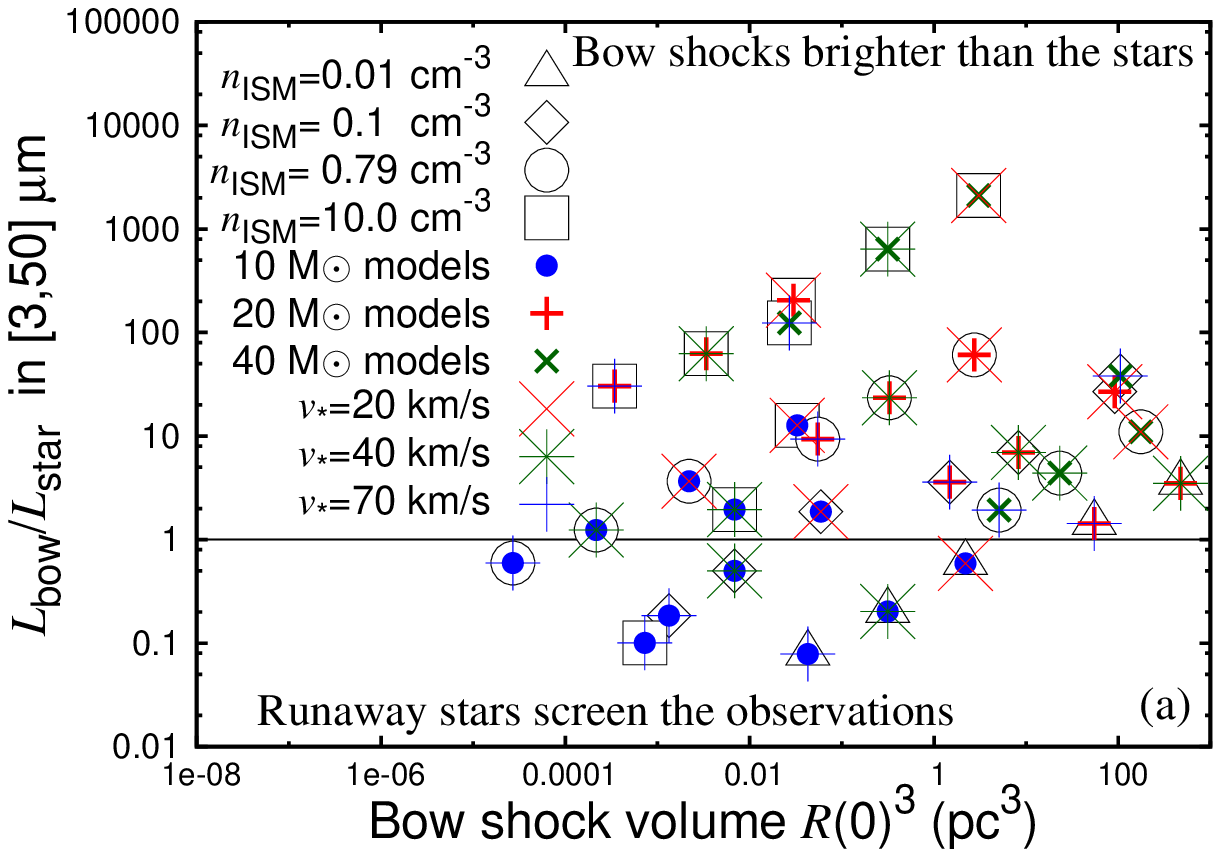}
	\end{minipage}	
	\begin{minipage}[b]{ 0.47\textwidth}
		\includegraphics[width=1.0\textwidth]{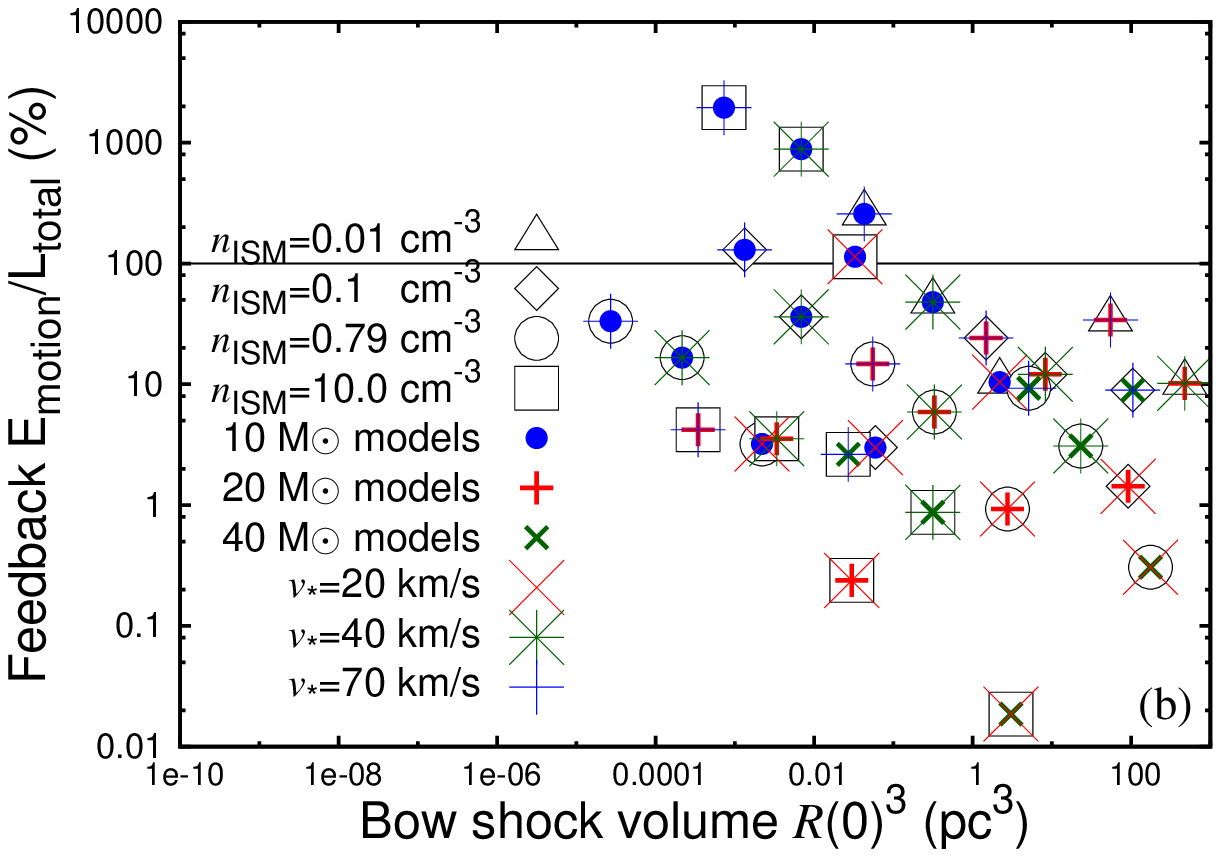}
	\end{minipage}		
	\caption{ 
	         Comparison between the bow shock luminosity of reprocessed starlight $L_{\rm bow}=L_{\rm IR}$ 
	         and the infrared stellar emission $L_{\star}$, both in the wavelength range $[3;50]\, \mu \rm m$ (a).
	         Ratio (in $\%$) of the energy deposited per unit time due to the motion of the 
	         bow shock with its losses per unit time because of optically-thin cooling (b).	         
		 }	
	\label{fig:obser}  
\end{figure*}

In Fig.~\ref{fig:lum1} we display the bow shocks luminosities as a 
function of the initial mass of the runaway star, its space velocity $v_{\star}$ 
and its ambient medium density $n_{\rm ISM}$.  
At a given density of the ISM, all of our models have luminosities from 
optically-thin gas radiation which \textcolor{black}{varies} with respect to the stellar mass loss are as described 
in Paper~I for the simulations with $n_{\rm ISM}\approx 0.79\, \rm 
cm^{-3}$. \textcolor{black}{ We can identify three major trends for the behaviour of 
the luminosity:
\begin{enumerate}
\item The total luminosity $L_{\rm total}$ decreases by at least an order of 
magnitude between the simulations with $v_{\star}=20$ to $70\, \rm km\, \rm 
s^{-1}$. For example, our $10\, \rm M_{\odot}$ \textcolor{black}{ZAMS} star, moving with velocity 
$v_{\star}=20\, \rm km\, \rm s^{-1}$ in an ISM of $n_{\rm ISM}\approx 0.01\, \rm 
cm^{-3}$, has $L_{\rm total}\approx 1.82 \times 10^{31}\, \rm erg\, \rm s^{-1}$  
whereas the same star, moving through the same ISM but with velocity 
$v_{\star}=70\, \rm km\, \rm s^{-1}$, has \textcolor{black}{$L_{\rm total}\approx 2.29 \times 
10^{30}\, \rm erg\, \rm s^{-1}$} (see models MS1020n0.01 and MS1070n0.01 in 
Fig~\ref{fig:lum1}a). This arises because if the space motion of the star increases 
the compression factor of the shell of shocked ISM gas in the bow shock, it also 
reduces its volume which decreases the emission by optical-thin 
radiative processes~\citep{meyer_mnras_2013}. 
\item The total bow shock luminosity \textcolor{black}{by optically-thin processes} increases by several orders of magnitude 
with $\dot{M}$. \textcolor{black}{For example, the bow shock produced by our $10\, \rm 
M_{\odot}$  \textcolor{black}{ZAMS star} moving with velocity $v_{\star}=40\, \rm km\, \rm s^{-1}$ in an ISM 
of $n_{\rm ISM}\approx 0.01\, \rm cm^{-3}$ has $L_{\rm total} \approx 8.68 
\times 10^{30}\, \rm erg\, \rm s^{-1}$ whereas our $20\, \rm 
M_{\odot}$  \textcolor{black}{ZAMS star} moving with the same speed through an identical medium has $L_{\rm 
total}\approx 5.38 \times 10^{33}\, \rm erg\, \rm s^{-1}$ (see models 
MS1040n0.01 and MS2040n0.01 in Fig~\ref{fig:lum1}a). 
}
In that sense, our results 
confirm that the bow shock volume which increases with $\dot{M}$ and decreases 
with $v_{\star}$ governs their luminosity by optically-thin cooling. 
\item  Finally, the bow shock luminosity decreases if the ambient medium density 
$n_{\rm ISM}$ increases. This happens because a larger $n_{\rm ISM}$ decreases 
the volume of the bow shock $\propto R(90)^{3}$ since $R(90)\propto 
1/\sqrt{n_{\rm ISM}}$, which has a stronger influence over the luminosity than 
the fact that the density in the post-shock region at the forward shock $\propto 
n_{\rm ISM}/4$ increases. For example, the bow shocks generated by our 
$20\, \rm M_{\odot}$  \textcolor{black}{ZAMS} star moving with velocity $v_{\star}=70\, \rm km\, \rm 
s^{-1}$ in an ISM of density $n_{\rm ISM}=0.01$, $0.1$, $0.79$ and $10.0\, \rm 
cm^{-3}$ have a bow shock luminosity $L_{\rm total}\approx 2.04 \times 10^{33}$, 
$ 2.59 \times 10^{33}$, \textcolor{black}{$ 3.72 \times 10^{33}$} and $ 5.62 \times 10^{33}\, \rm 
erg\, \rm s^{-1}$ respectively, see models MS2070n0.01, MS2070n0.1, MS2070 and 
MS2070n10 (Fig.~\ref{fig:grid_density}a-d). This further illustrates the 
dominant role of the bow shock volume on $L_{\rm total}$, which is governed by 
the compression of the shell and by the strength of its stellar wind momentum, 
i.e. $\dot{M}$ and $v_{\rm w}$. 
\end{enumerate}
}

The behaviour of the optically-thin emission originating from the shocked 
stellar wind $L_{\rm wind}$, the [O{\sc iii}] $\lambda \, 5007$ spectral line 
emission and the H$\alpha$ emission at fixed $n_{\rm ISM}$ are similar as described  
in~\citet{meyer_mnras_2013}.  The contribution of $L_{\rm wind}$ is smaller than 
$L_{\rm total}$ by several orders of magnitude for all models, e.g. our model 
MS1020n0.1 has $L_{\rm wind}/L_{\rm total} \approx 10^{-5}$. 
All our models have $L_{\rm H\alpha} < L_{[\rm O{\sc III}]} < L_{\rm
total}$ and the H$\alpha$ emission, the $[\rm O{\sc III}]$ spectral line 
emission and $L_{\rm wind}$ have variations which are similar to $L_{\rm ISM}$ 
with respect to $M_{\star}$, $v_{\star}$ and $n_{\rm ISM}$.  
%
%
%

Fig.~\ref{fig:lum3} shows the lightcurve of our model MS4070n10 computed over the 
whole simulation and plotted as a function of time with the color coding from 
Fig.~\ref{fig:lum1}. Very little variations of the emission are present at 
the beginning of the calculation up to a time of about $0.004\, \rm Myr$ and  
it remains almost constant at larger times.  
\textcolor{black}{
We infer that in the case of a bow shock producing a thin shell of density larger  
than about $10\, \rm cm^{-3}$,  the distortions of the global structure are largely 
irrelevant to the luminosity, which is dominated by the dense, shocked, cold ISM gas 
(see discussion in~\ref{sect:discussion}).
}
%
This is in accordance with the volume of 
the dense ISM gas trapped into the nebula (see large dotted black line in Fig.~\ref{fig:volume}).
The independence of $L_{\rm IR}$ with 
respect to the strong volume fluctuations of thin-shelled nebulae 
(Fig.~\ref{fig:lum3}) indicates that their spectral energy distributions is likely 
to be the appropriate tool to analyze them since it constitutes an observable which 
is not reliable to temporary effects.



\subsubsection{Infrared and X-rays luminosities}
\label{subsect:thermalisation}

\textcolor{black}{Not surprisingly, the infrared luminosity, which originates from} reprocessed starlight on dust grains penetrating 
the bow shocks, $L_{\rm IR}$, is larger than $L_{\rm total}$ by about $1-2$ 
orders of magnitude. This is possible because the reemission of starlight by 
dust grains is not taken into account in our simulations. 
\textcolor{black}{Unlike the optical luminosity, the infrared luminosity increases with $n_{\rm ISM}$, e.g. our models with 
$M_{\star}=10\, \rm M_{\odot}$ and $v_{\star}=20\, \rm km\, \rm s^{-1}$ have 
$L_{\rm IR} \approx 1.6\times 10^{33}$, $5.0\times 10^{33}$, 
$ 9.92\times 10^{33}$ and $ 3.43\times 10^{34}\, \rm erg\, \rm s^{-1}$ 
if $n_{\rm ISM}=0.01$, $0.1$, $0.79$ and $10.0\, \rm cm^{-3}$, respectively. 
Moreover, the ratio between $L_{\rm IR}$ and $L_{\rm total}$ increases with $n_{\rm ISM}$, 
e.g. $L_{\rm IR}$/$L_{\rm total} \approx 8.6$ and $ 144.1$} for the models MS2040n0.01 and 
MS2040n10, respectively. $L_{\rm IR}$ increases with $M_{\star}$ (Figs.~\ref{fig:lum1}a-d).
Particularly, we find that $L_{\rm IR} \gg L_{\rm H\alpha}$ and 
$L_{\rm IR} \gg L_{[\rm O{\sc III}]}$, and therefore we conclude that the infrared 
waveband is the best way to detect and observe bow shocks from massive main-sequence runaway 
stars regardless of $n_{\rm ISM}$ (see section~\ref{sect:observability}).

\textcolor{black}{
Note that, according to the prescription for the dust temperature,
\begin{equation}
    T_{\rm d}(r) = T_{\rm eff} \Big( \frac{R_{\star}}{2r} \Big)^{ \frac{2}{4+s} }, 
\end{equation}
where $T_{\rm eff}$ is the effective temperature of the moving star, $R_{\star}$ its 
radius and $s$ a parameter giving the slope of the opacity in the infrared 
regime~\citep[and references therein]{spitzer_1978,kuiper_aa_537_2012}. 
With $s=1$~\citep{decin_aa_456_2006}, it follows that the 
dust temperature of, e.g. our models with $n_{\rm ISM}=0.79\, \rm cm^{-3}$ is about 
$T_{\rm d}\le100\, \rm K$ in the bow shock. The Planck distribution of such temperatures 
would peak in the mid infrared and therefore one can expect that the dust continuum emission 
of the bow shocks lies in the wavelength range $3\, \leq\, \lambda\, \leq\, 50\, \mu \rm m$. 
In Fig.~\ref{fig:obser}a we compare the bow shocks and the stellar luminosities in this 
wavelength range, assuming that the $10$, $20$ and $40\, \rm M_{\odot}$  \textcolor{black}{ZAMS} runaway 
stars are black bodies. Most of the bow shocks generated by a star of initial mass   
$\ge 20\, \rm M_{\odot}$ are brighter in infrared than their central runaway star. This  
indicates that bow shocks can dominate the emission up to several orders of magnitudes 
for wavelengths $3\, \leq\, \lambda\, \leq\, 50\, \mu \rm m$ and that saturation effects 
of the observations are improbable for those stars. 
}

Several current and/or planned facilities are designed to observe at these wavelengths and may be able to detect bow 
shocks from runaway stars: 
\begin{enumerate}
\item First, the {\it James Webb Space Telescope} (JWST) which {\it Mid-Infrared Instrument}~\citep[MIRI,][]{swinyard_2004} 
observes in the infrared ($5$$-$$28\, \mu \rm m$) that roughly corresponds to our predicted waveband of dust 
continuum emission from stellar wind bow shocks of runaway OB stars.
\item Secondly, the {\it Stratospheric Observatory for Infrared Astronomy} (SOFIA) airborne 
facility which {\it Faint Object infraRed CAmera for the SOFIA Telescope}~\citep[FORCAST,][]{adams_2008} instrument 
detects photons in the $5.4$$-$$37\, \mu \rm m$ waveband.
\item Then, the proposed {\it Space Infrared Telescope for Cosmology and Astrophysics}~\citep[SPICA,][]{kaneda_2004}  
satellite would be the ideal tool the observe stellar wind bow shock, since it is planed to be mounted with 
(i) a far-infrared imaging spectrometer ($30$$-$$210\, \mu \rm m$), (ii) a mid-infrared coronograph 
($3.5/5$$-$$27\, \mu \rm m$) and (iii) a mid-infrared camera/spectrometer ($5$$-$$38\, \mu \rm m$). 
\item Finally, we should mention the proposed {\it The Mid-infrared E-ELT Imager and Spectrograph} (METIS) 
on the planned {\it European Extremely Large Telescope}~\citep[E-ELT,][]{brandl_2006}, that will be able to scan 
the sky in the $3$$-$$19\, \mu \rm m$ waveband.  
\end{enumerate}
Exploitation of the associated archives of these instruments in regions surroundings young stellar 
clusters and/or at the locations of previously detected bow-like 
nebulae~\citep{buren_apj_329_1988,vanburen_aj_110_1995, 
noriegacrespo_aj_113_1997,peri_aa_538_2012,2015arXiv150404264P} are research 
avenues to be explored.

Finally, we notice that the X-rays emission are much smaller than any other emission lines 
or bands, e.g. the model MS2070 has $L_{\rm X}/L_{\rm H\alpha} \approx 10^{-5}$, and it 
is consequently not a relevant waveband to observe our bow shocks.

\subsubsection{Feedback}
\label{subsect:feedback}

\textcolor{black}{
We compute the energy rate $\dot{E}_{\rm motion}$ deposited to the ISM by the stellar 
motion. It is estimated by multiplying the rate of volume of ISM swept-up with the 
bow shock per unit time $\dot{V} \approx \pi R(90)^{2} v_{\star}$  by the ISM 
kinetic energy density defined as $\epsilon_{\rm ISM} = \rho_{\rm ISM}\Delta 
v^{2}/2$, with $\Delta v$ is the changes in velocity across the shock. In the 
frame of reference of the moving star $\Delta v=||v_{\star}-v_{\mathrm gas}||$, 
and, 
\begin{equation}
    \epsilon_{\rm ISM} = \frac{1}{2} \rho_{\rm ISM} ||v_{\star}-v_{\mathrm gas}||^{2}, 
    \label{eq_energy}    
\end{equation}
where $v_{\mathrm gas}$ is the gas velocity  at the post-shock region at the forward shock. 
Since the Rankine-Hugoniot relation indicates that $v_{\mathrm gas}\approx v_{\star}/4$,  
then the relation $\dot{E}_{\rm motion} = \dot{V}\epsilon_{\rm ISM}$ reduces to, 
\begin{equation}
    \dot{E}_{\rm motion} = \frac{9}{32} \rho_{\rm ISM}v_{\star}^{3}\pi R(90)^{2}, 
    \label{eq_motion}    
\end{equation}
where $\rho_{\rm ISM}$ is the ISM gas mass density. 
}

The ratio $\dot{E}_{\rm motion}/L_{\rm total}$
is shown as a function of the bow shock volume in Fig.~\ref{fig:obser}b. 
\textcolor{black}{
The simulations with $M_{\star} \ge 20\, \rm M_{\odot}$ have a bow shock with 
$\dot{E}_{\rm motion}/L_{\rm total} \le 100 \%$ which indicates that their 
associated nebulae have energy losses by optically-thin radiative processes more 
important than the energy deposition by the stellar motion itself to the 
replenishing of the ISM. Our $10\, \rm M_{\odot}$ ZAMS star can produce 
bow shocks having $\dot{E}_{\rm motion}/L_{\rm total} \gg 100 \%$. However, 
since fast-moving $10\, \rm M_{\odot}$ stars are the most common Galactic 
runaway stars of our sample~\citep{eldridge_mnras_414_2011}, it is difficult to 
estimate which sub-population of runaway massive stars, and by which process, 
contributes the most to the Galactic feedback. A population synthesis study, 
beyond the scope of this work, is therefore necessary to assess this question. 
}
%

\begin{figure*}
\centering
	\begin{minipage}[b]{0.32\textwidth}
		\includegraphics[width=1.0\textwidth]{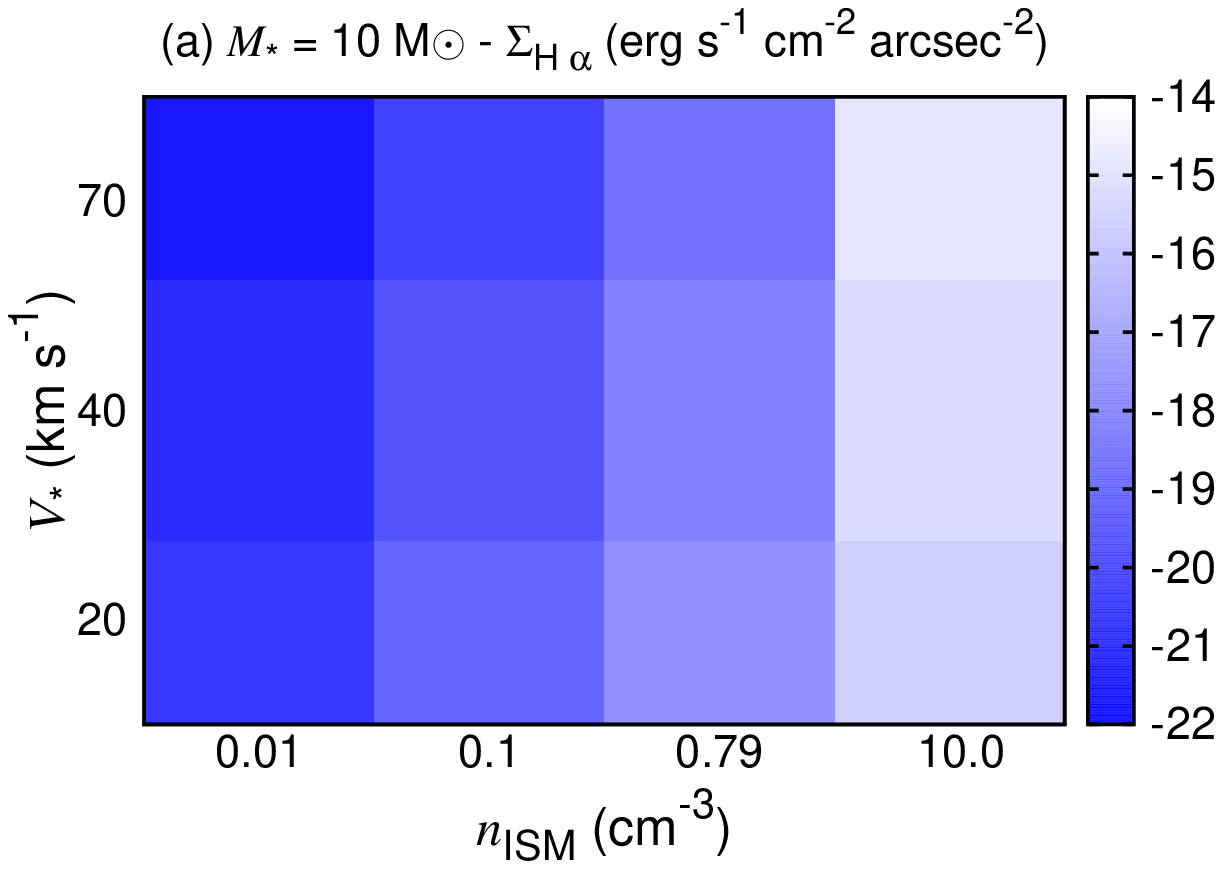}
	\end{minipage} 
	\begin{minipage}[b]{0.32\textwidth}
		\includegraphics[width=1.0\textwidth]{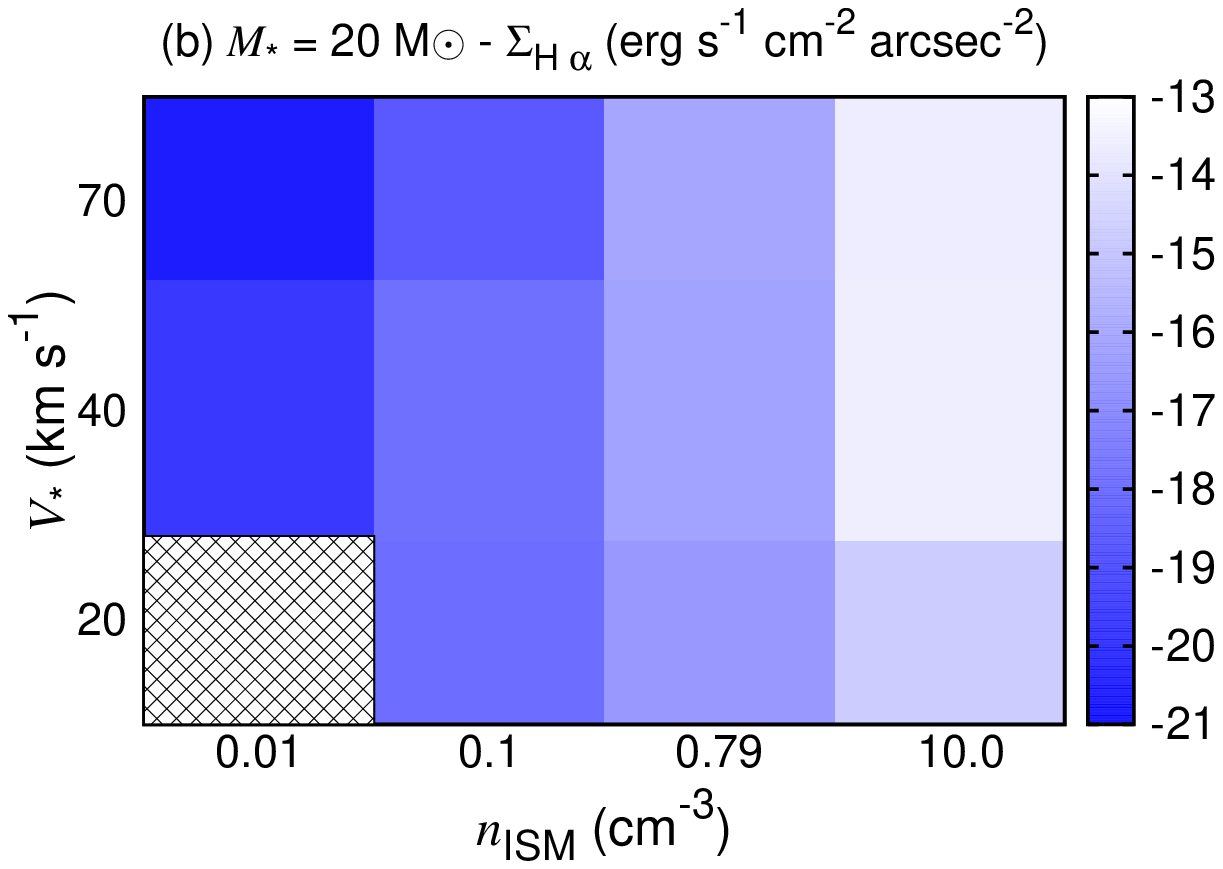}
	\end{minipage} 
	\begin{minipage}[b]{0.32\textwidth}
		\includegraphics[width=1.0\textwidth]{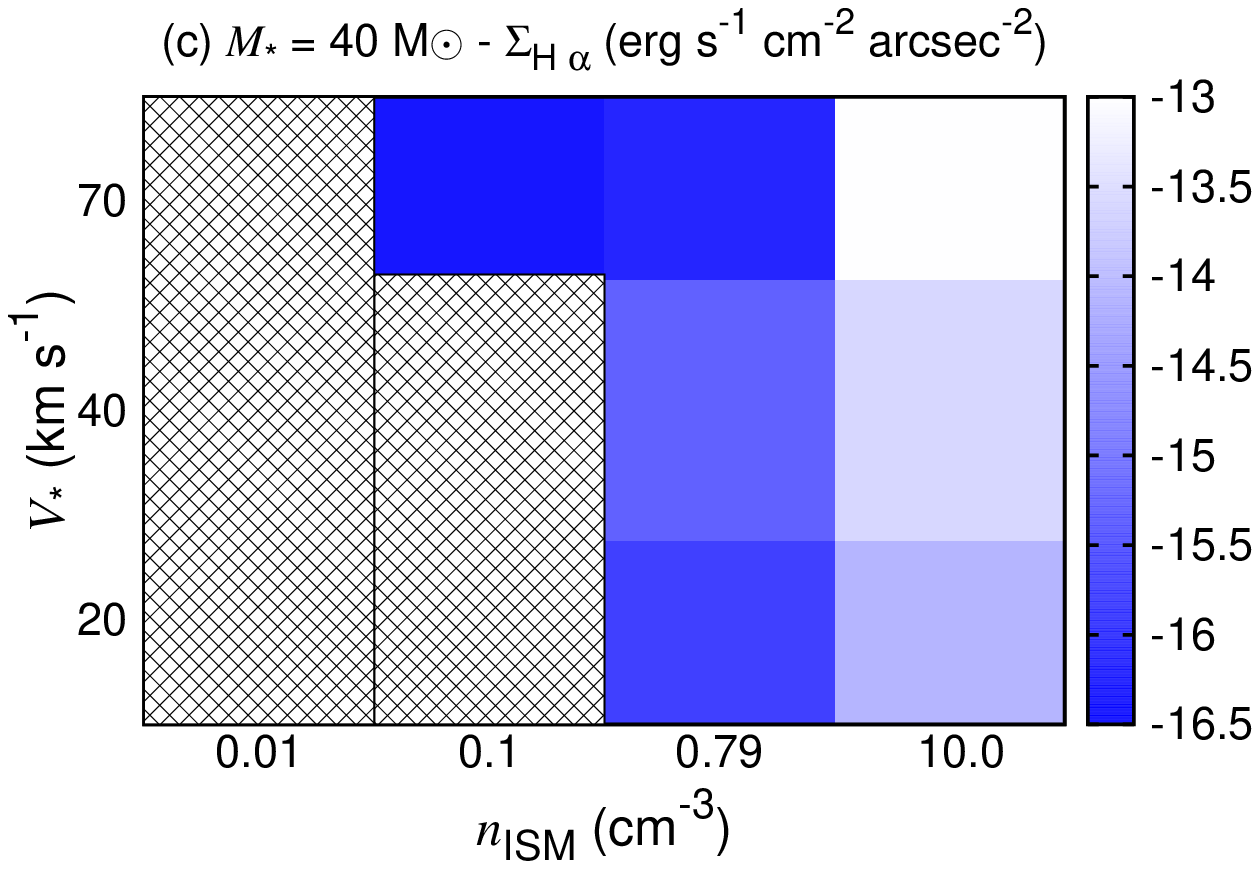}
	\end{minipage} 	\\
	\begin{minipage}[b]{0.32\textwidth}
		\includegraphics[width=1.0\textwidth]{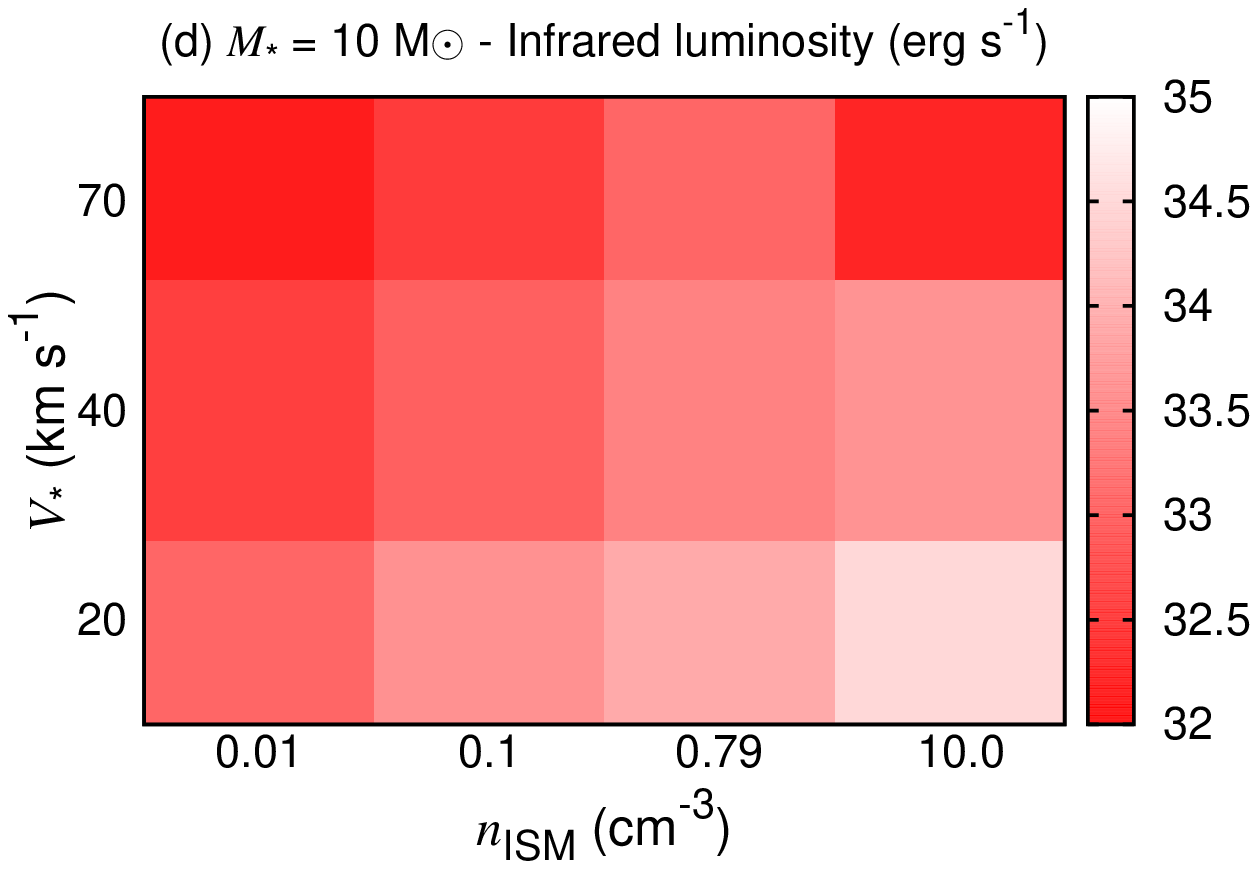}
	\end{minipage} 
	\begin{minipage}[b]{0.32\textwidth}
		\includegraphics[width=1.0\textwidth]{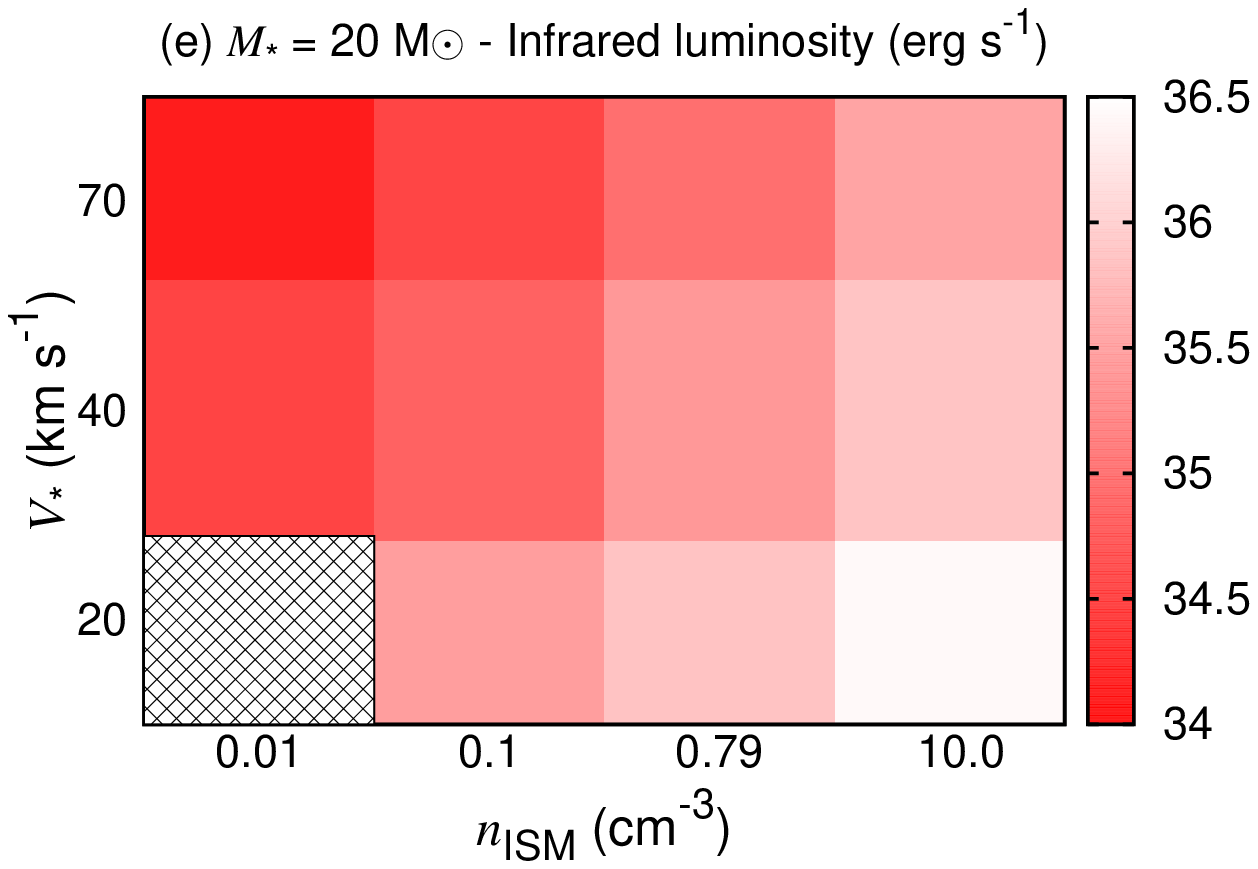}
	\end{minipage}
	\begin{minipage}[b]{0.32\textwidth}
		\includegraphics[width=1.0\textwidth]{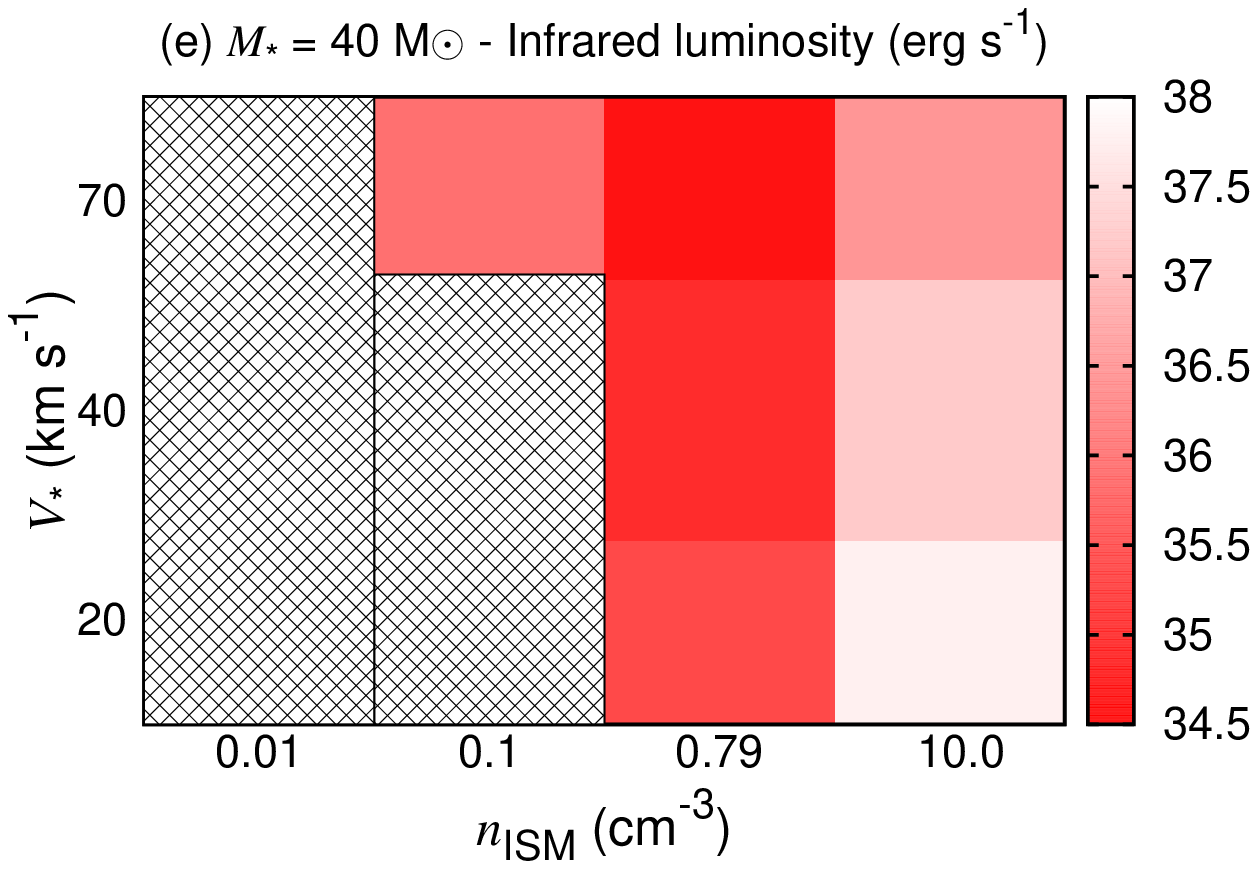}
	\end{minipage} 	
	\caption{ 
	Bow shock H$\alpha$ surface brightness (in $\rm erg\,\rm  s^{-1}\,\rm  cm^{-2}\,\rm  arcsec^{-2}$,  
	top blue panels) and infrared reprocessed starlight (in $\rm erg\, \rm s^{-1}$, bottom red panels) \textcolor{black}{in the logarithmic scale}. 
	We display these quantities for our models with $M_{\star}=10\, \rm M_{\odot}$ (left panels), $20\, \rm M_{\odot}$ (middle panels) 
	and $40\, \rm M_{\odot}$ (right panels). 
	The \textcolor{black}{hatched regions} indicate that the corresponding bow shock models are not included in our grid of 
	simulations, since the duration of the main-sequence phase of these stars does not allow to generate 
	bow shocks in a steady state at these ambient medium density. 
	On each plot the \textcolor{black}{horizontal axis} is the ambient medium density $n_{\rm ISM}$ (in $\rm cm^{-3}$) and the 
	\textcolor{black}{vertical axis} is the space velocity $v_{\star}$ (in $\rm km\, \rm s^{-1}$) of our runaway stars.
	}
	\label{fig:paving_lum}  
\end{figure*}


\subsection{Discussion}
\label{sect:discussion}

\subsubsection{The appropriated waveband to observe stellar wind bow shocks in the Galaxy}
\label{sect:observability}

In Fig.~\ref{fig:paving_lum} we show the H$\alpha$ surface brightness (in $\rm erg\,\rm 
s^{-1}\,\rm  cm^{-2}\,\rm  arcsec^{-2}$, panels a-c) and the 
infrared luminosity (in $\rm erg\, \rm s^{-1}$, panels d-f) for models with 
$M_{\star}=10\, \rm M_{\odot}$ (left panels), $20\, \rm M_{\odot}$ (middle panels) and $40\, \rm M_{\odot}$ 
(right panels). 
The surface brightness $\Sigma^{\rm max}_{\rm H \alpha}$ scales with 
$n^{2}$, see Appendix~A of Paper~I, therefore the lower the 
ISM background density of the star, i.e. the higher its Galactic latitude, 
the fainter the projected emission of the bow shocks and the lower the 
probability to observe them. 
The brightest bow shocks are generated both in infrared and H$\alpha$ by our 
most massive stars running in the denser regions of the ISM ($\rm n_{\rm ISM} = 
10.0\, \rm cm^{-3}$). The estimate of the infrared luminosity confirms our 
\textcolor{black}{earlier} result relative to bow shock models with $n_{\rm ISM}=0.79\, \rm 
cm^{-3}$ in the sense that the brightest bow shocks are produced by high-mass, 
stars (Paper~I) moving in a relatively dense ambient medium, 
i.e. within the Galactic plane (Fig.~\ref{fig:paving_lum}d-f). At H$\alpha$, these
bow shocks are associated to fast-moving stars ($v_{\star}=70\, \rm km\, \rm s^{-1}$)
producing the strongest shocks, whereas in infrared they are associated to slowly-moving 
stars ($v_{\star}=20\, \rm km\, \rm s^{-1}$) generating the largest nebulae.

\begin{figure*}
\centering
	\begin{minipage}[b]{ 0.29\textwidth}
		\includegraphics[width=1.0\textwidth]{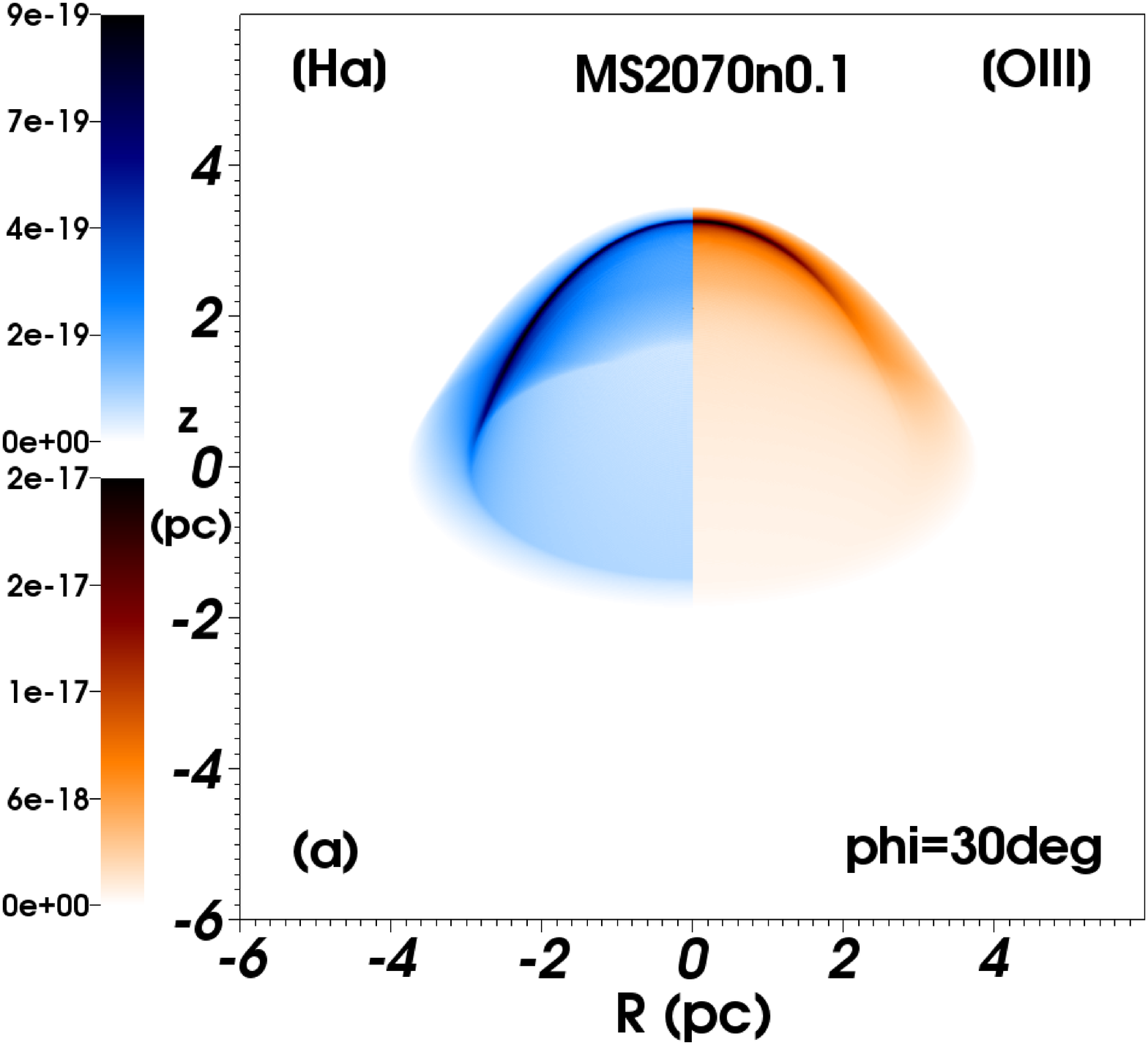}
	\end{minipage}
	\begin{minipage}[b]{ 0.29\textwidth}
		\includegraphics[width=1.0\textwidth]{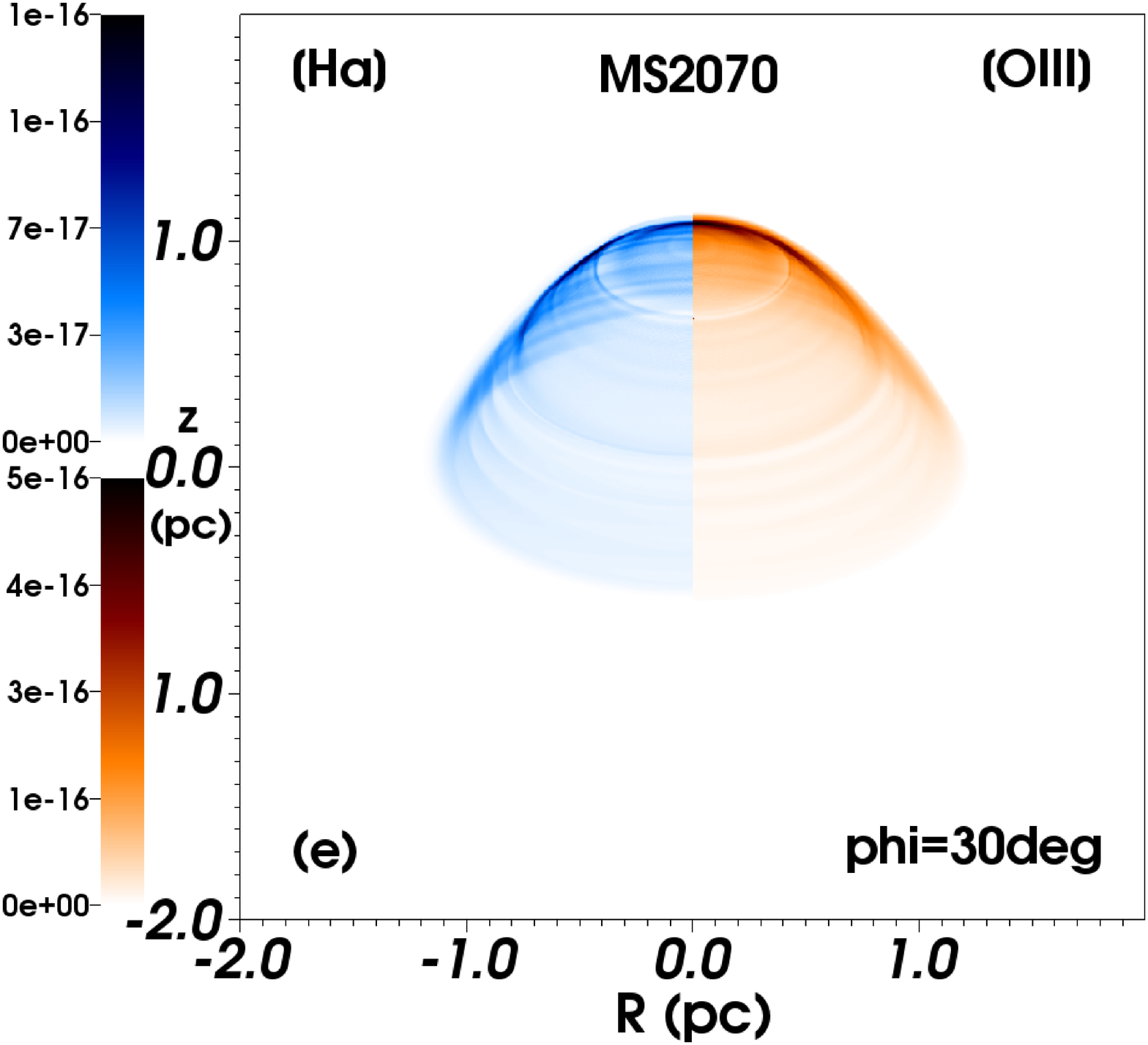}
	\end{minipage}
	\begin{minipage}[b]{ 0.29\textwidth}
		\includegraphics[width=1.0\textwidth]{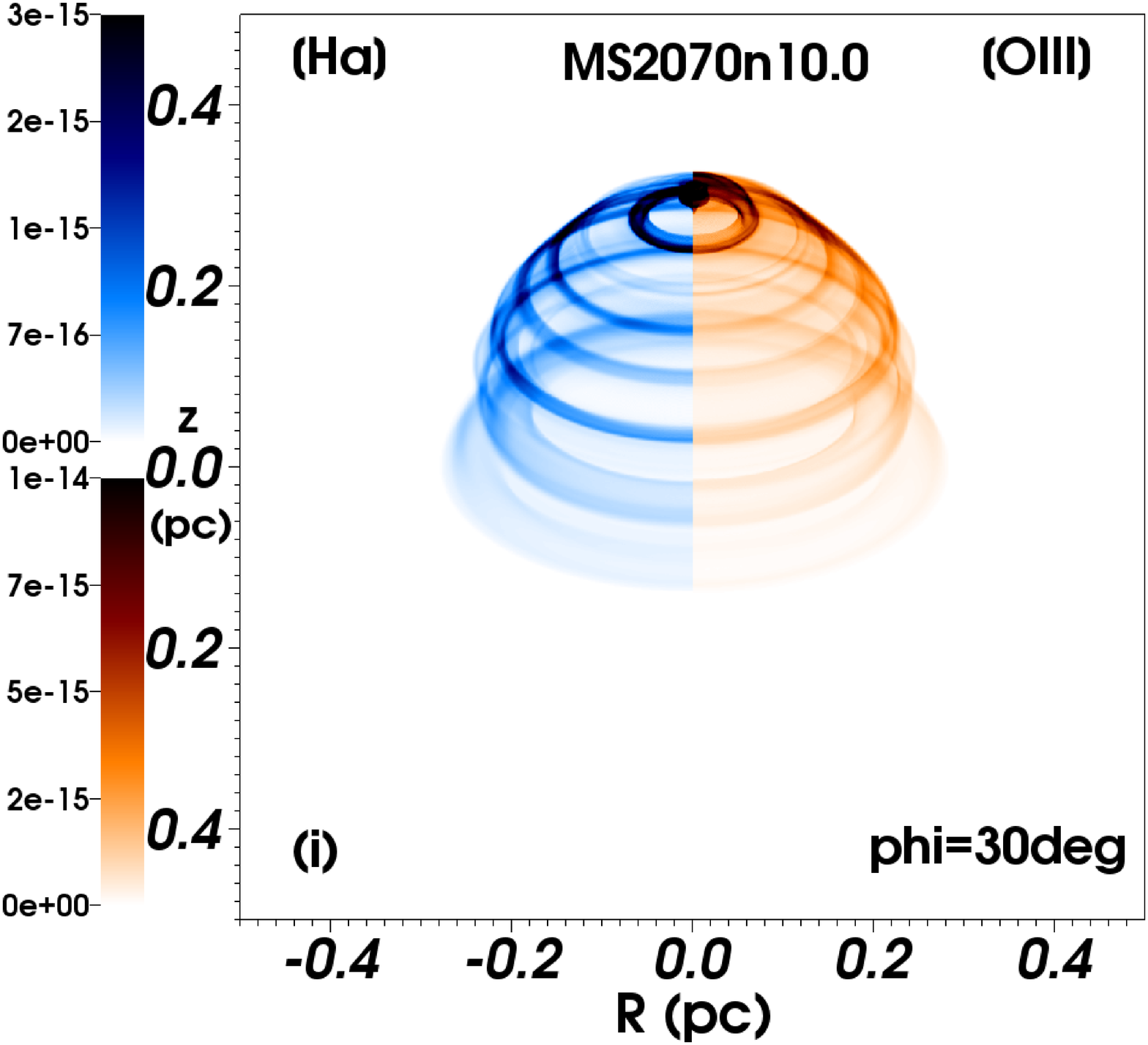}
	\end{minipage}\\
	\begin{minipage}[b]{ 0.29\textwidth}
		\includegraphics[width=1.0\textwidth]{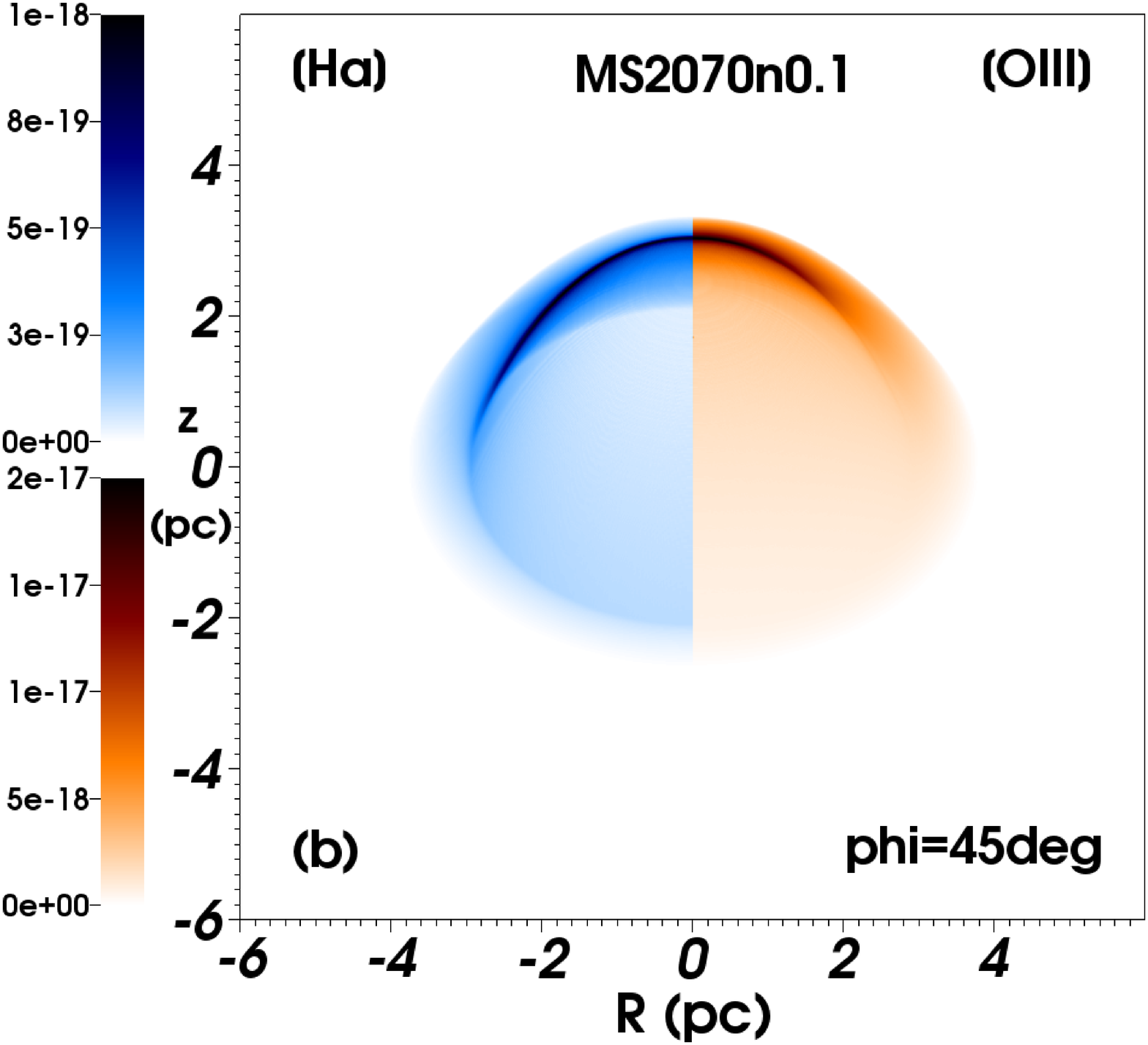}
	\end{minipage}
	\begin{minipage}[b]{ 0.29\textwidth}
		\includegraphics[width=1.0\textwidth]{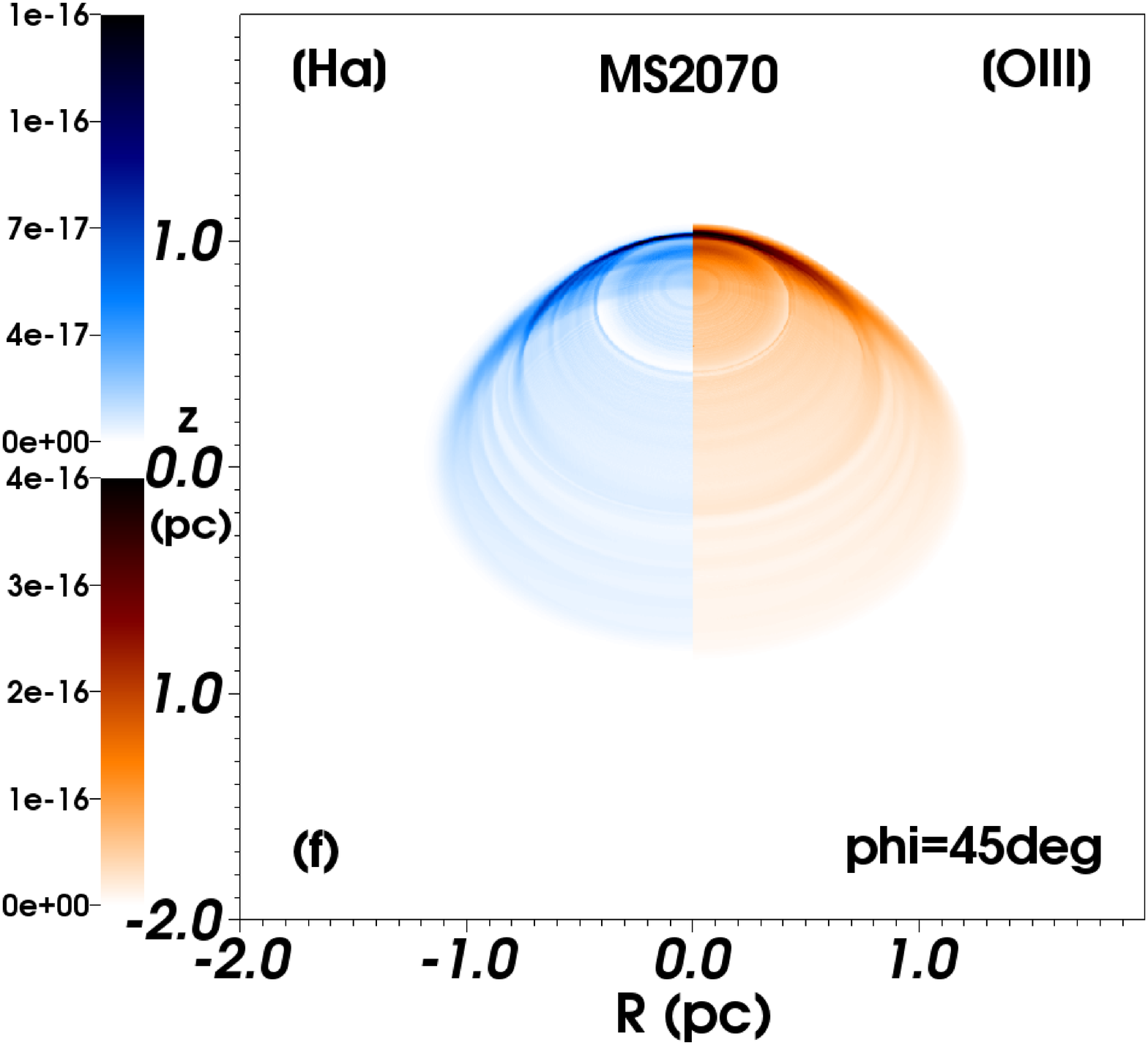}
	\end{minipage}
	\begin{minipage}[b]{ 0.29\textwidth}
		\includegraphics[width=1.0\textwidth]{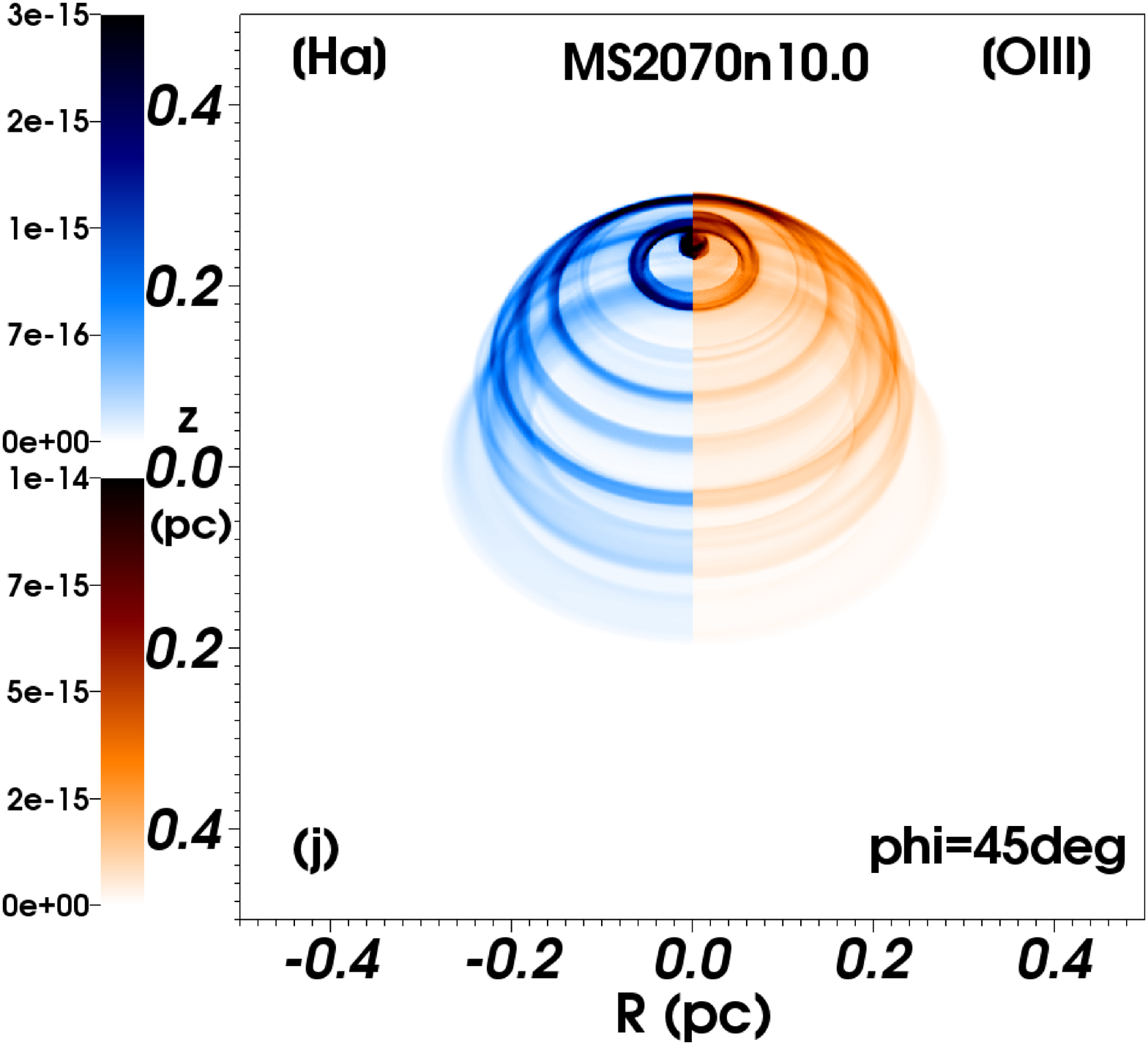}
	\end{minipage}\\	
	\begin{minipage}[b]{ 0.29\textwidth}
		\includegraphics[width=1.0\textwidth]{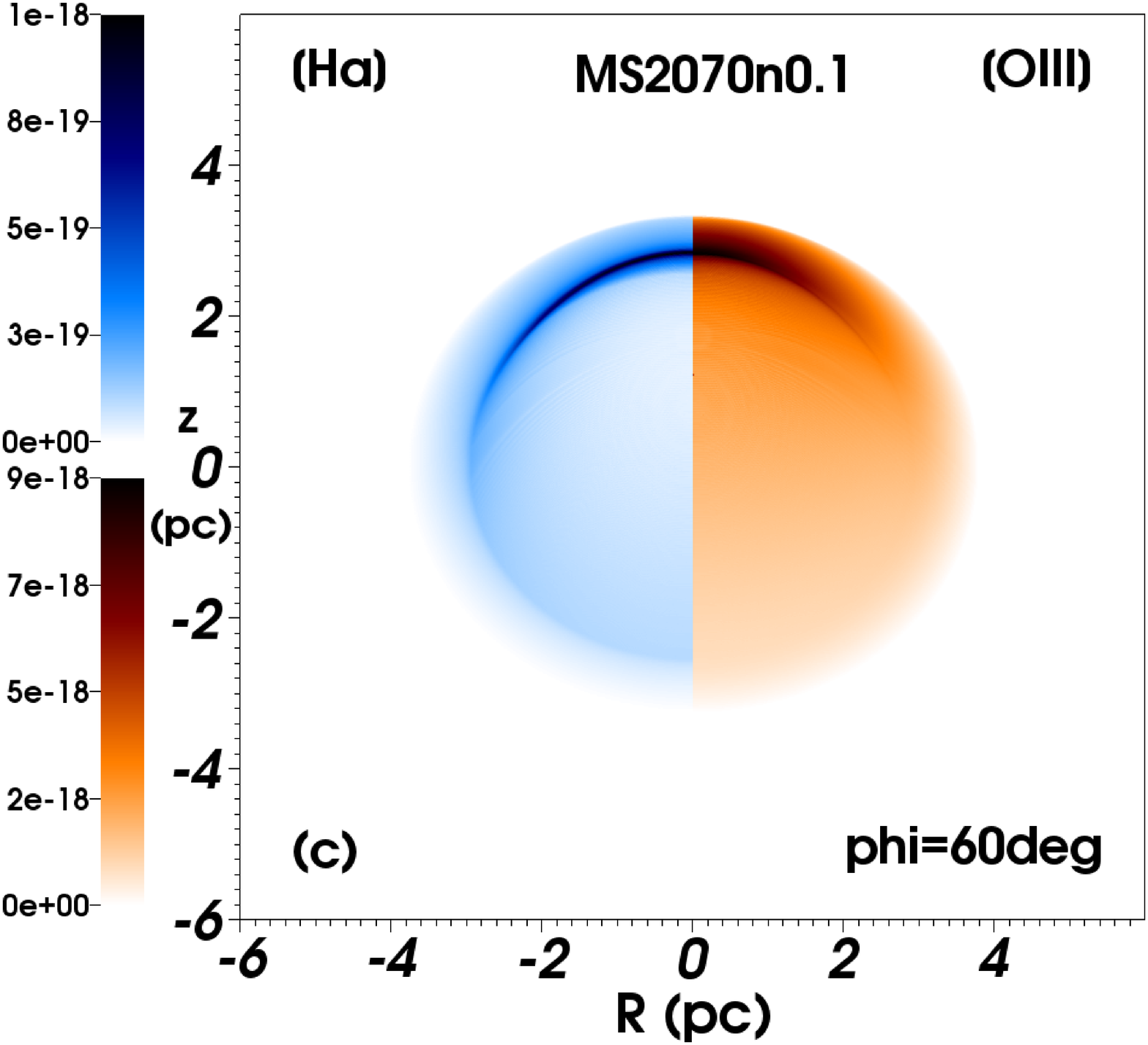}
	\end{minipage} 
	\begin{minipage}[b]{ 0.29\textwidth}
		\includegraphics[width=1.0\textwidth]{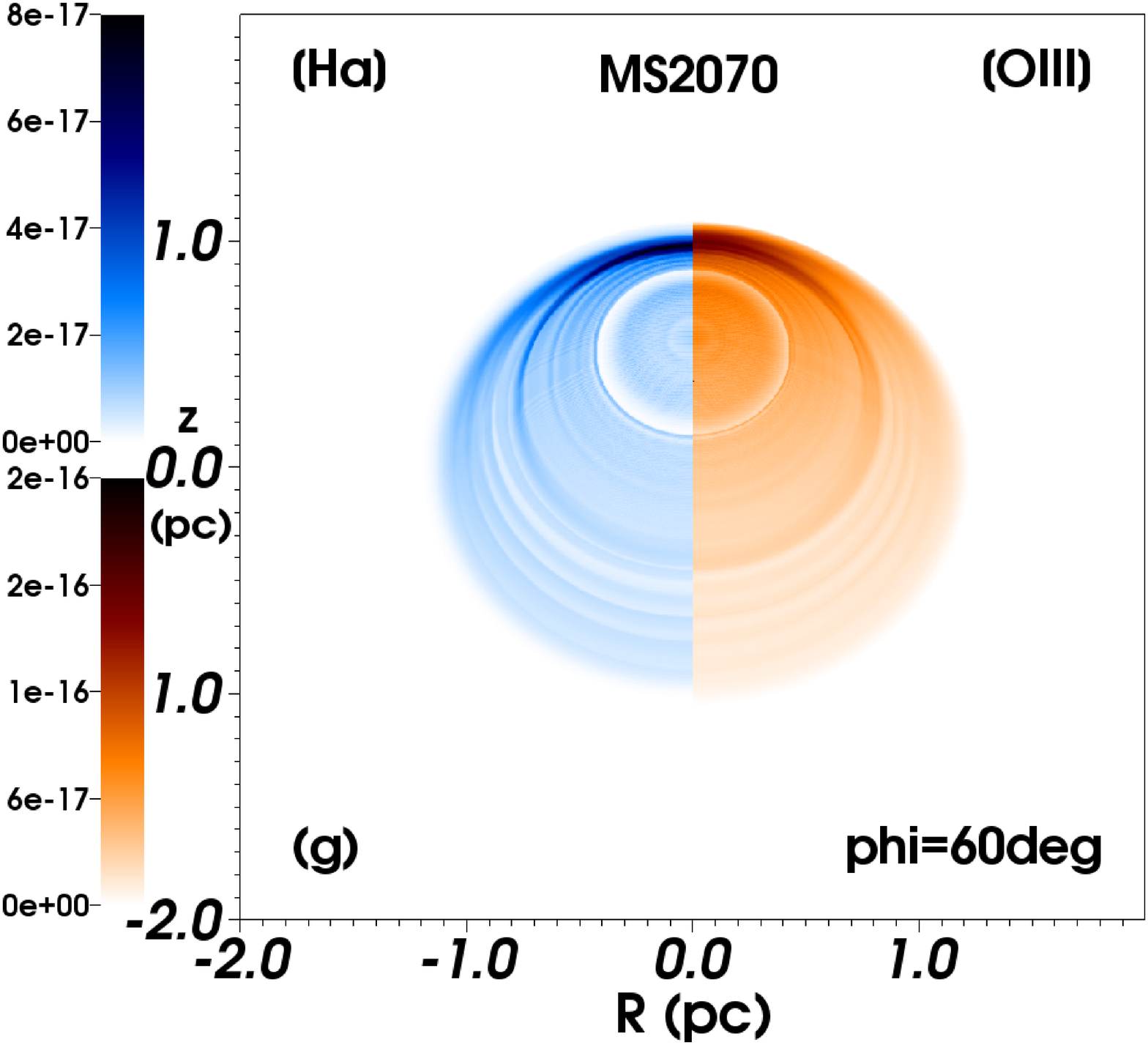}
	\end{minipage} 
	\begin{minipage}[b]{ 0.29\textwidth}
		\includegraphics[width=1.0\textwidth]{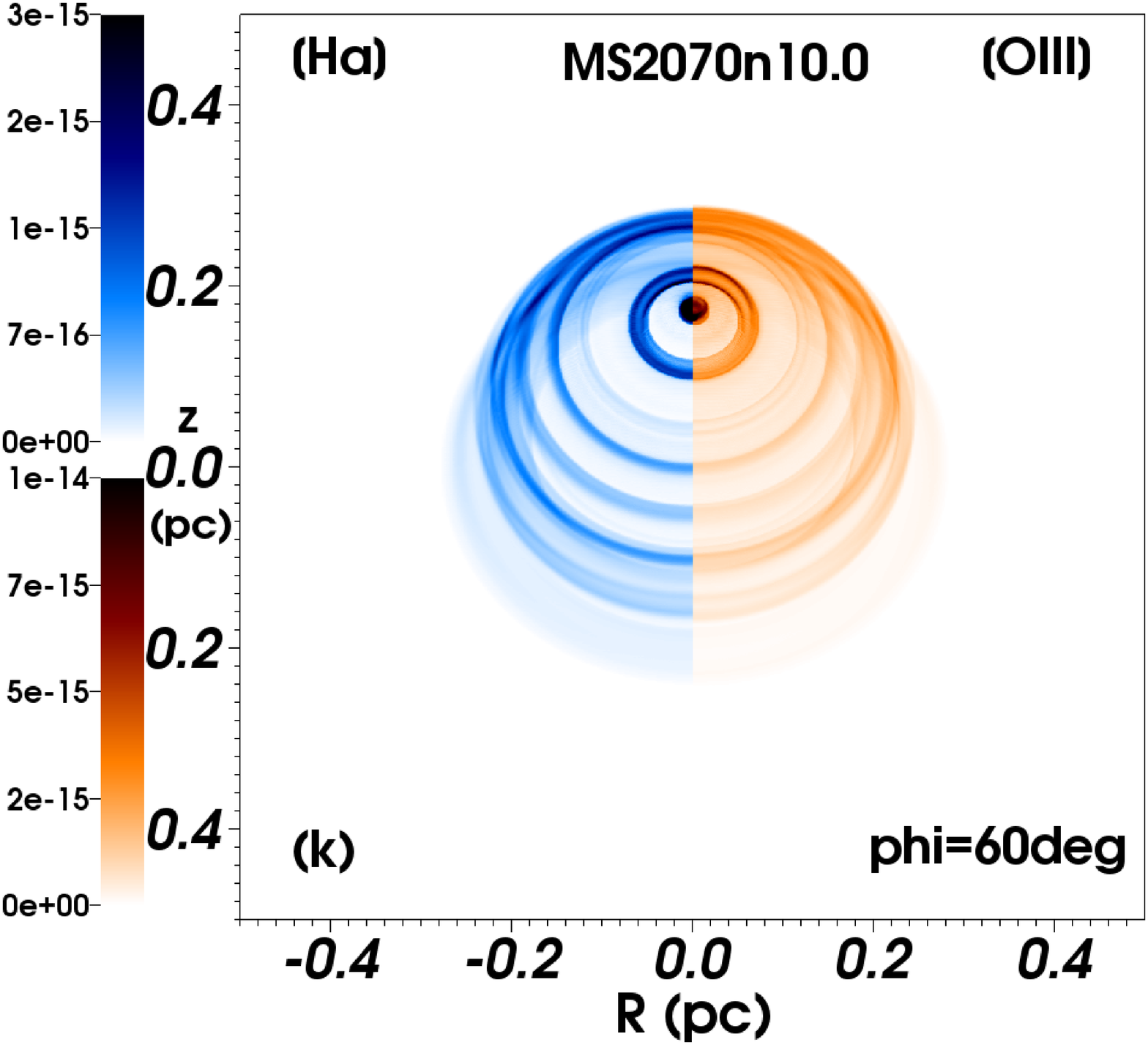}
	\end{minipage} \\	
	\begin{minipage}[b]{ 0.29\textwidth}
		\includegraphics[width=1.0\textwidth]{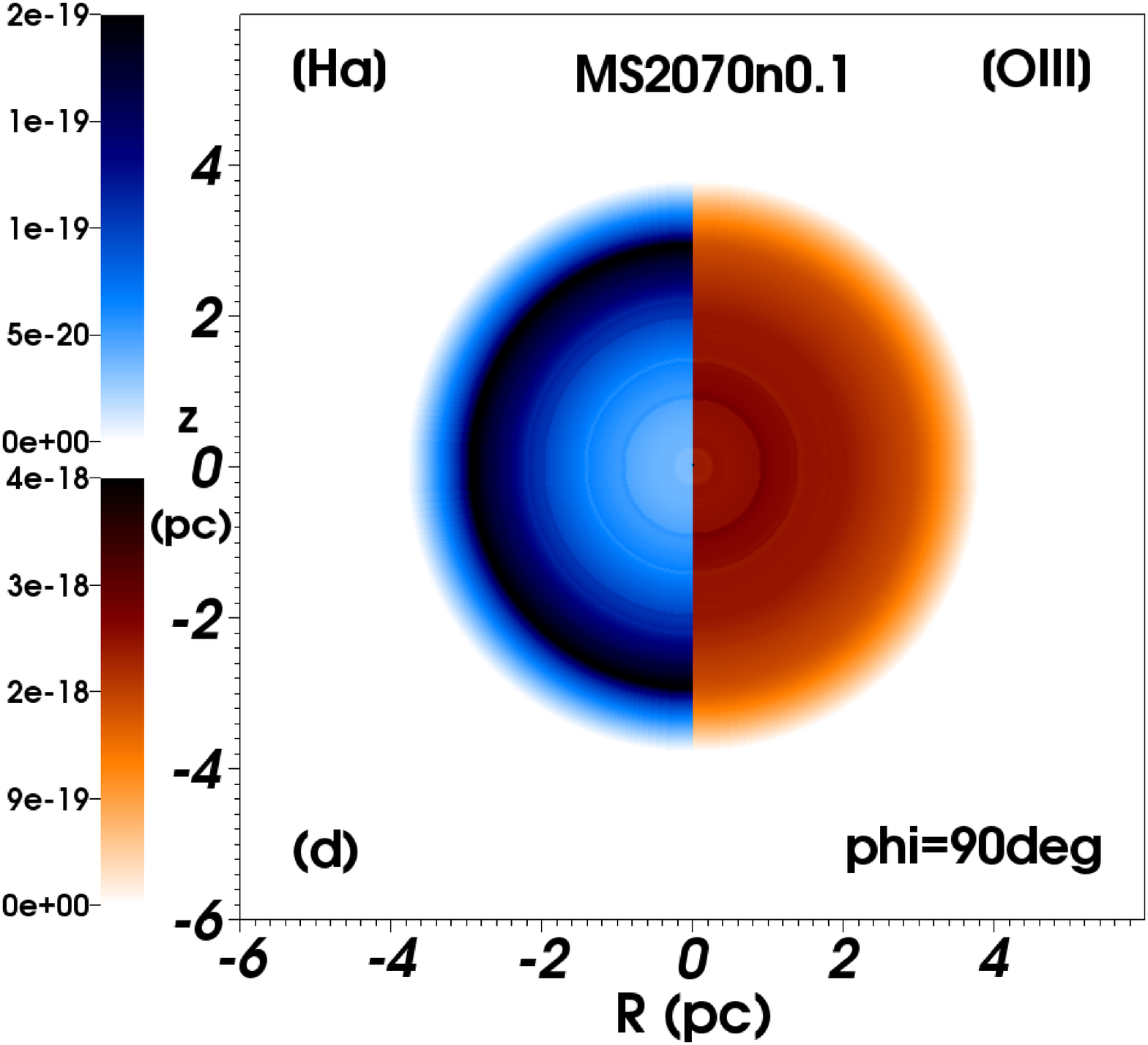}
	\end{minipage} 
	\begin{minipage}[b]{ 0.29\textwidth}
		\includegraphics[width=1.0\textwidth]{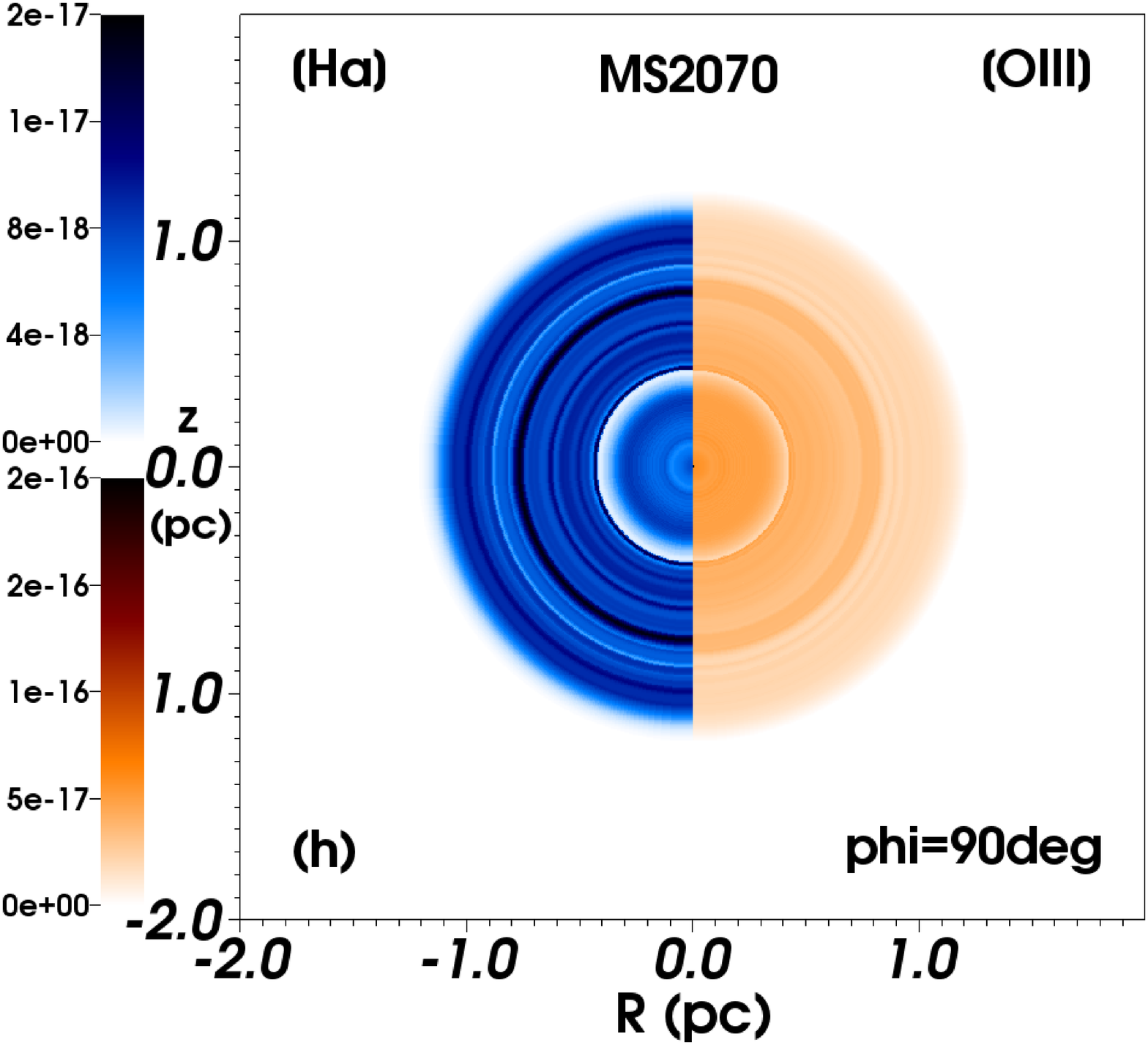}
	\end{minipage} 
	\begin{minipage}[b]{ 0.29\textwidth}
		\includegraphics[width=1.0\textwidth]{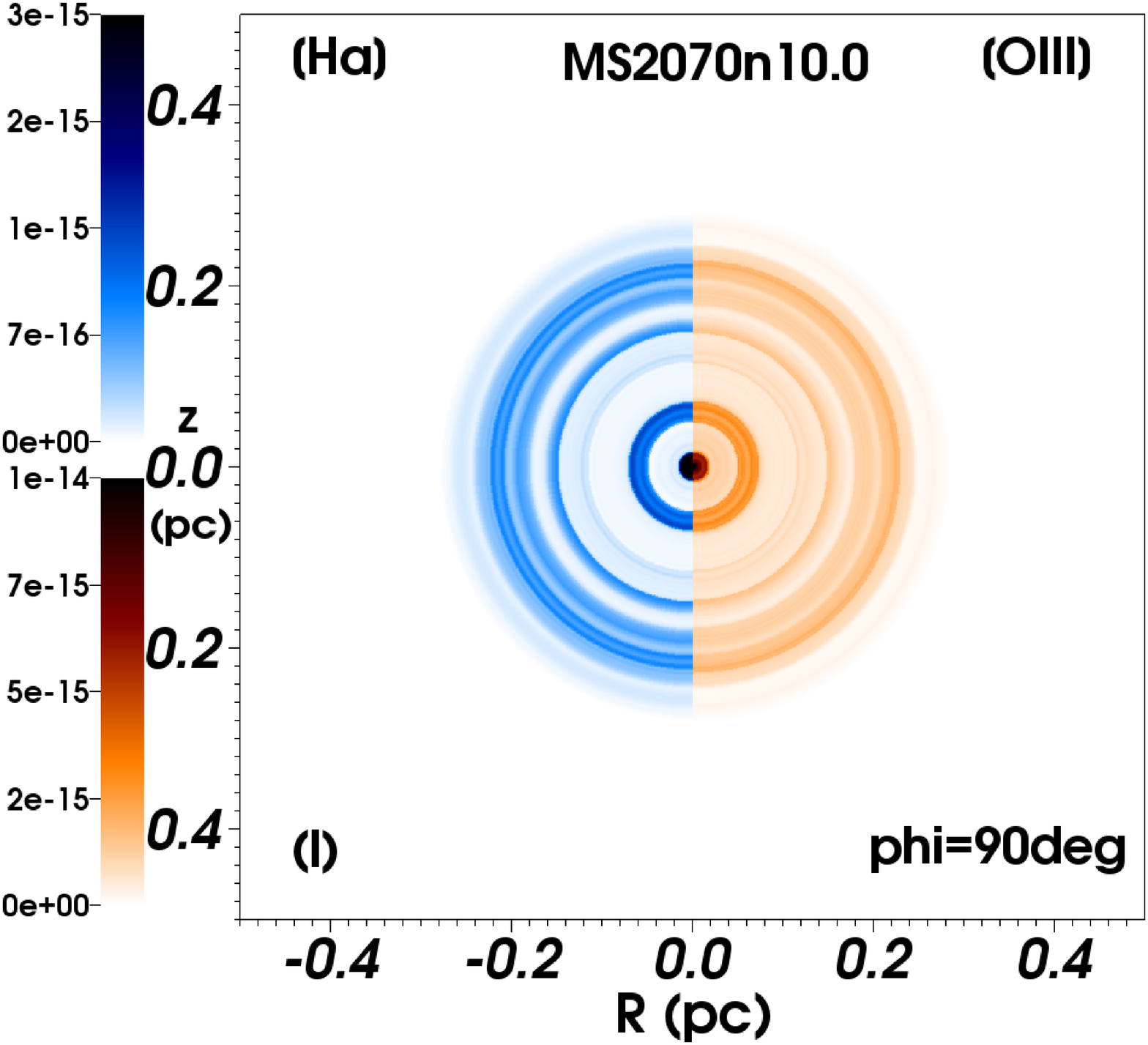}
	\end{minipage} \\	
	\caption{ 
	         H$\alpha$ surface brightness (left, in $\mathrm{erg}\, \mathrm{s}^{-1}\, 
	         \mathrm{cm}^{-2}\, \mathrm{arcsec}^{-2}$) and projected [O{\sc iii}] $\lambda \, 5007$ 
	         spectral line emission (right, in $\mathrm{erg}\, \mathrm{s}^{-1}\, 
	         \mathrm{cm}^{-2}\, \mathrm{arcsec}^{-2}$) of the bow shocks generated by a 
	         $20\, \rm M_{\odot}$  \textcolor{black}{ZAMS} star moving with velocity $v_{\star}=70\, \rm km\, \rm s^{-1}$ 
	         through in a medium with $n_{\rm ISM}=0.1$ (left column of panels a-d), $0.79$ 
	         (middle column of panels e-h) and $10\, \rm cm^{-3}$ (right column of panels i-l). 
	         The figures correspond to an inclination angle $\phi=30\degree$ (top line of panels a,e,i), 
	         $\phi=45\degree$ (second line of panels b,f,j), $\phi=60\degree$ (third line of panels c,g,k) and 
	         $\phi=90\degree$ (bottom line of panels d,h,l) with respect to the plane of the sky. 
	         Quantities are calculated excluding the undisturbed ISM and plotted in the linear scale, 
	         as a function of the inclination angle and the ambient medium density. 
		 }	
	\label{fig:maps1}  
\end{figure*}

\subsubsection{Synthetic optical emission maps}
\label{sect:maps}

In Fig.~\ref{fig:maps1} we plot synthetic H$\alpha$ and  [O{\sc iii}] $\lambda \, 5007$  emission maps of the bow shocks 
generated by our $20\, \rm M_{\odot}$  \textcolor{black}{ZAMS} star moving with velocity $70\, 
\mathrm{km}\, \mathrm{s}^{-1}$ moving through a medium with $n_{\rm ISM}=0.1$ 
(left column of panels), $0.79$ (middle column of panels) and $10.0\, \rm 
cm^{-3}$ (right column of panels). 
%
%
The region of maximum H$\alpha$ emission of the gas is located 
close to the apex of the bow shock and extended to its trail ($z\, 
\le 0$). This broadening of the emitting region is due to the high space 
velocity of the star, see Paper~I. Neither the shocked stellar wind 
nor the hot shocked ISM of the bow shock contributes significantly to these 
emission since the $\rm H\alpha$ emission coefficient $j_{\rm H\alpha} \propto T^{-0.9}$ and the contact 
discontinuity is the brightest part of the whole structure 
(Fig.~\ref{fig:maps1}a). The [O{\sc iii}] $\lambda \, 5007$ emission is maximum at 
the same location but, however, slightly different dependence on the 
temperature of the corresponding emission coefficient $j_{\rm [OIII]} \propto 
\exp(-1/T)/T^{1/2}$~\citep{dopita_aa_29_1973} induces a weaker extension of the 
emission to the tail of the structure (Fig.~\ref{fig:maps1}a). The unstable simulations with 
$v_{\star}\, \ge 40\, \mathrm{km}\, \mathrm{s}^{-1}$ and $n_{\rm ISM} \simeq 
10\, \rm cm^{-3}$ have ring-like artefacts which dominate the emission (see 
Fig.~\ref{fig:maps1}e-h and Fig.~\ref{fig:maps1}i-l). They are artificially generated by the 
over-dense regions of the shell that are rotated and mapped onto the Cartesian grid. A 
tri-dimensional unstable bow shock would have brighter clumps of matters 
sparsed around its layer of cold shocked ISM rather than regular 
rings~\citep{mohamed_aa_541_2012}. Regardless of the 
properties of their driving star, our bow shocks are brighter in large ambient medium, 
e.g. the model MS2070n0.1 has $\Sigma^{\rm max}_{[\rm H\alpha]} \approx 10^{-18} 
\, \mathrm{erg}\, \mathrm{s}^{-1}\, \mathrm{cm}^{-2}\, \mathrm{arcsec}^{-2}$ 
whereas the model MS2070n10 has $\Sigma^{\rm max}_{[\rm H\alpha]} \approx 
3\times 10^{-15} \, \mathrm{erg}\, \mathrm{s}^{-1}\, \mathrm{cm}^{-2}\, 
\mathrm{arcsec}^{-2}$. The projected [O{\sc iii}] $\lambda \, 5007$ emission 
behaves similarly.

\begin{figure}
\centering
	\includegraphics[width=0.45\textwidth,angle=0]{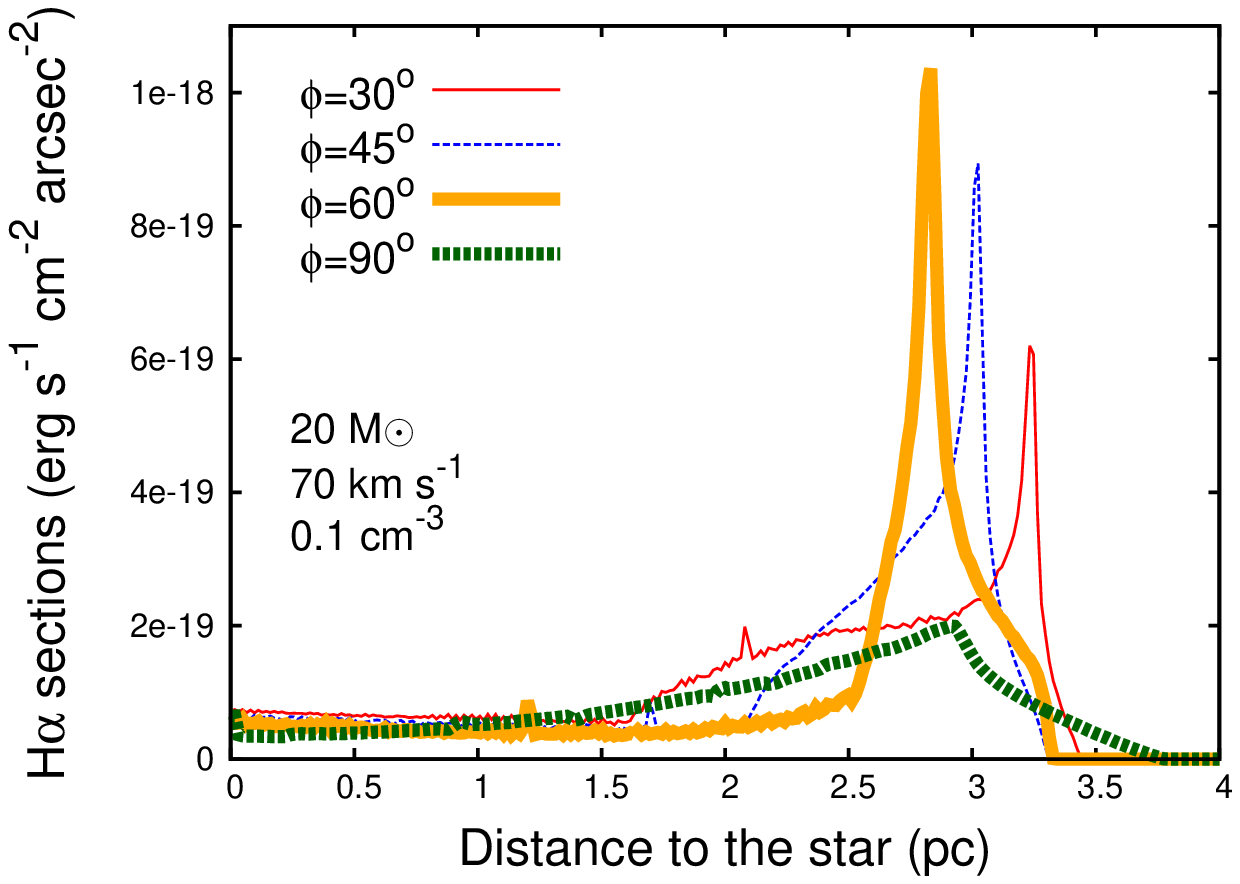}
	\caption{ 
	Cross-sections taken along the direction of motion of our $20\, \rm M_{\odot}$ 
	 \textcolor{black}{ZAMS} star moving with velocity $70\, \rm km\, \rm s^{-1}$ in an ambient medium of number density 
	$n_{\rm ISM}=0.1\, \rm cm^{-3}$. The data are plotted for inclination angles $\phi=30\degree$ (thin 
	solid red line), $\phi=45\degree$ (thin dotted blue line), $\phi=60\degree$ (thick solid orange line) and $\phi=90\degree$ 
	(thick dotted dark green line) through their H$\alpha$ surface brightness (see Fig.~\ref{fig:maps1}a-d). 
	The position of the star is located at the origin. 
		 }	
	\label{fig:profiles}  
\end{figure}

\begin{figure}
	\centering
	\begin{minipage}[b]{ 0.45\textwidth}
		\includegraphics[width=1.0\textwidth]{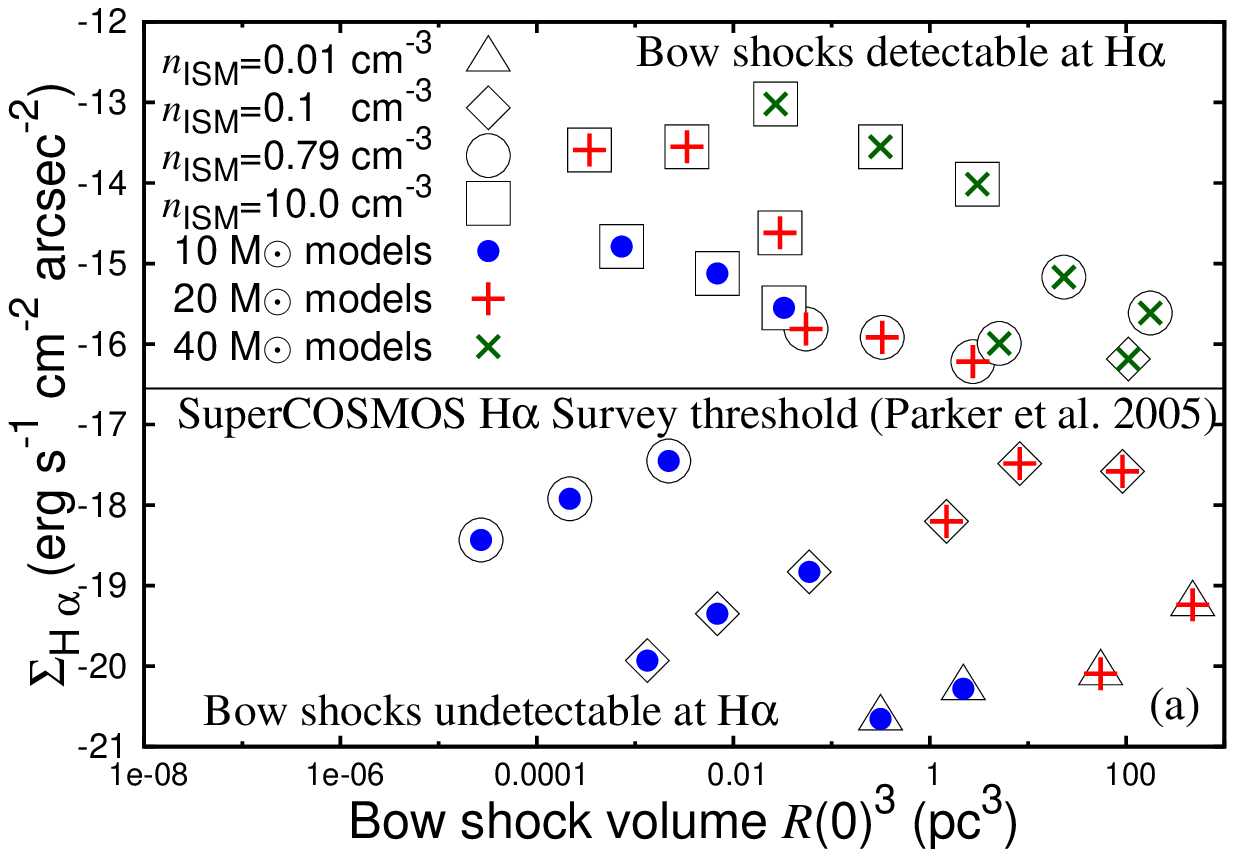}
	\end{minipage}		
	\begin{minipage}[b]{ 0.45\textwidth}
		\includegraphics[width=1.0\textwidth]{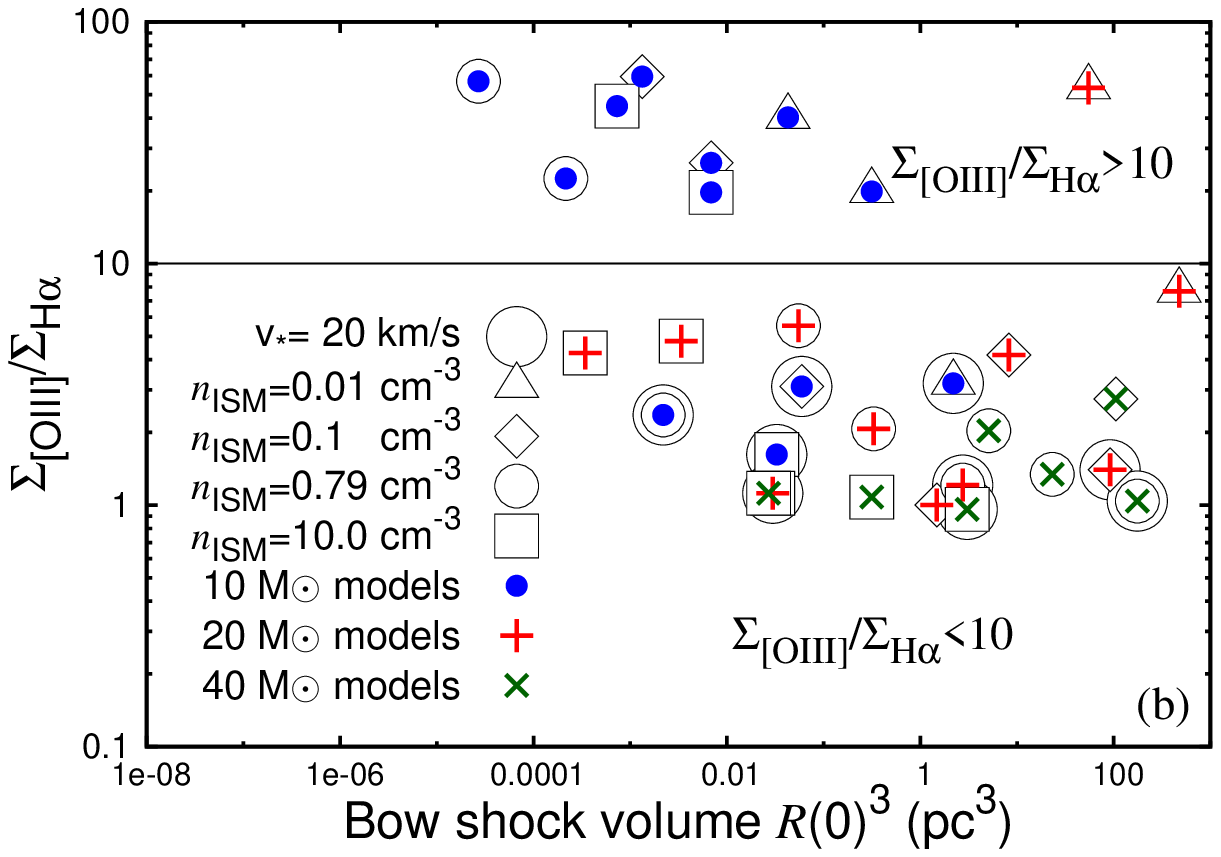}
	\end{minipage}	\\
	\caption{ 
	         Bow shock H$\alpha$ surface brightness (a) and ratio $\Sigma^{\rm max}_{[\rm O{\sc III}]}/
	         \Sigma^{\rm max}_{[\rm H\alpha]}$ (b) as a function of its volume $R(0)^{3}$ (in $\rm pc^{3}$). 
		 Upper panel shows the H$\alpha$ surface brightness as a function of the detection 
		 threshold of the SuperCOSMOS H$\alpha$ Survey (SHS) of $\Sigma_{\rm SHS}\approx 1.1-2.8 \times 10^{-17}\, 
	         \rm erg\, \rm s^{-1}\, \rm cm^{-2}\, \rm arcsec^{-2}$~\citep{parker_mnras_362_2005}. Lower panel plots 
	         the ratio $\Sigma^{\rm max}_{[\rm O{\sc III}]}/\Sigma^{\rm max}_{[\rm H\alpha]}$ of the same models. 
		 }	
	\label{fig:vol_vs_S_Ha}  
\end{figure}

In Fig.~\ref{fig:profiles} we show cross-sections of the H$\alpha$ surface 
brightness of the model MS2070n0.1. The cuts are taken along the symmetry axis 
of the figures and plotted as a function of the inclination angle $\phi$ with 
respect to the plane of the sky. The emission rises slightly as $\phi$ 
increases from  for $\phi=30\degree$ (thin red solid line) to $\phi=60\degree$ 
(thick solid orange line) since $\Sigma^{\rm max}_{[\rm H\alpha]}$ peaks at 
about $6\times 10^{-19}$ and about $10^{-18}\, \mathrm{erg}\, \mathrm{s}^{-1}\, 
\mathrm{cm}^{-2}\, \mathrm{arcsec}^{-2}$, respectively. The case with 
$\phi=90\degree$ is different since the emission decreases to about $\approx 
2\times 10^{-19}\, \rm erg\, \rm s^{-1}\, \rm cm^{-2}\, \rm arcsec^{-2}$ (see 
thick dotted green line in Fig.~\ref{fig:profiles}). 
The same is true for the [O{\sc iii}] emission since its dependence on the post-shock 
density is similar. \textcolor{black}{These differences arise because a 
line-of-sight corresponding to $\phi=60\degree$ intercepts a larger amount of dense, emitting 
material in the layer of shocked ISM than a line-of-sight corresponding to $\phi \le 30\degree$ or 
$\phi=90\degree$.} Large angles of inclination make the opening of the bow 
shocks larger (Fig.~\ref{fig:profiles}a-c, e-g, i-k) and the stand-off distance 
appears smaller (Fig.~\ref{fig:profiles}a-c). Note that bow shocks observed with 
a viewing angle of $\phi=90\degree$ do not resemble an arc-like shape but 
rather an overlapping of iso-emitting concentric circles (Fig.~\ref{fig:profiles}d,h,l).

\subsubsection{Bow shocks observability at H$\alpha$ and comparison with observations}
\label{sect:comp}

In Fig.~\ref{fig:vol_vs_S_Ha} we show our bow shocks' H$\alpha$ surface 
brightness (a) and their $\Sigma^{\rm max}_{[\rm O{\sc III}]}/\Sigma^{\rm 
max}_{[\rm H\alpha]}$ ratio (b), both as a function of the volume of emitting 
gas ($z\ge0$). The color coding of both panels  takes over the definitions 
adopted in Fig.~\ref{fig:axis_ratio}. The models with a $10\, \rm 
M_{\odot}$  \textcolor{black}{ZAMS star} have a volume smaller than about a few $\rm pc^{3}$ and have emission 
smaller than about $10^{-15}\, \rm erg\, \rm s^{-1}\, \rm cm^{-2}\, \rm 
arcsec^{-2}$. The models with $M_{\star}=20\, \rm M_{\odot}$ have larger volume 
at equal $n_{\rm ISM}$ and can reach surface brightness of about a few 
$10^{-14}\, \rm erg\, \rm s^{-1}\, \rm cm^{-2}\, \rm arcsec^{-2}$ if $n_{\rm 
ISM}=10\, \rm cm^{-3}$. Note that all models with $n_{\rm ISM} \ge 10.0\, \rm 
cm^{-3}$ produce emission larger than the diffuse emission sensitivity threshold 
of the {\it SuperCOSMOS H-Alpha Survey} (SHS) of $\Sigma_{\rm SHS} \approx 
1.1$-$2.8 \times 10^{−17}\, \rm erg\, \rm s^{-1}\, \rm cm^{-2}\, \rm 
arcsec^{-2}$~\citep{parker_mnras_362_2005} and such bow shocks should 
consequently be observed by this survey (see horizontal black line in 
Fig.~\ref{fig:vol_vs_S_Ha}a).

As discussed above, a significant fraction of our sample of bow shocks 
models have a H$\alpha$ surface brightness larger than the sensitivity limit of 
the SHS survey~\citep{parker_mnras_362_2005}. This 
remark can be extended to other (all-sky) H$\alpha$ observations campaigns, 
especially if their detection threshold is lower than the SHS. This is the 
case of, e.g. the {\it Virginia Tech Spectral-Line Survey} 
survey~\citep[VTSS,][]{dennison_aas_195_1999} and the {\it Wisconsin H-Alpha 
Mapper}~\citep[WHAM,][]{reynolds_pasa_15_1998} which provide us with images of 
diffuse sensitivity detection limit that allow the revelation of structures 
associated with sub-Rayleigh intensity. Consequently, one can expect to find 
optical traces of stellar wind bow shocks from OB stars in these data. According 
to our study, their driving stars are more likely to be of initial mass 
$M_{\star} \ge 20\, \rm M_{\odot}$ (Fig.~\ref{fig:vol_vs_S_Ha}a). 
This also implies that bow shocks {\it in the field} that are observed with such 
facilities are necessary produced by runaway stars of initial mass larger than 
$M_{\star} \ge 20\, \rm M_{\odot}$. Moreover, we find that the models involving 
an $10\, \rm M_{\odot}$ star and with $v_{\star}\, \ge 40\, \mathrm{km}\, 
\mathrm{s}^{-1}$ have $\Sigma^{\rm max}_{[\rm O{\sc III}]}/\Sigma^{\rm 
max}_{[\rm H\alpha]}>10$, whereas almost all of the other simulations do not 
satisfy this criterion (Fig.~\ref{fig:vol_vs_S_Ha}b).

\textcolor{black}{
Furthermore}, we find a similarity between some of the cross-sections taken along the 
symmetry axis of the H$\alpha$ surface brightness of our bow shock models 
(Fig.~\ref{fig:profiles}) and the measure of the radial brightness in emission 
measure of the bow shock generated by the runaway O star HD 57061~\citep[see fig.~5 
of][]{brown_aa_439_2005}. This observable and our model authorize a comparison 
since H$\alpha$ emission and emission measures have the same quadratic dependence 
on the gas number density. 
The emission measure profile of HD 57061 slightly increases from the star to the 
bow shock and steeply peaks in the region close to the contact discontinuity, before 
to decrease close to the forward shock of the bow shock and reach the ISM background 
emission. Our H$\alpha$ profile with $\phi=60\degree$ is consistent with (i) the above 
described variations and (ii) with the estimate of the inclination of the symmetry axis 
of HD 57061 with respect to the plane of the sky of about $75\degree$, see table~3 
of~\citet{brown_aa_439_2005}. Note that according to our simulations, the emission 
peaks in the region separating the hot from the cold shocked ISM gas.

\textcolor{black}{
\citet{brown_aa_439_2005} extracted a subset of 8 bow shocks at H$\alpha$ from 
the catalogue compiled by~\citet{vanburen_aj_110_1995}. The bow shocks of the 
stars HD149757 and HD158186 do not match any of our models. The O6.5V star 
HD17505 moving in a medium with $n_{\rm ISM} \approx 21\, \rm cm^{-3}$ is also 
uncompatible with the space of parameter covered by our study. The circumstellar 
nebulae of HD92206 ($R(0) \approx 3.67\, \rm pc$, $n_{\rm ISM} \approx0.007\, 
\rm cm^{-3}$, $v_{\star} \approx 40.5\, \rm km\, \rm s^{-1}$) and HD135240 
($R(0) \approx 3.50\, \rm pc$, $n_{\rm ISM}\approx0.21\, \rm cm^{-3}$, 
$v_{\star} \approx 32.5\, \rm km\, \rm s^{-1}$) have some of their properties 
similar to our models MS2040n0.01/MS2070n0.01 ($R(0) \approx 7.80$-$3.80\, \rm 
pc$, $n_{\rm ISM}=0.01\, \rm cm^{-3}$, $v_{\star}=40$-$70\, \rm km\, \rm 
s^{-1}$) and MS2020n0.1/MS2040n0.1 ($R(0) \approx 3.51$-$2.02\, \rm pc$, $n_{\rm 
ISM}=0.1\, \rm cm^{-3}$, $v_{\star}=20$-$40\, \rm km\, \rm s^{-1}$) but do not 
properly fit them. The bow shock of HD57061 ($R(0) \approx 7.56\, \rm pc$, $n_{\rm 
ISM}\approx 0.07\, \rm cm^{-3}$, $v_{\star} \approx 55.8\, \rm km\, \rm s^{-1}$) 
matches particularly our model MS2040 with $R(0) \approx 7.8\, \rm pc$, $n_{\rm 
ISM}=0.01\, \rm cm^{-3}$ and $v_{\star}=40\, \rm km\, \rm s^{-1}$, and therefore 
constitute a good candidate for future tailored numerical simulations. 
More detailed simulations and subsequent post-processing of the corresponding data, 
e.g. including the effects of the extinction of the ISM on the infrared emission 
are necessary for a more detailed discussion of these results.
}

\subsubsection{Implication for the evolution of supernova remnants generated by massive runaway stars}
\label{sect:pre_sn}

Massive stars evolve and die \textcolor{black}{as} supernovae, a sudden and strong release of matter, energy 
and momentum taking place inside the ISM pre-shaped by their past stellar 
evolution~\citep{langer_araa_50_2012}. In the case of a runaway progenitor, the 
circumstellar medium at the pre-supernova phase can be a bow shock nebula with 
which the shock wave interacts \textcolor{black}{before expanding further} into the unperturbed 
ISM~\citep{brighenti_mnras_270_1994}. The subsequent growing supernova remnant 
develops asymmetries since it is braked by the mass at the apex of the bow 
shock but expands freely in the cavity driven by the star in the opposite  
direction~\citep{borkowski_apj_400_1992}. If the progenitor is slightly 
supersonic, the bow shock is mainly shaped during the main-sequence phase of the 
star; whereas if the progenitor is a fast-moving star then the bow shock is 
essentially made of material from the last pre-supernova evolutionary phase. In 
the Galactic plane ($n_{\rm ISM}=0.79\, \rm cm^{-3}$) such asymmetries arise if 
the apex of the bow shock accumulates at least $1.5\, \rm M_{\odot}$ of shocked 
material~\citep{meyer_mnras_450_2015}.

\begin{figure}
	\centering
	\begin{minipage}[b]{ 0.45\textwidth}
		\includegraphics[width=1.0\textwidth]{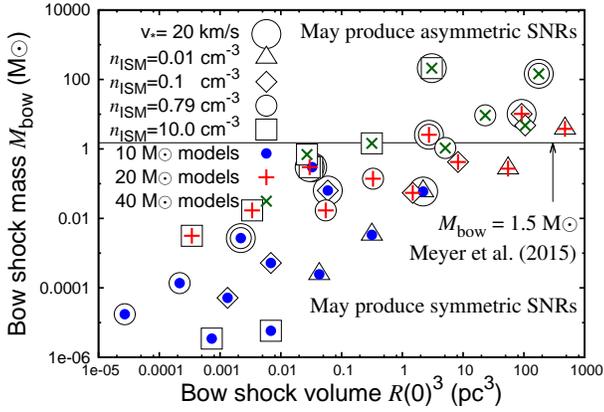}
	\end{minipage}	\\	
	\caption{ 
	         Bow shocks mass as a function of the bow shock volume. 
		 The figure shows the mass $M_{\rm bow}$ (in $M_{\odot}$) trapped in the $z \ge 0$ 
		 region of the bow shock as a function of its volume $R(0)^{3}$ (in $\rm pc^{3}$).
		 The dots distinguish between models (i) as a function of the ISM ambient medium with 
		 $n_{\rm ISM}=0.01$ (triangles), $0.1$ (diamonds), $0.79$ (circles) and $10\, \rm cm^{-3}$ 
		 (squares), and (ii) as a function of the initial mass of the star with $10$ (blue dots), 
		 $20$ (red plus signs) and $40\, \rm M_{\odot}$ (green crosses).
		 The thin horizontal black line corresponds to $M_{\rm bow} = 1.5\, \rm M_{\odot}$, i.e. the 
		 condition to produce an asymmetric supernova remnant if 
		 $n_{\rm ISM}=0.79\, \rm cm^{-3}$~\citep{meyer_mnras_450_2015}.  
		 }	
	\label{fig:mass}  
\end{figure}

In Fig.~\ref{fig:mass} we present the mass trapped into the $z\ge0$ region of our 
bow shock models as a function of their volume. As in Fig.~\ref{fig:vol_vs_S_Ha} 
the figure distinguishes the initial mass and the ambient medium density of each 
models. Amongst our bow shock simulations, 9 models have $M_{\rm bow}  \gtrsim  
1.5\, \rm M_{\odot}$ and 4 of them are generated by the runaway stars which 
asymmetric supernova remnant studied in detail in~\citet{meyer_mnras_450_2015}. 
The other models with $v_{\star} \le 40\, \rm km\, \rm s^{-1}$ may produce 
asymmetric remnants because they will explode inside their main-sequence wind 
bubble. The model MS4070n0.1 has $v_{\star}=70\, \rm km\, \rm s^{-1}$ which 
indicates that the main-sequence bow shock will be advected downstream by the 
rapid stellar motion and the surroundings of the progenitor at the pre-supernova 
phase is made of, e.g. red supergiant material. Consequently, its shock wave may 
be unaffected by the presence of the circumstellar medium. We leave the 
examination via hydrodynamical simulations of this conjecture for future 
works. Interestingly, we notice that most of the potential progenitors of 
asymmetric supernova remnants are moving in a low density medium $n_{\rm ISM} 
\le 0.1\, \rm cm^{-3}$, i. e. in the rather high latitude regions of the Milky 
Way. This is consistent with the interpretation of the elongated shape of, e.g. 
Kepler's supernova remnant as the consequence of the presence of a massive bow 
shock at the time of the 
explosion~\citep{velazquez_apj_649_2006,toledoray_mnras_442_2014}.

\subsubsection{The influence of the interstellar magnetic field on the shape of supernovae remnants}

\textcolor{black}{
An alternative explanation for the asymmetrical shape of supernova remnants can 
be found in the influence of the interstellar magnetic field. Although the 
interstellar magnetic field does not influence the shape of an expanding supernova 
blast wave directly~\citep{manchester_aa_171_1987} it can influence the shape and size of the 
wind-blown bubble, as suggested by~\citet{arnal_aa_254_1992} and shown numerically 
by~\citet{vanmarle_2015}. 
Such magnetic fields slow the expansion of the  wind-blown 
bubble in the direction perpendicular to the direction of the field and, 
depending on the field strength can stop the expansion in that direction 
completely. The end result is an elongated, ellipsoid bubble, which in turn, 
would influence the expansion of the supernova remnant. 
}

\textcolor{black}{
As shown by~\citet{vanmarle_2015}, the interstellar magnetic field would have to 
be fairly strong (beyond about $20\, \mu \rm G$) to enable it to constrain the wind-bubble 
sufficiently that the scale of the bubble would be reduced to that of a bow 
shock. However, such field-strengths are not unreasonable as field strengths of 
up to $60\, \mu \rm G$, have been observed in the galactic core~\citep{rand_apj_343_1989,
ohno_mnras_262_1993,frick_mnras_325_2001,opher_natur_462_2009,shabala_mnras_405_2010,
fletcher_mnras_412_2011,heerikhuisen_apj_738_2011,
vallee_newar_55_2011} and fields  that are stronger 
than that by an order of magnitude can be found inside molecular 
clouds~\citep{crutcher_apj_515_1999}. 
}

\textcolor{black}{
There are two tests that can be used to distinguish whether the bubble into 
which a supernova remnant expands has been constrained by stellar motion, or by 
an interstellar magnetic field. One: the magnetic field in the galaxy tends to 
be aligned with the spiral arms.~\citet{gaensler_apj_493_1998} showed that supernova 
remnants tend to be aligned with the galactic disk. This would seem to support 
the theory that it is the magnetic field, rather than a bow shock,  that 
constrained the wind expansion. However, the correlation is not very strong. 
Two: the shape of the  supernova remnant itself. Van Marle (2015) showed 
that a supernova remnant, expanding inside a magnetically constrained bubble, 
first collides with the outer edge along the minor axis of the ellipsoid bubble, 
while, in the direction along the major axis the expansion can 
continue uninterrupted. This produces a barrel-like supernova remnant. When 
expanding inside a bow shock, the collision would first occur at the front of 
the bow shock, creating a parabolic shape with free expansion only possible 
along the tail of the bow shock. 
}

\section{Conclusion}
\label{section:cc}


Our bow shock simulations indicate that no structural difference arise when 
changing the density of the background ISM in which the stars move, i.e. their 
internal organisation is similar as described 
in~\citet{comeron_aa_338_1998} and Paper~I. 
The same is true for their radiative properties, governed by line cooling 
such as the [O{\sc iii}] $\lambda \, 5007$ line and showing faint H$\alpha$ emission, both 
principally originating from outer region of shocked 
ISM gas. 
We also find that their X-rays signature is fainter by several orders of 
magnitude than their H$\alpha$ emission, and, consequently, it is not a good 
waveband to search for such structures. 
%

The best way to observe bow shocks remains their infrared emission of starlight 
reprocessed by shocked ISM dust~\citep{meyer_mnras_2013}. We find that the 
brightest infrared bow shocks, i.e. the most easily observable ones, are produced by 
high-mass ($M_{\star} \approx 40\, \rm M_{\odot}$) stars moving with a slow 
velocity ($v_{\star}\approx 20\, \rm km\, \rm s^{-1}$) in the relatively dense 
regions ($n_{\rm ISM}\approx 10\, \rm cm^{-3}$) of the ISM, whereas the brightest 
H$\alpha$ structures are produced by these stars when moving rapidly  
($v_{\star}\approx 70\, \rm km\, \rm s^{-1}$). Thin-shelled bow shocks have 
mid-infrared luminosities which does not report the time-variations of their unstable 
structures. This indicates that spectral energy distributions of stellar wind 
bow shocks are the appropriate tool to analyze them since they do not depend 
on the temporary effects that affect their density field. 
\textcolor{black}{We find that bow shocks from Galactic runaway stars 
have emission peaking in the wavelength range $3\, \leq\, \lambda\, \leq\, 50\, \mu \rm m$.  
Interestingly, the circumstellar material can be up to several oders of magnitude brighter 
than the star and dominates the emission, especially if $M_{\star} \ge 20\, \rm M_{\odot}$}.

A detailed analysis of our grid of simulations indicates that the H$\alpha$ surface 
brightness of Galactic stellar wind bow shocks increases if their angle of inclination  
with respect to the plane of the sky increases up to $\phi = 60\degree$, however, 
edge-on viewed bow shocks are particularly faint. We find that all bow shocks generated 
by a $40\, \rm M_{\odot}$  \textcolor{black}{ZAMS} runaway star could be observed with 
Rayleigh-sensitive H$\alpha$ facilities and that bow shocks observed {\it in the field} 
by means of these facilities should have an initial mass larger than about $20\, \rm M_{\odot}$. 
Furthermore, all of our bow shocks generated 
by a $10\, \rm M_{\odot}$  \textcolor{black}{ZAMS} star moving with $v_{\star}\ge 40\, \rm km\, 
\rm s^{-1}$ have a line ratio $\Sigma^{\rm max}_{[\rm O{\sc III}]}/\Sigma^{\rm max}_{[\rm H\alpha]}>10$. 
Our study suggests that slowly-moving stars of ZAMS mass $M_{\star} \ge 20\, \rm 
M_{\odot}$ moving in a medium of $n_{\rm ISM }\ge 0.1\, \rm cm^{-3}$ generate 
massive bow shocks, i.e. are susceptible to induce asymmetries in their subsequent 
supernova shock wave. 
This study will be enlarged, e.g. estimating observability of red supergiant stars. 


\section*{Acknowledgements}

\textcolor{black}{
We thank the anonymous referee for numerous comments which greatly improved both 
the quality and the presentation of the paper. 
}
D.~M.-A.~Meyer thanks P.~F.~Velazquez, F.~Brighenti and L.~Kaper for their advices, 
and  F.~P.~Wilkin for useful comments on stellar wind bow shocks which partly 
motivated this work. 
This study was conducted within the Emmy Noether research group on
"Accretion Flows and Feedback in Realistic Models of Massive Star Formation" 
funded by the German Research Foundation under grant no. KU 2849/3-1.  
A.-J.~van~Marle acknowledges support from FWO, grant G.0227.08, KU Leuven 
GOA/2008, 04 and GOA/2009/09. 
The authors gratefully acknowledge the computing time granted by the John von
Neumann Institute for Computing (NIC) and provided on
the supercomputer JUROPA at J\" ulich Supercomputing Centre (JSC). 
%


\bibliographystyle{mn2e}

\footnotesize{
\bibliography{grid}

\begin{thebibliography}{}

\end{thebibliography}


\begin{thebibliography}{}

\bibitem[\protect\citeauthoryear{{Adams}, {Herter}, {Gull}, {Schoenwald},
  {Keller}, {Berthoud}, {Stacy}, {Nikola} \& {Henderson}}{{Adams}
  et~al.}{2008}]{adams_2008}
{Adams} J.~D.,  {Herter} T.~L.,  {Gull} G.~E.,  {Schoenwald} J.,  {Keller}
  L.~D.,  {Berthoud} M.,  {Stacy} G.~J.,  {Nikola} T.,    {Henderson} C.~P.,
  2008, in Society of Photo-Optical Instrumentation Engineers (SPIE) Conference
  Series Vol.~7014 of Society of Photo-Optical Instrumentation Engineers (SPIE)
  Conference Series, {FORCAST: the first light instrument for SOFIA}.
p.~2

\bibitem[\protect\citeauthoryear{{Arnal}}{{Arnal}}{1992}]{arnal_aa_254_1992}
{Arnal} E.~M.,  1992, \aap, 254, 305

\bibitem[\protect\citeauthoryear{{Arnaud}}{{Arnaud}}{1996}]{arnaud_aspc_101_1996}
{Arnaud} K.~A.,  1996, in {Jacoby} G.~H.,  {Barnes} J.,  eds, Astronomical Data
  Analysis Software and Systems V Vol.~101 of Astronomical Society of the
  Pacific Conference Series, {XSPEC: The First Ten Years}.
p.~17

\bibitem[\protect\citeauthoryear{{Asplund}, {Grevesse}, {Sauval} \&
  {Scott}}{{Asplund} et~al.}{2009}]{asplund_araa_47_2009}
{Asplund} M.,  {Grevesse} N.,  {Sauval} A.~J.,    {Scott} P.,  2009, \araa, 47,
  481

\bibitem[\protect\citeauthoryear{{Blaauw}}{{Blaauw}}{1961}]{blaauw_bain_15_1961}
{Blaauw} A.,  1961, \bain, 15, 265

\bibitem[\protect\citeauthoryear{{Blondin} \& {Koerwer}}{{Blondin} \&
  {Koerwer}}{1998}]{blondin_na_57_1998}
{Blondin} J.~M.,  {Koerwer} J.~F.,  1998, \na, 3, 571

\bibitem[\protect\citeauthoryear{{Borkowski}, {Blondin} \&
  {Sarazin}}{{Borkowski} et~al.}{1992}]{borkowski_apj_400_1992}
{Borkowski} K.~J.,  {Blondin} J.~M.,    {Sarazin} C.~L.,  1992, \apj, 400, 222

\bibitem[\protect\citeauthoryear{{Brandl}, {Lenzen}, {Venema}, {K{\"a}ufl},
  {Finger}, {Glasse}, {Brandner} \& {Stuik}}{{Brandl}
  et~al.}{2006}]{brandl_2006}
{Brandl} B.,  {Lenzen} R.,  {Venema} L.,  {K{\"a}ufl} H.-U.,  {Finger} G.,
  {Glasse} A.,  {Brandner} W.,    {Stuik} R.,  2006, in Society of
  Photo-Optical Instrumentation Engineers (SPIE) Conference Series Vol.~6269 of
  Society of Photo-Optical Instrumentation Engineers (SPIE) Conference Series,
  {MIDIR/T-OWL: the thermal/mid-IR instrument for the E-ELT}.
p.~20

\bibitem[\protect\citeauthoryear{{Brighenti} \& {D'Ercole}}{{Brighenti} \&
  {D'Ercole}}{1994}]{brighenti_mnras_270_1994}
{Brighenti} F.,  {D'Ercole} A.,  1994, \mnras, 270, 65

\bibitem[\protect\citeauthoryear{{Brighenti} \& {D'Ercole}}{{Brighenti} \&
  {D'Ercole}}{1995}]{brighenti_mnras_277_1995}
{Brighenti} F.,  {D'Ercole} A.,  1995, \mnras, 277, 53

\bibitem[\protect\citeauthoryear{{Brott}, {de Mink}, {Cantiello}, {Langer}, {de
  Koter}, {Evans}, {Hunter}, {Trundle} \& {Vink}}{{Brott}
  et~al.}{2011}]{brott_aa_530_2011a}
{Brott} I.,  {de Mink} S.~E.,  {Cantiello} M.,  {Langer} N.,  {de Koter} A.,
  {Evans} C.~J.,  {Hunter} I.,  {Trundle} C.,    {Vink} J.~S.,  2011, \aap,
  530, A115

\bibitem[\protect\citeauthoryear{{Brown} \& {Bomans}}{{Brown} \&
  {Bomans}}{2005}]{brown_aa_439_2005}
{Brown} D.,  {Bomans} D.~J.,  2005, \aap, 439, 183

\bibitem[\protect\citeauthoryear{{Comer\'{o}n} \& {Kaper}}{{Comer\'{o}n} \&
  {Kaper}}{1998}]{comeron_aa_338_1998}
{Comer\'{o}n} F.,  {Kaper} L.,  1998, \aap, 338, 273

\bibitem[\protect\citeauthoryear{{Comer{\'o}n} \& {Pasquali}}{{Comer{\'o}n} \&
  {Pasquali}}{2007}]{comeron_aa_467_2007}
{Comer{\'o}n} F.,  {Pasquali} A.,  2007, \aap, 467, L23

\bibitem[\protect\citeauthoryear{{Cowie} \& {McKee}}{{Cowie} \&
  {McKee}}{1977}]{cowie_apj_211_1977}
{Cowie} L.~L.,  {McKee} C.~F.,  1977, \apj, 211, 135

\bibitem[\protect\citeauthoryear{{Crutcher}, {Roberts}, {Troland} \&
  {Goss}}{{Crutcher} et~al.}{1999}]{crutcher_apj_515_1999}
{Crutcher} R.~M.,  {Roberts} D.~A.,  {Troland} T.~H.,    {Goss} W.~M.,  1999,
  \apj, 515, 275

\bibitem[\protect\citeauthoryear{{de Jager}, {Nieuwenhuijzen} \& {van der
  Hucht}}{{de Jager} et~al.}{1988}]{dejager_aas_72_1988}
{de Jager} C.,  {Nieuwenhuijzen} H.,    {van der Hucht} K.~A.,  1988, \aaps,
  72, 259

\bibitem[\protect\citeauthoryear{{Decin}, {Hony}, {de Koter}, {Justtanont},
  {Tielens} \& {Waters}}{{Decin} et~al.}{2006}]{decin_aa_456_2006}
{Decin} L.,  {Hony} S.,  {de Koter} A.,  {Justtanont} K.,  {Tielens}
  A.~G.~G.~M.,    {Waters} L.~B.~F.~M.,  2006, \aap, 456, 549

\bibitem[\protect\citeauthoryear{{Dennison}, {Simonetti} \&
  {Topasna}}{{Dennison} et~al.}{1999}]{dennison_aas_195_1999}
{Dennison} B.,  {Simonetti} J.~H.,    {Topasna} G.~A.,  1999, in American
  Astronomical Society Meeting Abstracts Vol.~31 of Bulletin of the American
  Astronomical Society, {The Virginia Tech H-alpha and [SII] Imaging Survey of
  the Northern Sky}.
p.~1455

\bibitem[\protect\citeauthoryear{{Diaz-Miller}, {Franco} \&
  {Shore}}{{Diaz-Miller} et~al.}{1998}]{diazmiller_apj_501_1998}
{Diaz-Miller} R.~I.,  {Franco} J.,    {Shore} S.~N.,  1998, \apj, 501, 192

\bibitem[\protect\citeauthoryear{{Dopita}}{{Dopita}}{1973}]{dopita_aa_29_1973}
{Dopita} M.~A.,  1973, \aap, 29, 387

\bibitem[\protect\citeauthoryear{{Dyson}}{{Dyson}}{1975}]{dyson_ass_35_1975}
{Dyson} J.~E.,  1975, \apss, 35, 299

\bibitem[\protect\citeauthoryear{{Eldridge}, {Langer} \& {Tout}}{{Eldridge}
  et~al.}{2011}]{eldridge_mnras_414_2011}
{Eldridge} J.~J.,  {Langer} N.,    {Tout} C.~A.,  2011, \mnras, 414, 3501

\bibitem[\protect\citeauthoryear{{Fletcher}, {Beck}, {Shukurov}, {Berkhuijsen}
  \& {Horellou}}{{Fletcher} et~al.}{2011}]{fletcher_mnras_412_2011}
{Fletcher} A.,  {Beck} R.,  {Shukurov} A.,  {Berkhuijsen} E.~M.,    {Horellou}
  C.,  2011, \mnras, 412, 2396

\bibitem[\protect\citeauthoryear{{Frick}, {Stepanov}, {Shukurov} \&
  {Sokoloff}}{{Frick} et~al.}{2001}]{frick_mnras_325_2001}
{Frick} P.,  {Stepanov} R.,  {Shukurov} A.,    {Sokoloff} D.,  2001, \mnras,
  325, 649

\bibitem[\protect\citeauthoryear{{Gaensler}}{{Gaensler}}{1998}]{gaensler_apj_493_1998}
{Gaensler} B.~M.,  1998, \apj, 493, 781

\bibitem[\protect\citeauthoryear{{Garcia-Segura}, {Mac Low} \&
  {Langer}}{{Garcia-Segura} et~al.}{1996}]{garciasegura_1996_aa_305}
{Garcia-Segura} G.,  {Mac Low} M.-M.,    {Langer} N.,  1996, \aap, 305, 229

\bibitem[\protect\citeauthoryear{{Gull} \& {Sofia}}{{Gull} \&
  {Sofia}}{1979}]{gull_apj_230_1979}
{Gull} T.~R.,  {Sofia} S.,  1979, \apj, 230, 782

\bibitem[\protect\citeauthoryear{{Gvaramadze} \& {Bomans}}{{Gvaramadze} \&
  {Bomans}}{2008}]{gvaramadze_aa_490_2008}
{Gvaramadze} V.~V.,  {Bomans} D.~J.,  2008, \aap, 490, 1071

\bibitem[\protect\citeauthoryear{{Heerikhuisen} \& {Pogorelov}}{{Heerikhuisen}
  \& {Pogorelov}}{2011}]{heerikhuisen_apj_738_2011}
{Heerikhuisen} J.,  {Pogorelov} N.~V.,  2011, \apj, 738, 29

\bibitem[\protect\citeauthoryear{{Heger}, {Woosley} \& {Spruit}}{{Heger}
  et~al.}{2005}]{heger_apj_626_2005}
{Heger} A.,  {Woosley} S.~E.,    {Spruit} H.~C.,  2005, \apj, 626, 350

\bibitem[\protect\citeauthoryear{{Henney}, {Arthur}, {de Colle} \&
  {Mellema}}{{Henney} et~al.}{2009}]{henney_mnras_398_2009}
{Henney} W.~J.,  {Arthur} S.~J.,  {de Colle} F.,    {Mellema} G.,  2009,
  \mnras, 398, 157

\bibitem[\protect\citeauthoryear{{Holzer} \& {Axford}}{{Holzer} \&
  {Axford}}{1970}]{holzer_araa_8_1970}
{Holzer} T.~E.,  {Axford} W.~I.,  1970, \araa, 8, 31

\bibitem[\protect\citeauthoryear{{Hoogerwerf}, {de Bruijne} \& {de
  Zeeuw}}{{Hoogerwerf} et~al.}{2001}]{hoogerwerf_aa_365_2001}
{Hoogerwerf} R.,  {de Bruijne} J.~H.~J.,    {de Zeeuw} P.~T.,  2001, \aap, 365,
  49

\bibitem[\protect\citeauthoryear{{Hummer}}{{Hummer}}{1994}]{hummer_mnras_268_1994}
{Hummer} D.~G.,  1994, \mnras, 268, 109

\bibitem[\protect\citeauthoryear{{Huthoff} \& {Kaper}}{{Huthoff} \&
  {Kaper}}{2002}]{huthoff_aa_383_2002}
{Huthoff} F.,  {Kaper} L.,  2002, \aap, 383, 999

\bibitem[\protect\citeauthoryear{{Kaneda}, {Nakagawa}, {Onaka}, {Matsumoto},
  {Murakami}, {Enya}, {Kataza}, {Matsuhara} \& {Yui}}{{Kaneda}
  et~al.}{2004}]{kaneda_2004}
{Kaneda} H.,  {Nakagawa} T.,  {Onaka} T.,  {Matsumoto} T.,  {Murakami} H.,
  {Enya} K.,  {Kataza} H.,  {Matsuhara} H.,    {Yui} Y.~Y.,  2004, in {Mather}
  J.~C.,  ed., Optical, Infrared, and Millimeter Space Telescopes Vol.~5487 of
  Society of Photo-Optical Instrumentation Engineers (SPIE) Conference Series,
  {Development of space infrared telescope for the SPICA mission}.
pp 991--1000

\bibitem[\protect\citeauthoryear{{Kaper}, {van Loon}, {Augusteijn},
  {Goudfrooij}, {Patat}, {Waters} \& {Zijlstra}}{{Kaper}
  et~al.}{1997}]{kaper_apj_475_1997}
{Kaper} L.,  {van Loon} J.~T.,  {Augusteijn} T.,  {Goudfrooij} P.,  {Patat} F.,
   {Waters} L.~B.~F.~M.,    {Zijlstra} A.~A.,  1997, \apjl, 475, L37

\bibitem[\protect\citeauthoryear{{Kudritzki}, {Pauldrach}, {Puls} \&
  {Abbott}}{{Kudritzki} et~al.}{1989}]{kudritzki_aa_219_1989}
{Kudritzki} R.~P.,  {Pauldrach} A.,  {Puls} J.,    {Abbott} D.~C.,  1989, \aap,
  219, 205

\bibitem[\protect\citeauthoryear{{Kuiper}, {Klahr}, {Beuther} \&
  {Henning}}{{Kuiper} et~al.}{2012}]{kuiper_aa_537_2012}
{Kuiper} R.,  {Klahr} H.,  {Beuther} H.,    {Henning} T.,  2012, \aap, 537,
  A122

\bibitem[\protect\citeauthoryear{{Langer}}{{Langer}}{2012}]{langer_araa_50_2012}
{Langer} N.,  2012, \araa, 50, 107

\bibitem[\protect\citeauthoryear{{Manchester}}{{Manchester}}{1987}]{manchester_aa_171_1987}
{Manchester} R.~N.,  1987, \aap, 171, 205

\bibitem[\protect\citeauthoryear{{Meyer}, {Gvaramadze}, {Langer}, {Mackey},
  {Boumis} \& {Mohamed}}{{Meyer} et~al.}{2014}]{meyer_mnras_2013}
{Meyer} D.~M.-A.,  {Gvaramadze} V.~V.,  {Langer} N.,  {Mackey} J.,  {Boumis}
  P.,    {Mohamed} S.,  2014, \mnras, 439, L41

\bibitem[\protect\citeauthoryear{{Meyer}, {Langer}, {Mackey}, {Vel{\'a}zquez}
  \& {Gusdorf}}{{Meyer} et~al.}{2015}]{meyer_mnras_450_2015}
{Meyer} D.~M.-A.,  {Langer} N.,  {Mackey} J.,  {Vel{\'a}zquez} P.~F.,
  {Gusdorf} A.,  2015, \mnras, 450, 3080

\bibitem[\protect\citeauthoryear{{Meyer}, {Mackey}, {Langer}, {Gvaramadze},
  {Mignone}, {Izzard} \& {Kaper}}{{Meyer} et~al.}{2014}]{meyer}
{Meyer} D.~M.-A.,  {Mackey} J.,  {Langer} N.,  {Gvaramadze} V.~V.,  {Mignone}
  A.,  {Izzard} R.~G.,    {Kaper} L.,  2014, \mnras, 444, 2754

\bibitem[\protect\citeauthoryear{{Mignone}, {Bodo}, {Massaglia}, {Matsakos},
  {Tesileanu}, {Zanni} \& {Ferrari}}{{Mignone}
  et~al.}{2007}]{mignone_apj_170_2007}
{Mignone} A.,  {Bodo} G.,  {Massaglia} S.,  {Matsakos} T.,  {Tesileanu} O.,
  {Zanni} C.,    {Ferrari} A.,  2007, \apjs, 170, 228

\bibitem[\protect\citeauthoryear{{Mignone}, {Zanni}, {Tzeferacos}, {van
  Straalen}, {Colella} \& {Bodo}}{{Mignone}
  et~al.}{2012}]{migmone_apjs_198_2012}
{Mignone} A.,  {Zanni} C.,  {Tzeferacos} P.,  {van Straalen} B.,  {Colella} P.,
     {Bodo} G.,  2012, \apjs, 198, 7

\bibitem[\protect\citeauthoryear{{Mohamed}, {Mackey} \& {Langer}}{{Mohamed}
  et~al.}{2012}]{mohamed_aa_541_2012}
{Mohamed} S.,  {Mackey} J.,    {Langer} N.,  2012, \aap, 541, A1

\bibitem[\protect\citeauthoryear{{Neugebauer}, {Habing}, {van Duinen},
  {Aumann}, {Baud}, {Beichman}, {Beintema}, {Boggess}, {Clegg}, {de Jong},
  {Emerson}, {Gautier}, {Gillett}, {Harris}, {Hauser}, {Houck} \&
  {Jennings}}{{Neugebauer} et~al.}{1984}]{neugebauer_278_apj_1984}
{Neugebauer} G.,  {Habing} H.~J.,  {van Duinen} R.,  {Aumann} H.~H.,  {Baud}
  B.,  {Beichman} C.~A.,  {Beintema} D.~A.,  {Boggess} N.,  {Clegg} P.~E.,  {de
  Jong} T.,  {Emerson} J.~P.,  {Gautier} T.~N.,  {Gillett} F.~C.,  {Harris} S.,
   {Hauser} M.~G.,  {Houck} J.~R.,    {Jennings} R.,  1984, \apjl, 278, L1

\bibitem[\protect\citeauthoryear{{Noriega-Crespo}, {van Buren}, {Cao} \&
  {Dgani}}{{Noriega-Crespo} et~al.}{1997}]{noriegacrespo_aj_114_1997}
{Noriega-Crespo} A.,  {van Buren} D.,  {Cao} Y.,    {Dgani} R.,  1997, \aj,
  114, 837

\bibitem[\protect\citeauthoryear{{Noriega-Crespo}, {van Buren} \&
  {Dgani}}{{Noriega-Crespo} et~al.}{1997}]{noriegacrespo_aj_113_1997}
{Noriega-Crespo} A.,  {van Buren} D.,    {Dgani} R.,  1997, \aj, 113, 780

\bibitem[\protect\citeauthoryear{{Ohno} \& {Shibata}}{{Ohno} \&
  {Shibata}}{1993}]{ohno_mnras_262_1993}
{Ohno} H.,  {Shibata} S.,  1993, \mnras, 262, 953

\bibitem[\protect\citeauthoryear{{Opher}, {Bibi}, {Toth}, {Richardson},
  {Izmodenov} \& {Gombosi}}{{Opher} et~al.}{2009}]{opher_natur_462_2009}
{Opher} M.,  {Bibi} F.~A.,  {Toth} G.,  {Richardson} J.~D.,  {Izmodenov} V.~V.,
     {Gombosi} T.~I.,  2009, \nat, 462, 1036

\bibitem[\protect\citeauthoryear{{Osterbrock} \& {Bochkarev}}{{Osterbrock} \&
  {Bochkarev}}{1989}]{osterbrock_1989}
{Osterbrock} D.~E.,  {Bochkarev} N.~G.,  1989, \sovast, 33, 694

\bibitem[\protect\citeauthoryear{{Parker}, {Phillipps}, {Pierce}, {Hartley},
  {Hambly}, {Read} \& {MacGillivray}}{{Parker}
  et~al.}{2005}]{parker_mnras_362_2005}
{Parker} Q.~A.,  {Phillipps} S.,  {Pierce} M.~J.,  {Hartley} M.,  {Hambly}
  N.~C.,  {Read} M.~A.,    {MacGillivray} 2005, \mnras, 362, 689

\bibitem[\protect\citeauthoryear{{Peri}, {Benaglia}, {Brookes}, {Stevens} \&
  {Isequilla}}{{Peri} et~al.}{2012}]{peri_aa_538_2012}
{Peri} C.~S.,  {Benaglia} P.,  {Brookes} D.~P.,  {Stevens} I.~R.,
  {Isequilla} N.~L.,  2012, \aap, 538, A108

\bibitem[\protect\citeauthoryear{{Peri}, {Benaglia} \& {Isequilla}}{{Peri}
  et~al.}{2015}]{2015arXiv150404264P}
{Peri} C.~S.,  {Benaglia} P.,    {Isequilla} N.~L.,  2015, \aap, 578, A45

\bibitem[\protect\citeauthoryear{{Peters}, {Banerjee}, {Klessen}, {Mac Low},
  {Galv{\'a}n-Madrid} \& {Keto}}{{Peters} et~al.}{2010}]{peters_apj_711_2010}
{Peters} T.,  {Banerjee} R.,  {Klessen} R.~S.,  {Mac Low} M.-M.,
  {Galv{\'a}n-Madrid} R.,    {Keto} E.~R.,  2010, \apj, 711, 1017

\bibitem[\protect\citeauthoryear{{Petrovic}, {Langer}, {Yoon} \&
  {Heger}}{{Petrovic} et~al.}{2005}]{petrovic_aa_435_2005}
{Petrovic} J.,  {Langer} N.,  {Yoon} S.-C.,    {Heger} A.,  2005, \aap, 435,
  247

\bibitem[\protect\citeauthoryear{{Petrovic}, {Pols} \& {Langer}}{{Petrovic}
  et~al.}{2006}]{petrovic_aa_450_2006}
{Petrovic} J.,  {Pols} O.,    {Langer} N.,  2006, \aap, 450, 219

\bibitem[\protect\citeauthoryear{{Rand} \& {Kulkarni}}{{Rand} \&
  {Kulkarni}}{1989}]{rand_apj_343_1989}
{Rand} R.~J.,  {Kulkarni} S.~R.,  1989, \apj, 343, 760

\bibitem[\protect\citeauthoryear{{Reynolds}, {Tufte}, {Haffner}, {Jaehnig} \&
  {Percival}}{{Reynolds} et~al.}{1998}]{reynolds_pasa_15_1998}
{Reynolds} R.~J.,  {Tufte} S.~L.,  {Haffner} L.~M.,  {Jaehnig} K.,
  {Percival} J.~W.,  1998, \pasa, 15, 14

\bibitem[\protect\citeauthoryear{{Rozyczka}, {Tenorio-Tagle}, {Franco} \&
  {Bodenheimer}}{{Rozyczka} et~al.}{1993}]{rozyczka_mnras_261_1993}
{Rozyczka} M.,  {Tenorio-Tagle} G.,  {Franco} J.,    {Bodenheimer} P.,  1993,
  \mnras, 261, 674

\bibitem[\protect\citeauthoryear{{Shabala}, {Mead} \& {Alexander}}{{Shabala}
  et~al.}{2010}]{shabala_mnras_405_2010}
{Shabala} S.~S.,  {Mead} J.~M.~G.,    {Alexander} P.,  2010, \mnras, 405, 1960

\bibitem[\protect\citeauthoryear{{Spitzer}}{{Spitzer}}{1962}]{spitzer_1962}
{Spitzer} L.,  1962, {Physics of Fully Ionized Gases}

\bibitem[\protect\citeauthoryear{{Spitzer}}{{Spitzer}}{1978}]{spitzer_1978}
{Spitzer} L.,  1978, {Physical processes in the interstellar medium}

\bibitem[\protect\citeauthoryear{{Swinyard}, {Rieke}, {Ressler}, {Glasse},
  {Wright}, {Ferlet} \& {Wells}}{{Swinyard} et~al.}{2004}]{swinyard_2004}
{Swinyard} B.~M.,  {Rieke} G.~H.,  {Ressler} M.,  {Glasse} A.,  {Wright} G.~S.,
   {Ferlet} M.,    {Wells} M.,  2004, in {Mather} J.~C.,  ed., Optical,
  Infrared, and Millimeter Space Telescopes Vol.~5487 of Society of
  Photo-Optical Instrumentation Engineers (SPIE) Conference Series,
  {Sensitivity estimates for the mid-infrared instrument (MIRI) on the JWST}.
pp 785--793

\bibitem[\protect\citeauthoryear{{Thun}, {Kuiper}, {Schmidt} \& {Kley}}{{Thun}
  et~al.}{2016}]{2016arXiv160107799T}
{Thun} D.,  {Kuiper} R.,  {Schmidt} F.,    {Kley} W.,  2016, ArXiv
  e-prints:1601.07799

\bibitem[\protect\citeauthoryear{{Toledo-Roy}, {Esquivel}, {Vel{\'a}zquez} \&
  {Reynoso}}{{Toledo-Roy} et~al.}{2014}]{toledoray_mnras_442_2014}
{Toledo-Roy} J.~C.,  {Esquivel} A.,  {Vel{\'a}zquez} P.~F.,    {Reynoso} E.~M.,
   2014, \mnras, 442, 229

\bibitem[\protect\citeauthoryear{{Vall{\'e}e}}{{Vall{\'e}e}}{2011}]{vallee_newar_55_2011}
{Vall{\'e}e} J.~P.,  2011, \nar, 55, 23

\bibitem[\protect\citeauthoryear{{van Buren} \& {McCray}}{{van Buren} \&
  {McCray}}{1988}]{buren_apj_329_1988}
{van Buren} D.,  {McCray} R.,  1988, \apjl, 329, L93

\bibitem[\protect\citeauthoryear{{van Buren}, {Noriega-Crespo} \& {Dgani}}{{van
  Buren} et~al.}{1995}]{vanburen_aj_110_1995}
{van Buren} D.,  {Noriega-Crespo} A.,    {Dgani} R.,  1995, \aj, 110, 2914

\bibitem[\protect\citeauthoryear{{van Marle}, {Decin} \& {Meliani}}{{van Marle}
  et~al.}{2014}]{vanmarle_aa_561_2014}
{van Marle} A.~J.,  {Decin} L.,    {Meliani} Z.,  2014, \aap, 561, A152

\bibitem[\protect\citeauthoryear{{van Marle}, {Langer} \&
  {Garc{\'{\i}}a-Segura}}{{van Marle} et~al.}{2007}]{vanmarle_aa_469_2007}
{van Marle} A.~J.,  {Langer} N.,    {Garc{\'{\i}}a-Segura} G.,  2007, \aap,
  469, 941

\bibitem[\protect\citeauthoryear{{van Marle}, {Meliani}, {Keppens} \&
  {Decin}}{{van Marle} et~al.}{2011}]{vanmarle_apj_734_2011}
{van Marle} A.~J.,  {Meliani} Z.,  {Keppens} R.,    {Decin} L.,  2011, \apjl,
  734, L26

\bibitem[\protect\citeauthoryear{{van Marle}, {Meliani} \& {Marcowith}}{{van
  Marle} et~al.}{2015}]{vanmarle_2015}
{van Marle} A.~J.,  {Meliani} Z.,    {Marcowith} A.,  2015, \aap, 584, A49

\bibitem[\protect\citeauthoryear{van Veelen}{van Veelen}{2010}]{vanveelen_phd}
van Veelen B.,  2010, PhD thesis, Utrecht University, Netherland

\bibitem[\protect\citeauthoryear{{Vel{\'a}zquez}, {Vigh}, {Reynoso},
  {G{\'o}mez} \& {Schneiter}}{{Vel{\'a}zquez}
  et~al.}{2006}]{velazquez_apj_649_2006}
{Vel{\'a}zquez} P.~F.,  {Vigh} C.~D.,  {Reynoso} E.~M.,  {G{\'o}mez} D.~O.,
  {Schneiter} E.~M.,  2006, \apj, 649, 779

\bibitem[\protect\citeauthoryear{{Villaver}, {Manchado} \&
  {Garc{\'{\i}}a-Segura}}{{Villaver} et~al.}{2012}]{villaver_apj_748_2012}
{Villaver} E.,  {Manchado} A.,    {Garc{\'{\i}}a-Segura} G.,  2012, \apj, 748,
  94

\bibitem[\protect\citeauthoryear{{Vishniac}}{{Vishniac}}{1994}]{vishniac_apj_428_1994}
{Vishniac} E.~T.,  1994, \apj, 428, 186

\bibitem[\protect\citeauthoryear{{Weaver}, {McCray}, {Castor}, {Shapiro} \&
  {Moore}}{{Weaver} et~al.}{1977}]{weaver_apj_218_1977}
{Weaver} R.,  {McCray} R.,  {Castor} J.,  {Shapiro} P.,    {Moore} R.,  1977,
  \apj, 218, 377

\bibitem[\protect\citeauthoryear{{Wiersma}, {Schaye} \& {Smith}}{{Wiersma}
  et~al.}{2009}]{wiersma_mnras_393_2009}
{Wiersma} R.~P.~C.,  {Schaye} J.,    {Smith} B.~D.,  2009, \mnras, 393, 99

\bibitem[\protect\citeauthoryear{{Wilkin}}{{Wilkin}}{1996}]{wilkin_459_apj_1996}
{Wilkin} F.~P.,  1996, \apjl, 459, L31

\bibitem[\protect\citeauthoryear{{Woosley}, {Heger} \& {Weaver}}{{Woosley}
  et~al.}{2002}]{woosley_rvmp_74_2002}
{Woosley} S.~E.,  {Heger} A.,    {Weaver} T.~A.,  2002, Reviews of Modern
  Physics, 74, 1015

\bibitem[\protect\citeauthoryear{{Wright}, {Eisenhardt}, {Mainzer}, {Ressler},
  {Cutri}, {Jarrett}, {Kirkpatrick}, {Padgett}, {McMillan}, {Skrutskie} \&
  {Stanford}}{{Wright} et~al.}{2010}]{wright_aj_140_2010}
{Wright} E.~L.,  {Eisenhardt} P.~R.~M.,  {Mainzer} A.~K.,  {Ressler} M.~E.,
  {Cutri} R.~M.,  {Jarrett} T.,  {Kirkpatrick} J.~D.,  {Padgett} D.,
  {McMillan} R.~S.,  {Skrutskie} M.,    {Stanford} S.,  2010, \aj, 140, 1868

\bibitem[\protect\citeauthoryear{{Yoon} \& {Langer}}{{Yoon} \&
  {Langer}}{2005}]{yoon_443_aa_2005}
{Yoon} S.-C.,  {Langer} N.,  2005, \aap, 443, 643

\end{thebibliography}
}


\end{document}